\documentclass[12pt,showkeys]{report}

\usepackage[francais]{babel}
\usepackage[latin1]{inputenc} 
\usepackage[T1]{fontenc} 

\usepackage{graphicx}
\usepackage{latexsym}
\usepackage{amsfonts}
\usepackage{amsmath}
\usepackage{comment}

\newcommand{\be}{\begin{equation}}
\newcommand{\ee}{\end{equation}}
\newcommand{\ba}{\begin{eqnarray}}
\newcommand{\ea}{\end{eqnarray}}
\newcommand{\pl}{\left\{}
\newcommand{\pr}{\right\}}
\newcommand{\al}{\left|}
\newcommand{\ar}{\right|}
\newcommand{\rr}{\right)}
\newcommand{\rl}{\left(}

\begin{document}
\begin{titlepage}


\includegraphics[width=30mm]{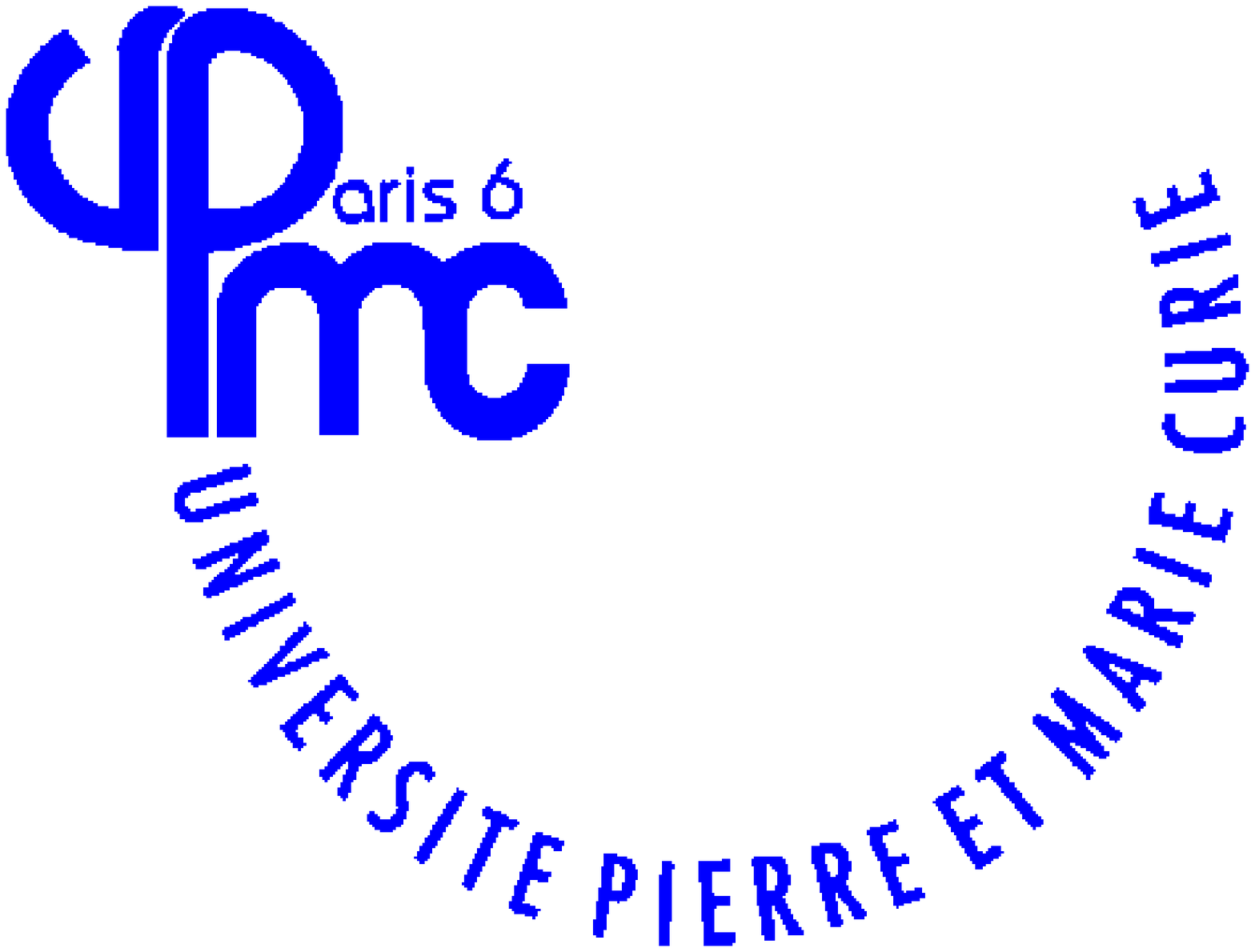}
\hfill
\includegraphics[width=20mm]{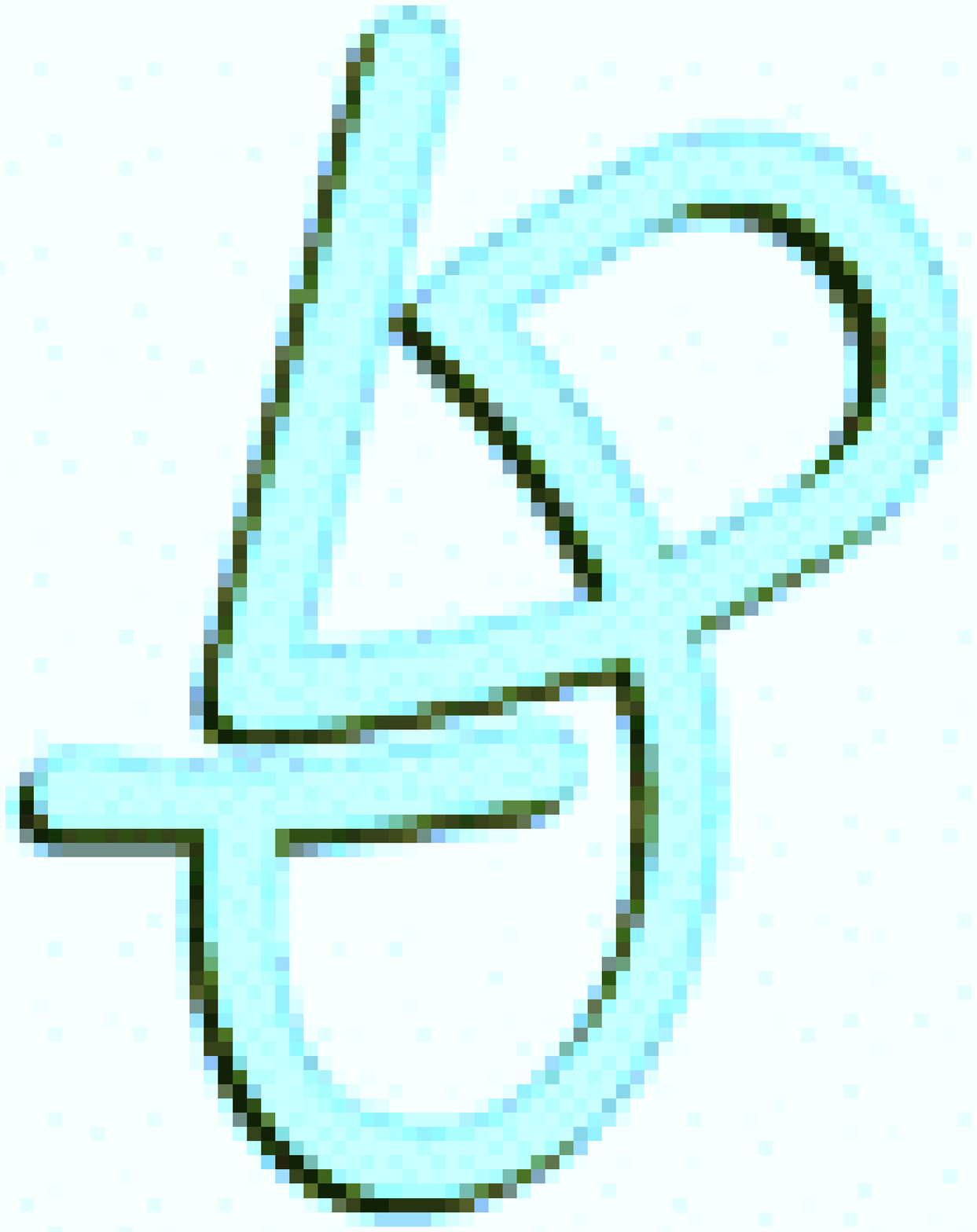}

\begin{center}
\fbox{{\bf UNIVERSITÉ PIERRE ET MARIE CURIE PARIS VI}}

\vskip1cm

{\bf THÈSE}
\vskip1cm

Sp\'ecialit\'e: {\bf  PHYSIQUE TH\'EORIQUE}\\

Pr\'esentée\\
 pour obtenir le grade de
\vskip0.5cm
\large {\bf Docteur de l'Université Paris VI}

\vskip0.75cm

par

\vskip0.75cm

{\sc \bf Maria-Cristina Timirgaziu}

\vskip0.5cm

Sujet: \\

\Large{\bf Aspects des th\'eories des cordes non-supersym\'etriques ou avec brisure de supersym\'etrie}

\end{center}

Soutenue le 17 d\'ecembre 2004 devant la commission d'examen:\\

$\begin{array}{lll}
\mbox{MM}.& \mbox{Pierre Bin\'etruy}, &\\
 &    \mbox{Philippe Brax}, & \mbox{rapporteur},\\
  &  \mbox{Emilian Dudas}, & \mbox{directeur de th\`ese},\\
   & \mbox{Michael Joyce}, & \\
  &  \mbox{Jihad Mourad}, & \mbox{directeur de th\`ese},\\
   & \mbox{Augusto Sagnotti}, & \mbox{rapporteur}.
   \end{array}$

\end{titlepage}

\newpage
$\ $
\thispagestyle{empty}
\newpage

\tableofcontents

\chapter*{Introduction}
\addcontentsline{toc}{chapter}{Introduction}

 Un des principaux d\'efis pour les physiciens th\'eoriciens d'aujourd'hui est la description quantique de la gravitation. Toutes les autres interactions, \'electromagn\'etique, faible et forte sont d\'ecrites par le Mod\`ele Standard, bas\'e sur la th\'eorie quantique des champs et qui est verifi\'e exp\'erimentalement avec une tr\`es bonne pr\'ecision. La tentative de description de la gravitation par une th\'eorie quantique des champs se solde par un \'echec, la th\'eorie r\'esultante \'etant perturbativement nonrenormalizable. 
 
 La th\'eorie des cordes est un des candidats actuels \`a la description quantique de la gravitation. Elle postule que les objets fondamentaux ne sont pas des particules ponctuelles comme dans la th\'eorie des champs, mais des cordes, c'est \`a dire des objets \`a une dimension. La longueur des cordes doit \^etre suffisamment petite pour expliquer le fait qu'on ne les observe pas aux \'energies actuelles. L'extension spatiale des cordes permet la r\'egularisation des interactions gravitationnelles, d'une mani\`ere similaire \`a la r\'egularisation de la th\'eorie de Fermi par l'introduction de l'\'echange des bosons W et Z. 
 
  En th\'eorie des cordes les particules \'el\'ementaires du Mod\`ele Standard sont obtenues comme des diff\'erents  \'etats d'excitation des cordes. Les seuls processus d'interaction des cordes sont celui de fusion de deux cordes dans une corde et celui de s\'eparation d'une corde en deux cordes. Ils contiennent toutes les interactions du Mod\`ele Standard plus la gravitation. La th\'eorie des cordes ne se contente pas d'offrir une description quantique de la gravitation, mais elle a comme ambition l'unification de toutes les interactions dans une seule th\'eorie. Le seul param\`etre de la th\'eorie des cordes est l'\'echelle de masse, $M_S$( ou l'\'echelle de longueur, $l_S$). La th\'eorie des cordes pourraient expliquer les valeurs des masses des particules \'el\'ementaires, qui apparaissent  dans le  Mod\`ele Standard comme des param\`etres sans une justification th\'eorique. D'autre part, il n'est pas facile d'obtenir des pr\'edictions \`a basse \'energie de la th\'eorie des cordes. La coh\'erence quantique de la th\'eorie des cordes impose que la dimension de l'espace-temps soit dix. Mais nous percevons un espace-temps quadri-dimensionel, ce qui implique que les six dimensions suppl\'ementaires doivent \^etre compactifi\'ees \`a une \'echelle assez petite pour qu'elles soient invisibles aux \'energies atteintes actuellement. Le nombre tr\`es grand des compactifications possibles rend difficile la question du contact avec la r\'ealit\'e. 
  
  Historiquement la th\'eorie des cordes est apparue vers la fin des ann\'ees 1960 comme une possible description des int\'eractions fortes\cite{veneziano}. Mais tr\`es vite elle a \'et\'e abandonn\'ee au profit de la chromodynamique quantique. Un des d\'efauts de la th\'eories des cordes en tant que th\'eorie des interactions fortes \'etait l'existence d'un \'etat de masse nulle et de spin 2 dans le spectre, d\'ecrivant une particule sans \'equivalent dans le spectre des hadrons. 
  
  En 1974 Scherk et Schwarz proposent d'interpr\'eter cette particule comme le graviton et la th\'eorie des cordes devient un candidat \`a l'unification des interactions. Mais les mod\`eles \`a int\'er\^et ph\'enom\'enologique, comme la th\'eorie de type I, pr\'esentaient des anomalies gravitationnelles et de jauge. 
  
  Dix ans plus tard Green et Schwarz proposent un m\'ecanisme\cite{gs} d'annulation des anomalies qui impose que le groupe de jauge de la th\'eorie de type I soit SO(32). Une deuxi\`eme possibilit\'e pour le groupe de jauge qui permet l'annulation des anomalies est $E_8\times E_8$, mais elle est incompatible avec la th\'eorie des cordes ouvertes. Gross, Harvey, Martinec and Rohm montrent une ann\'ee plus tard que ce groupe de jauge, ainsi que $SO(32)$ peuvent \^etre r\'ealis\'es dans une th\'eorie de cordes ferm\'ees, la corde h\'et\'etrotique\cite{heterotique}. 
  
  Une dizaine d'ann\'ees plus tard l'introductions des D-branes\cite{branes} par Polchinski a rendu possible la conjecture des dualit\'es entre diff\'erentes th\'eories des cordes qui apparaissent comme des diff\'erentes r\'egions dans l'espace des modules d'une th\'eorie unique, la M-th\'eorie, ouvrant la voie \`a une formulation nonperturbative de la th\'eorie des cordes. 
  
  Un des d\'efis actuels de la th\'eorie des cordes est d'obtenir des pr\'edictions \`a basse \'energie qui pourraient valider la th\'eorie des cordes comme une description de la nature. En particulier il est n\'ecessaire d'obtenir le Mod\`ele Standard comme limite de basse \'energie de la th\'eorie des cordes et expliquer pourquoi la nature a choisi cette compactification parmi toutes les possibilit\'es. Les mod\`eles actuels pr\'esentent des caract\'eristiques prometteuses, comme la brisure de la supersym\'etrie, la pr\'esence des fermions chiraux, le bon nombre de g\'en\'erations. Mais ces mod\`eles restent \`a am\'eliorer. 
  
   R\'ecemment une autre voie a \'et\'e explor\'ee, les pr\'edictions en cosmologie de la th\'eorie des cordes. La th\'eorie des cordes est un cadre naturel pour formuler des questions li\'ees \`a la cosmologie, comme le probl\`eme de la constante cosmologique ou la singularit\'e du Big Bang. Les dix derni\`eres  ann\'ees des mod\`eles explicites ont \'et\'e propos\'es qui pourraient r\'esoudre certains des probl\`emes de la cosmologie standard, comme le mod\`ele du pre-Big Bang, l'univers ekpyrotic et l'inflation des branes. 
   
   La supersym\'etrie appara\^it de mani\`ere naturelle en th\'eorie des cordes et elle permet d'\'eliminer le tachyon pr\'esent dans le spectre des cordes bosoniques. Les th\'eories des cordes supersym\'etriques sont bien connues, mais pour que la th\'eorie des cordes soit une extension du Mod\`ele Standard il faut que la supersym\'etrie soit bris\'ee. G\'en\'eralement la brisure de supersym\'etrie reintroduit des tachyons dans le spectre, toutefois il existe des mod\`eles de cordes nonsupersym\'etriques sans tachyons et qui pr\'esentent toutes les propri\'et\'es int\'eressantes des cordes supersym\'etriques, comme l'absence d'anomalies, la pr\'esence des fermions chiraux et la possibilit\'e d'avoir un groupe de jauge int\'eressant.
   
   Les cordes nonsupersym\'etriques sont particuli\`erement int\'eressantes pour la cosmologie, car la brisure de supersym\'etrie entra\^ine une red\'efinition du vide de la th\'eorie, l'espace de Minkowski n'\'etant plus une solution. Les solutions des cordes nonsupersym\'etriques peuvent d\'ependre du temps et d\'ecrire, donc,  une \'evolution cosmologique. De plus, la stabilisation des modules des cordes, n\'ecessaire pour avoir une cosmologie statisfaisante, est, au moins partiellement, li\'ee \`a la brisure de la supersym\'etrie. 
   Le probl\`eme principal des solutions d\'ependantes du temps est leur stabilit\'e. 
  
  {\bf Plan de la th\`ese.} Le manuscrit de cette th\`ese est organis\'e en 
  cinq
 parties:
  
  \begin{description}
\item[$\bullet$ ] Le premier chapitre est une introduction g\'en\'erale \`a la th\'eorie des cordes bosoniques et fermioniques, aux interactions des cordes, ainsi qu'\`a la notion de $D$-brane. Les dualit\'es entre les diff\'erentes th\'eories des cordes sont pr\'esent\'ees et une discussion des annulation des anomalies cl\^ot le chapitre.

\item[$\bullet$ ] Le deuxi\`eme chapitre traite de la construction des orientifolds et orbifolds. 

\item[$\bullet$ ] Le troisi\`eme chapitre presente le m\'ecanisme de Scherk-Schwarz de brisure de supersym\'etrie.

\item[$\bullet$ ]Le quatri\`eme chapitre pr\'esente d'abord une introduction au mod\`ele standard de la cosmologie et aux probl\`emes non-r\'esolus de ce mod\`ele. Ensuite l'inflation est introduite comme une possible solution aux probl\`emes de la cosmologie standard.

\item[$\bullet$ ]Le dernier chapitre est d\'edi\'e aux alternatives de l'inflation qui s'inspirent de la th\'eorie des cordes et aux solutions d\'ependantes du temps des th\'eories nonsupersym\'etriques des cordes.

\end{description}


\chapter{Notions introductives}
\section{La corde bosonique}
\subsection{La corde bosonique classique}

Une corde est un objet unidimensionel. En mouvement elle d\'ecrit une surface d'univers $\footnote{L'\'equivalent de la ligne d'univers pour une particule ponctuelle.}$ parametris\'ee par deux coordonn\'ees: une coordonn\'ee de type temps, $\tau$,
et la coordonn\'ee qui d\'ecrit la corde, $\sigma$. La surface d'univers d'une codre bosonique , plong\'ee dans un espace-temps D dimensionel, est d\'ecrite par D fonctions scalaires $X^\mu(\tau,\sigma), \mu=0...D-1$. L'action la plus simple, independante de la parametrisation, qui d\'ecrit un tel objet est proportionelle \`a l'aire de la surface d'univers et s'appelle l'action de Nambu-Goto\cite{nambu}\cite{goto}$\footnote{L'analogue de l'action proportionelle au temps propre le long de la ligne d'univers pour une particule ponctuelle.}$ :

\be S_{NG} = -{1\over 2 \pi \alpha '}\int_M d^2\sigma \sqrt{-\gamma},  \ee
o\` u $\gamma$ est le determinant de la m\'etrique induite sur la surface d'univers, $\gamma_{ab}=\partial_a X^\mu \partial_b X^\nu g_{\mu\nu}$, $M$ est la surface de l'univers, $\sigma^ a =(\tau,\sigma)$, $\partial_a={\partial \over \partial \sigma ^ a}$ et $g_{\mu\nu}$ $\footnote{Par la suite on va considerer la propagation des cordes dans l'espace-temps de Minkowski, donc $g_{\mu\nu}=\eta_{\mu\nu}$.}$ est la m\'etrique de l'espace-temps. La constante $\alpha '$, appell\' ee  param\` etre de Regge, est reli\'ee \`a la tension, $T$, de la corde :

\be T= {1 \over 2 \pi \alpha '}.\ee

 En plus de l'invariance sous les reparametrisations,  $X'^\mu(\tau',\sigma')=X^\mu(\tau,\sigma)$, l'action de Nambu-Goto est invariante sous le groupe de Poincar\'e \`a D dimensions :
 
 \be X'^\mu(\tau,\sigma)=\Lambda^\mu_\nu X^\nu(\tau,\sigma)+a^\mu,\ee
 o\` u $\Lambda^\mu_\nu$ est une transformation de Lorentz et $a^\mu$ une translation.

   La pr\'esence de la racine carr\'e rend la quantification difficile, mais l'action de Nambu-Goto peut \^ etre simplifi\'ee en introduisant une m\'etrique independante, $h_{ab}(\tau,\sigma)$, sur la surface d'univers. La nouvelle action qu'on obtient de cette mani\`ere s'appelle l'action de Polyakov$\footnote{Bien quelle ait \'et\'e trouv\'ee pour la premi\`ere fois par Brink, Di Vecchia, Howe, Deser et Zumino , c'est Polyakov qui a mis en \'evidence son utilit\'e pour la quantification.}$ :
 
 \be S_P= -{1\over 4 \pi \alpha '} \int_M d^2\sigma (-h)^{1\over 2}h^{ab}\partial_a X^\mu \partial_b X_\mu\label{actionpolyakov}. \ee

 Bien \'evidemment l'action de Polyakov est classiquement \'equivalente \`a celle de Nambu-Goto et pour s'en convaincre il suffit d'\'eliminer la m\'etrique $h_{ab}$ de l'action de Polyakov \`a l'aide de son equation de mouvement :
 
 \be \delta_h S_P=0 \Rightarrow \gamma_{ab}={1\over 2}h_{ab}h^{cd}\gamma_{cd}\label{eqh}.\ee 
 
  L'equation ($\ref{eqh}$) se r\'e\'ecrit :
  \be h_{ab}(-h)^{-1/ 2}=\gamma_{ab}(-\gamma)^{-1/2}\label{proph},\ee
  ce qui implique que $h_{ab}\propto \gamma_{ab}$ et l'\'equivalence des deux actions s'en suit. L'\'equation ($\ref{proph}$) determine $h_{ab}$ \` a une transformation de Weyl pr\`es ce qui fait que l'action de Polyakov poss\`ede une invariance de plus par rapport \`a l'action de Nambu-Goto. En r\'esum\'e les sym\'etries classiques de l'action de Polyakov sont$\footnote{Il y a un autre terme qui est compatible avec les sym\'etries de l'action de Polyakov, ${1\over 4\pi} \int d^2\sigma \sqrt{-h}R$, avec $R$ le scalaire de Ricci de la m\'etrique $h_{ab}$. Ce terme joue un r\^ole important dans les interactions des cordes, mais ne contribue pas aux \'equations de mouvement, car en 2 dimensions $\sqrt{-h}R $ est une d\'eriv\'e totale.}$ :
  
  \begin{itemize}
 
  \item[i.)]{l'invariance sous le groupe de Poincar\'e D dimensionel :
  $$X'^\mu(\tau,\sigma)=\Lambda^\mu_\nu X^\nu(\tau,\sigma)+a^\mu,$$
  $$h'_{ab}(\tau,\sigma)=h_{ab}(\tau,\sigma). $$ }
  
  \item[ii.)]{l'invariance sous les diff\'eomorphismes(ou reparam\'etrisations) :
  $$X'^\mu(\tau',\sigma')=X^\mu(\tau,\sigma),$$
$${\partial\sigma'^c\over\partial\sigma^a}{\partial\sigma'^d\over\partial\sigma^b} h'_{cd}(\tau',\sigma')=h_{ab}(\tau,\sigma).$$ }
  
  \item[iii.)]{l'invariance de Weyl:
  $$X'^\mu(\tau,\sigma)=X^\mu(\tau,\sigma),$$
  $$h'_{ab}(\tau,\sigma)=\mbox{exp}(2\omega(\tau,\sigma))h_{ab}(\tau,\sigma). $$  
    }
  \end{itemize}
  
  La variation de l'action par rapport \`a $h_{ab}$ d\'efinit le tenseur \'energie-impulsion :
  \be T^{ab}=-4\pi(-\gamma)^{-1/2}{\delta\over \delta h_{ab}}S_P,\ee 
  et, par cons\'equent, l'\'equation de mouvement de $h_{ab}$ s'\'ecrit :
  
  \be T_{ab}={1\over \alpha'}(\partial_a X^\mu \partial_b X_\mu - {1\over 2}h_{ab}  \partial^c X^\mu \partial_c X_\mu )  =0.\ee
  
  L'invariance sous les transformations de Weyl implique l'annulation de la trace du tenseur \'energie-impulsion, $T^a_a=0$, et l'invariance sous les reparam\'etrisations a comme cons\'equence sa conservation: $\nabla_aT^{ab}=0$. 
  
  Int\'eressons-nous maintenant aux \'equations du mouvement des champs $X^\mu$. En variant l'action de Polyakov par rapport \`a $X^\mu$ on obtient :
  
  \ba \delta S_P&=&{1\over 2\pi \alpha'}\int_{-\infty}^{\infty}d\tau\int_0^l d\sigma (-h)^{1/2}\nabla ^2X_\mu \delta X^\mu \nonumber\\&-&{1\over 2\pi \alpha'}\int_{-\infty}^{\infty}d\tau  (-h)^{1/2} \delta X^\mu \partial^\sigma X_\mu  \mid^{\sigma=l}_{\sigma=0},
  \ea
  o\`u $l$ est la longueur de la corde. Le terme de bord s'annule dans les situations suivantes:
  
  \begin{itemize}
   \item[1.] $X^\mu(\tau,0) =X^\mu(\tau,l), \partial^\sigma X^\mu(\tau,0)=\partial^\sigma X^\mu(\tau,l), h_{ab}(\tau,0)=h_{ab}(\tau,l) $. \\ Ces conditons de p\'eriodicit\'e d\'efinissent la corde ferm\'ee. 
     \item[2.] $\partial^\sigma X^\mu(\tau,0)=\partial^\sigma X^\mu(\tau,l)=0$. \\ Ce sont ce qu'on appelle les conditions de Neumann et elles d\'efinissent   une corde ouverte avec les extr\'emit\'es libres.\\
  \item[3.] $\delta X^\mu(\tau,0)=\delta X^\mu(\tau,l)=0$.\\
     Ces conditions, appel\'ees conditions de Dirichlet, brisent l'invariance de Poincar\'e. Elles d\'efinissent une corde ouverte avec les extr\'emit\'es fix\'ees.
     \item[4.] On peut \'egalement choisir une condition de Dirichlet pour une extr\'emi\'e de la corde ouverte et une condition de Neumann pour l'autre extr\'emit\'e. Par exemple: $\partial^\sigma X^\mu(\tau,0)= 0$ et  $\delta X^\mu(\tau,l)=0$.
        \end{itemize}

    L'\'equation de mouvement prend alors la forme :
    
    \be \nabla ^2 X^\mu =0.\ee   
    
    Pour simplifier les \'equations du mouvement on peut faire un choix de jauge convenable. L'invariance sous les reparam\'etrisations permet de fixer 2 des 3 composantes de la m\'etrique bi-dimensionelle $h_{ab}$. La m\'etrique peut alors \^etre mise sous la forme :
    
    \be h_{ab}=e^{2\Lambda}\eta_{ab}\label{conforme}. \ee
    
    Dans cette jauge, appell\'ee jauge conforme, l'action devient :
    
    \be S=-{1\over 4\pi \alpha'}\int d^2\sigma \eta^{ab}\partial_a X^\mu \partial_b X_\mu \ee
    et l'\'equation de mouvement des champs $X^\mu $ devient l'\'equation d'onde \`a 2 dimensions :
    \be \Box X^\mu=(\partial^2_\sigma-\partial^2_\tau)X^\mu=0\label{eqonde}.\ee
    
     Les contraintes $T_{ab}=0$ prennent aussi une forme tr\`es simple $\footnote{Nous avons utilis\'e les notations $'=\partial_\sigma, \dot{}=\partial_\tau$ et $X^2=X^\mu X_\mu $.}$ :     
     \ba T_{10}=T_{01}={1\over \alpha'}\dot{X}\cdot X'=0,\nonumber\\
     \\
     T_{00}=T_{11}={1\over 2\alpha'}(\dot{X}^2+X'^2)=0, \ea
      ou exprim\'es autrement ${1\over 2}(\dot{X}\pm X')^2=0$.

     La solution g\'en\'erale de l'\'equation (\ref{eqonde}) peut \^etre \'ecrite comme la somme de deux fonctions arbitraires :
     
     \be X^\mu(\tau,\sigma)=X^\mu_L(\sigma^+)+X^\mu_R(\sigma^-),\ee
    avec $\sigma^+=\tau+\sigma, \sigma^-=\tau-\sigma$, les coordonn\'ees du c\^one de lumi\`ere sur la surface d'univers. $X^\mu_L$ d\'ecrit les modes qui se propagent \`a gauche ($L$ pour "left") et $X^\mu_R$ les modes qui se propagent \`a droite ($R$ pour "right"). 
     
     Dans les coordonn\'ees du c\^one de lumi\`ere les contraintes se r\'e\'ecrivent :
     
     \ba &&T_{++}={1\over \alpha'}\partial_+X\cdot \partial_+X=\dot{X_L}^2=0,\nonumber \\
     &&T_{--}={1\over \alpha'}\partial_-X\cdot \partial_-X=\dot{X_R}^2=0,\nonumber \\
     &&T_{+-}=T_{-+}=0. \label{contraintes}
     \ea

     La solution g\'en\'erale de l'\'equation de mouvement (\ref{eqonde}) en tenant compte des conditions de p\'eriodicit\'e pour les cordes ferm\'ees, $X^\mu(\tau,\sigma)= X^\mu(\tau,\sigma + 2\pi)$   
   est :
   \ba X^\mu_R={1\over 2}x^\mu+\alpha'p^\mu(\tau-\sigma)+i\sqrt{{\alpha'\over 2}}\sum_{n\neq 0}{1\over n}\alpha_n^\mu e^{-2in(\tau-\sigma)},\nonumber\\  
     X^\mu_L={1\over 2}x^\mu+\alpha'p^\mu(\tau+\sigma)+i\sqrt{{\alpha'\over 2}}\sum_{n\neq 0}{1\over n}\bar{\alpha}_n^\mu e^{-2in(\tau+\sigma)}, \label{xlxr}\ea
     o\`u   $\alpha_n^\mu $ sont les modes de Fourier, $p_\mu$ est la quantit\'e de mouvement du centre de masse de la corde et $x^\mu$ la position du centre de masse. $X^\mu$ \'etant des fonctions r\'eelles il en resulte que $x^\mu$ et $p^\mu$ sont r\'eelles et \'egalement :
      
      \be \alpha_{-n}^\mu=(\alpha_n^\mu)^\dag \quad,\quad \bar{\alpha}_{-n}^\mu=(\bar{\alpha}_n^\mu)^\dag.\ee
    
    Pour une corde ouverte de longueur $\pi$ la solution de l'\'equation du mouvement avec les conditions aux bords $X'^\mu\mid_{\sigma=0,\pi}=0$ est :
    
    \be X^\mu=x^\mu+2\alpha'p^\mu   \tau + i\sqrt{2\alpha'}\sum_{n\neq 0}\mbox{cos}(n\sigma){\alpha_n^\mu \over n}e^{-in\tau}\label{open}.\ee
    
    Par la suite les contraintes (\ref{contraintes}) doivent \^etre impos\'ees aux solutions des \'equations de mouvement. Leurs modes de Fourier, appell\'es operateurs de Virasoro, sont pour la corde ferm\'ee :
    
    \ba &&L_m={T\over 2}\int_0^\pi e^{-2im\sigma}\dot{X}_R^2\ d\sigma={1\over 2}\sum_{n=-\infty}^\infty
\alpha_{m-n}\cdot\alpha_n,\nonumber\\
&&\bar{L}_m={T\over 2}\int_0^\pi e^{2im\sigma}\dot{X}_L^2\ d\sigma={1\over 2}\sum_{n=-\infty}^\infty
\bar{\alpha}_{m-n}\cdot\bar{\alpha}_n,\ea  
    avec $\alpha^\mu_0=\bar{\alpha}^\mu_0=\sqrt{{\alpha'\over 2}}p^\mu$.
    Le syst\`eme  doit respecter les contraintes : $ L_n=0, \bar{L}_n=0, \forall n \in Z$.
    En particulier les contraintes \ba L_0={1\over2}\alpha_0^2+\sum_{n=1}^\infty
\alpha_{-n}\cdot\alpha_n=0, \nonumber\\\bar{L}_0={1\over2}\alpha_0^2+\sum_{n=1}^\infty\bar{\alpha}_{-n}\cdot\bar{\alpha}_n=0,\ea permettent de trouver l'expression de la masse d'un mode de la corde :

\be M^2=-p^\mu p_\mu={1\over 2}(M_L^2+M_R^2)={2\over \alpha'}\sum_{n=1}^\infty
(\alpha_{-n}\cdot\alpha_n+\bar{\alpha}_{-n}\cdot\bar{\alpha}_n).\ee
    
       Pour les cordes ouvertes il y a un seul type d'op\'erateurs de Virasoro d\'efinis par :   
       
       \be L_m=T\int_0^\pi(e^{im\sigma}T_{++}+e^{-im\sigma}T_{--})d\sigma ={T\over 4}\int_{-\pi}^\pi e^{im\sigma}(\dot{X}+X')^2d\sigma={1\over 2}\sum_{n=-\infty}^\infty
\alpha_{m-n}\cdot\alpha_n,\ee   
o\`u $\alpha^\mu_0=\sqrt{2\alpha'}p^\mu$. Les contraintes de Virasoro sont dans ce cas $L_n=0, \forall n\in Z$ et la masse d'un mode de la corde ouverte s'exprime :

\be M^2={1\over \alpha'}\sum_{n=1}^\infty\alpha_{-n}\cdot\alpha_n.\ee

 A noter que l'Hamiltonian de la corde ferm\'ee s'\'ecrit :
 
 \be H={1\over 2}\sum_{n=-\infty}^\infty(\alpha_{-n}\cdot\alpha_n  +
          \bar{\alpha}_{-n}\cdot\bar{\alpha}_n)\ee
       et donc $H=L_0+\bar{L}_0$.
       
       Pour la corde ouverte il devient : $H={1\over 2}\sum_{n=-\infty}^\infty\alpha_{-n}\cdot\alpha_n =L_0$.


  \subsection{Quantification de la corde bosonique}

  La quantification de la corde bosonique peut \^etre effectu\'ee de plusieures mani\`eres distinctes, mais \'equivalentes. Deux exemples seront expos\'es ici :
  
  \begin{itemize}
  \item[1.] {\bf La quantification covariante} dans laquelle l'invariance de Lorentz est manifeste, mais qui pr\'esente le d\'esavantage de la pr\'esence des \'etats de norme n\'egative, qui ne sont pas physiques.\\
  \item[2.] {\bf La quantification dans la jauge du c\^one de lumi\`ere} qui est une approche o\`u, comme son nom l'indique, des nouvelles restrictions de jauge sont impos\'ees ce qui a l'avantage d'\'eliminer les \'etats de norme n\'egative, mais l'invariance de Lorentz n'est plus manifeste. Comme on le verra par la suite la condition que la th\'eorie soit invariante de Lorentz fixe la dimension de l'espace-temps.
  \end{itemize}
  
  Dans la premi\`ere approche on consid\`ere les fonctions $X^\mu(\tau,\sigma)$ comme des op\'erateurs quantiques. Par cons\'equent ils obeissent aux r\'elations de commutation habituelles :
  
  \be [X^\mu(\tau,\sigma),P^\nu(\tau,\sigma')]=i\eta^{\mu\nu}\delta(\sigma-\sigma')\label{comm1},\ee
  o\`u $P^\mu={\delta S_P\over \delta \dot{X}^\mu}=T\dot{X}^\mu$ est le moment conjugu\'e.
  
  La relation (\ref{comm1}) d\'etermine les commutateurs pour les modes de Fourier de $X^\mu$ $\footnote{Pour la corde ouverte les $\bar{\alpha}^\mu_n$ sont absentes.}$ :
  
  \ba &&[x^\mu,p^\nu]=i\eta^{\mu\nu},\nonumber\\
      &&[\alpha^\mu_m,\alpha_n^\nu]=[\bar{\alpha}^\mu_m,\bar{\alpha}_n^\nu]=m \delta_{m+n}\eta^{\mu\nu},\nonumber\\      
&&[\bar{\alpha}^\mu_m,\alpha_n^\nu]=0.\ea
        
         Les op\'erateurs $\alpha_m^\mu$ sont reli\'es aux op\'erateurs canoniques de l'oscillateur harmonique par : $a^\mu_m={1\over \sqrt{m}}\alpha^\mu_m, a^{\dag\mu}_m={1\over \sqrt{m}}\alpha^\mu_{-m}, \forall m>0 \Rightarrow$
         
        \be [a^\mu_m,a^{\dag\nu}_n]=\delta_{m,n}\eta^{\mu,\nu}.\ee

        Les modes de fr\'equence n\'egative, $\alpha_m,\  m>0$ sont des op\'erateurs d'annihilation et les modes de fr\'equence positive $\alpha_{-m}, \ m>0$  sont des op\'erateurs de cr\'eation. L'op\'erateur nombre pour le mode $m,\ m>0$ est alors $N_m=:\alpha_m\alpha_{-m}:=\alpha_{-m}\alpha_m$ $\footnote{Le symbole :: d\'enote l'ordre normale qui consite \`a placer les modes de fr\'equence n\'egative \`a droite des modes de fr\'equence positive.}$. 
        
        L'\'etat fondamental $\mid\!0;p^\mu\!\!>$ est d\'efini par :
    \ba \alpha^\mu_m    \mid\!0;p^\mu\!\!>=0, \quad m>0,\nonumber\\
    \hat{p}^\mu\mid\!0;p^\mu\!\!>=p^\mu\mid\!0;p^\mu \!\!>.\ea
    
    Il est alors facile de voir que des \'etats de norme n\'egative apparaissent. Comme $[\alpha_m^0,\alpha^0_{-m}]=-m$ il en r\'esulte que tous les \'etats de la forme $\alpha^0_{-m}\mid\!\!0\!\!>$ sont de norme n\'egative: $<\!\!0\!\!\mid \!\!\alpha^0_m\alpha^0_{-m}\!\!\mid\!\!0\!\!>=-m<\!\!0\!\!\mid\!\!0\!\!>\ <0$. N\'eanmoins on peut esp\'erer que les contraintes de Virasoro \'elimineront ces \'etats de l'espace de Hilbert. Les op\'erateurs de Virasoro quantiques sont d\'efinis par leurs expressions classiques, mais avec l'ordre normal :
    
    \be L_m={1\over 2}\sum_{n=-\infty}^{+\infty}:\alpha_{m-n}\cdot \alpha_n:,\ee
     et de m\^eme pour les $\bar{L}_m$.
    Ceci ne pose pas de probl\`eme pour $m\neq0$ car, dans ce cas, $\alpha_{m-n}$ et $\alpha_n$ commutent. Ceci n'est plus le cas pour $L_0$ et pour r\'esoudre l'ambiguit\'e on va inclure une constante \`a d\'eterminer dans toutes les formules contenant $L_0$. Par cons\'equent un \'etat physique, $\mid\!\phi\!\!>$, de la corde ouverte, par exemple, devrait satisfaire a priori les conditions suivantes : 
    
    \ba &&L_n\mid\!\phi\!\!>=0,\quad n\neq 0,\nonumber\\
   && (L_0-a)\mid\!\phi\!\!>=0.\ea  
   
   Les op\'erateurs $L_n$ satsifont l'alg\'ebre de Virasoro :
   
   \be [L_m,L_n]=(m-n)L_{m+n}+ {c\over 12}m(m^2-1)\delta_{m+n}.\ee
   
   La constante $c$ s'appelle charge centrale et elle est \'egale \`a la dimension de l'espace-temps pour la corde bosonique. Le terme  ${c\over 12}m(m^2-1)\delta_{m+n}$ appara\^it comme un effet quantique et est responsable du fait qu'on ne peut pas implementer dans la th\'eorie quantique toutes les contraintes classiques. En effet :
   
   \be <\!\!\phi\!\mid [L_m,L_{-m}]\mid \!\phi\!\!>=<\!\!\phi\!\mid 2mL_0\mid \!\phi\!\!>  +{c\over 12}m(m^2-1)<\!\!\phi\!\mid \!\phi\!\!> ,  \ee
               donc on ne peut pas  imposer  $L_m\mid\!\phi\!\!>=0,\  \forall m$. 
 Le maximum des conditions qu'on peut imposer sur les \'etats physiques est :
  \ba L_m\mid\!\phi\!\!>=0,\quad m> 0,\nonumber\\
    (L_0-a)\mid\!\phi\!\!>=0.\label{poz}\ea  
   pour la codre ouverte. En fait les conditions (\ref{poz}) englobent toutes les contraintes, car en vertu de la relation $L_{-m}=L_m^\dag$ on a :
   
   \be <\phi'\mid L_m \mid \phi>=0, \quad \forall m\neq 0.\ee
   
   Pour la corde fem\'ee il faut imposer \'egalement les contraintes $\bar{L}_m\mid\!\!\phi\!\!>=0, \ m>0$. Les op\'erateurs $\bar{L}_m$ satisfont aussi une alg\`ebre de Virasoro et commutent avec les $L_m$. 
   
   Avec la nouvelle forme de la contrainte g\'en\'er\'ee par $L_0$ l'op\'erateur de masse pour la corde ouverte devient :
   
   \be M^2={1\over \alpha'}(N-a),\label{openmass}\ee
   avec $N=\sum_{m>0}N_m=\sum_{m>0}\alpha_{-m}\cdot\alpha_m$.   
   
   Pour la corde ferm\'ee les conditions $(L_0-a)\mid\!\phi\!\!>=0$ et 
   $(\bar{L}_0-a)\mid\!\phi\!\!>=0$ impliquent $(L_0-\bar{L}_0)\mid\!\phi\!\!>=0$ ou plus simplement $N=\bar{N}$ $\footnote{On rappelle que $\alpha_0^\mu=\bar{\alpha}_0^\mu=\sqrt{2\alpha'}p^\mu$.}$. C'est ce qu'on appelle la condition de raccordement des niveaux(level matching).  L'op\'erateur de masse est donn\'e par :
   
   \be M^2= {2\over \alpha'}(N+\bar{N}-2a).  \ee
   
   Il peut \^etre montr\'e que dans le cas o\`u la dimension de l'espace-temps est $D=26$ et $a=1$ les \'etats de norme n\'egative d\'ecouplent\cite{string}. Nous allons prouver le fait que $D=26$ et $a=1$ dans le cadre de    
    la quantification dans la jauge de lumi\`ere. L'approche consiste \`a choisir une jauge dans laquelle les contraintes de Virasoro peuvent \^etre explicitement r\'esolues pour que la th\'eorie soit d\'ecrite en terme des \'etats physiques uniquement. Pour la quantification covariante nous avons choisi une jauge dans laquelle la m\'etrique sur la surface de l'univers est conformement plate :
   
   \be ds^2=e^{2\Lambda}(d\sigma^2-d\tau^2)=e^{2\Lambda}d\sigma^+ d\sigma^-,\ee
   mais ceci ne fixe pas compl\`etement la jauge car toutes les transformations du type $\sigma^+\rightarrow \tilde{\sigma}^+(\sigma^+),\ \sigma^-\rightarrow\tilde{\sigma}^-(\sigma^-) $ peuvent \^etre compens\'ees par une transformation de Weyl et ne changent pas la jauge.  La transformation de $\tilde{\tau}$ :
   
   \be  \tilde{\tau}={1\over 2}\left(\tilde{\sigma}^+(\tau+\sigma)+\tilde{\sigma}^-(\tau-\sigma)  \right),\ee  
    implique que $\tilde{\tau}$ satisfait l'\'equation d'onde \`a 2 dimensions :
    
    \be (\partial_{\sigma}^2-\partial_{\tau}^2)\tilde{\tau}=0.\ee   
      
      Rappelons que les champs $X^\mu$ satisfont \'egalement cette \'equation, on peut donc faire une reparametrisation de mani\`ere \`a ce que $\tilde{\tau}$ soit \'egal \`a un des $X^\mu$. La jauge du c\^one de lumi\`ere correspond au choix $ \tilde{\tau}=X^+/p^+ + \mbox{const.}$ ou autrement :
      \be X^+(\tau,\sigma)=x^+ + 2\alpha'p^+ \tau.\ee    
  Les coordonn\'ees du c\^one de lumi\`ere de l'espace-temps sont d\'efinies par: $X^\pm={1\over \sqrt{2}} (X^0\pm X^{D-1}) $ . 
  
   Ayant fix\'e $X^+$ on peut exprimer $X^-$ en fonction des coordonn\'ees transverses $X^i, \ i=1,..., D-2$  \`a partir des contraintes de Virasoro, $(\dot{X}\pm X')^2=0$   :
   
   \be \dot{X}^-\pm {X^-}'={1\over 4 \alpha' p^+}(\dot{X}^i\pm {X^i}') ,\ee                      
                                   ce qui d\'etermine les $\alpha^-_m$ :
  
  \be  \alpha^-_m= {1\over 2\alpha' p^+}\sum_{n=1}^\infty:\alpha_{m-n}^i\alpha_{n,i}:-a\delta_m.\ee
  
  Dans la jauge du c\^one de lumi\`ere toutes les excitations de la corde sont g\'en\'er\'ees par les oscillateurs transverses $\alpha_m^i$(et les $\tilde{\alpha}_m^i$ pour les cordes ferm\'ees).
  Les diff\'erents \'etats des cordes sont obtenus en agissant avec les op\'erateurs de cr\'eation  $\alpha_m^i, \  m<0$ sur l'\'etat fondamental $\mid \! 0;p\!\!>$. Par exemple le premier \'etat excit\'e
  de la corde ouverte est $\alpha_{-1}^i \mid \! 0;p\!\!>$ qui est un vecteur avec $D-2$ composantes. Il appartient \`a la repr\'esentation 
  vectorielle du petit groupe $SO(D-2)$.
  Sous une transformation de Lorentz un vecteur avec une polarisation transverse peut aqu\'erir \'egalement une polarisation longitudinale, sauf s'il est non massif. Pour que la th\'eorie soit invariante de Lorentz il faut donc que cet l'\'etat $\alpha_{-1}^i \mid \! 0;p\!\!>$  
  soit de masse nulle. En utilisant la formule de l'op\'erateur de masse pour les cordes ouvertes (\ref{openmass}) on trouve :
  
  \be M^2= {1-a\over \alpha'}=0 \Rightarrow a=1.\ee
  
  On peut voir mainenant que l'\'etat fondamental de la corde ouverte, de masse $M^2= -{a\over \alpha'}$ est un tachyon. Ceci est \'egalement le cas pour les cordes ferm\'ees dont l'\'etat fondamental a la masse $
  M^2= -{4a\over \alpha'}$. Le premier \'etat excit\'e des cordes ferm\'ees est donn\'e par $\alpha_{-1}^i\bar{\alpha}_{-1}^j \mid \! 0;p\!\!>$. Ceci est un tenseur de masse nulle, $M^2={2\over \alpha'}(2-2a)=0$, qui se d\'ecompose en repr\'esentations irreductibles de 
  $SO(D-2)$: un tenseur antisym\'etrique, $B_{\mu\nu}$, un tenseur sym\'etrique de trace nulle, $g_{\mu\nu}$, correspondant \`a une particule de spin deux et sans masse, le graviton, et un champ scalaire, $\phi$, appel\'e dilaton. 
  
  Nous avons vu que pour respecter l'invariance de Lorentz la constante $a$ doit \^etre \'egale \`a 1. Mais cette constante est reli\'ee \`a la dimension de l'espace-temps. Pour voir cela on peut calculer $a$ \`a partir de la formule :
  
  \be {1\over 2}\sum_{i=1}^{D-2}\sum_{n=-\infty}^\infty \alpha^i_{-n}\alpha^i_n={1\over 2}\sum_{i=1}^{D-2}\sum_{n=-\infty}^\infty :\alpha^i_{-n}\alpha^i_n  : +{D-2\over 2}\sum_{n=1}^\infty n.\ee
 
 La somme divergente $\sum_{n=1}^\infty n$ peut \^etre calcul\'ee par la m\'ethode de r\'egularisation de la fonction zeta $\footnote{La formule g\'en\'erale est $\sum_{n\geq 0}(n+a)=\zeta(-1,a)=-{1\over 12}(6a^2-6a+1)$.}$ et vaut $-1/12$. Par cons\'equent $a=-{D-2\over 24}=1$, donc la dimension de l'espace temps est 26. 
 
 \newpage
   
\section{Cordes supersym\'etriques}

 Les cordes bosoniques ne permettent pas la description des fermions de l'espace-temps et, par cons\'equent, ne sont pas de bons candidats pour une description de la r\'ealit\'e. En plus, l'\'etat fondamental pour les cordes bosoniques, ouvertes ou ferm\'ees, est tachyonique, ce qui indique une instabilit\'e. Introduire des d\'egr\'ees de libert\'e fermioniques pourrait r\'esoudre ce probl\'eme, car la masse n\'egative de l'\'etat fondamental provient de l'\'energie du point z\'ero des oscillateurs bosoniques et les  d\'egr\'ees de libert\'e fermioniques apportent, en principe, une contribution de signe oppos\'e.
 
 Le proc\'ed\'e de construction de la corde fermionique consiste  \`a introduire des spineurs sur la  surface d'univers, qui seront les superpartenaires des champs $X^\mu$. La supersym\'etrie sur la surface d'univers est manifeste, mais, en revanche, la supersym\'etrie de l'espace-temps n'est pas garantie. Dans le formalisme de Ramond\cite{ramond}, Neveu et Schwarz\cite{ns} la supersym\'etrie de l'espace-temps est obtenue apr\`es troncation du spectre par la projection dite GSO (Gliozzi, Scherk et Olive)\cite{gso}. Il existe  un deuxi\`eme formalisme, de Green et Schwarz\cite{gs}, dans lequel la supersym\'etrie de l'espace temps est manifeste, mais pas celle de la surface d'univers. Le premier formalisme sera adopt\'e dans la suite. 
   
 \subsection{La corde fermionique classique} L'action de Polyakov (\ref{actionpolyakov}) pour la corde bosonique
 repr\'esente $D$ champs scalaires $X^\mu$ coupl\'es \`a la m\'etrique, $h_{ab}$ \`a deux dimensions. La g\'en\'eralisation supersym\'etrique de cette action necessite l'introduction des partenaires supersym\'etriques de $X^\mu$ et $h_{ab}$.
  
 En $D$ dimensions les champs $X^\mu$ repr\'esentent $D$ degr\'es de libert\'e bosoniques. En introduisant $D$ spineurs de Majorana sur la surface d'univers, $\psi^\mu$, on obtient $D$ degr\'es de libert\'e fermioniques "on-shell"$\footnote{Un spineur de Dirac sur la surface d'univers repr\'esente 4 degr\'es de libert\'e fermioniques. La condition de Majorana et l'\'equation de mouvement divisent chaqune le nombre de degr\'es de libert\'e par 2.}$. "Off-shell" il faut introduire $D$ champs scalaires auxiliares $F^\mu$ pour former un multiplet scalaire avec supersym\'etrie $N=(1,1)$ en dimension deux. Le gravitino, partenaire supersym\'etrique du $h_{ab}$, est un spineur-vecteur de Majorana, repr\'esentant 2 degr\'es de libert\'e fermioniques "off-shell"$\footnote{Deux des quatre degr\'es de libert\'e sont \'elimin\'es par les transformations de supersym\'etrie.}$. $h_{ab}$ repr\'esentant un seul degr\'e de libert\'e bosonique $\footnote{Deux des trois degr\'es de libert\'e de la m\'etrique pouvant \^etre \'elimin\'es par les reparametrisations \`a 2 dimensions comme on l'a vu pr\'ec\'edement.}$ il faut introduire un champ scalaire auxiliare. "On-shell" le graviton et le gravitino suffisent.
 
  La g\'en\'eralisation supersym\'etrique de l'action (\ref{actionpolyakov}) est : 
  
  \ba S=-{1\over 4\pi\alpha'}\int d^2\sigma \sqrt{-h}\left\{h^{ab}\partial_a X^\mu\partial_b X_\mu +i\bar{\psi}^\mu\gamma^a\nabla_a \psi_\mu  \right\}- \nonumber\\
-{1\over 4\pi\alpha'}\int d^2\sigma \sqrt{-h} \ i\chi_a\gamma^\beta \gamma^a \psi^\mu \left(\partial_\beta X_\mu+{i\over 4}\chi_\beta \psi_\mu\right) , \label{actionsusy}  \ea
 o\`u $\chi^a$ est le gravitino et $\gamma^a$ sont les matrices de Dirac \`a 2 dimensions :
 
 \ba \gamma^0=\sigma_2=\left(\begin{array}{cc} 0 & -i\\
 i & 0 \end{array} \right), \quad 
 \gamma^1=i\sigma_1=\left(\begin{array}{cc} 0 & i\\
 i & 0 \end{array} \right) .
  \ea

  L'action (\ref{actionsusy}) est invariante sous les transformations de supersym\'etrie :
 
 \ba \delta h_{ab}&=&i\epsilon(\gamma_a\chi_b+\gamma_b\chi_a),\nonumber\\
 \delta \chi_a&=&2\nabla_a\epsilon,\nonumber\\
 \delta \psi^\mu&=&\gamma^a(\partial_aX^\mu-{i\over 2}\chi_a\psi^\mu)\epsilon,\nonumber\\
 \delta X^\mu&=&i\epsilon \psi^\mu,\ea
 o\`u $\epsilon$ est un spineur de Majorana qui param\'etrise la supersym\'etrie.
 
  La jauge superconforme, l'\'equivalent supersym\'etrique de la jauge conforme (\ref{conforme}) du cas bosonique, est d\'efinie par les conditions :
  
  \ba h_{ab}&=&e^{2\Lambda}\eta_{ab},\nonumber\\
  \chi_a&=&\gamma_a\lambda,\ea 
   avec $\lambda={1\over 2}\gamma^c \chi_c$. Dans cette jauge  (\ref{actionsusy}) devient l'action de $D$ champs scalaires et $D$ champs fermioniques libres :
 
 \be S=-{1\over 4\pi \alpha' }\int d^2\sigma (\partial^a X^\mu \partial_a X_\mu + i\bar{\psi}^\mu\gamma^a\partial_a\psi_\mu).\ee
 
 Les champs $X^\mu$ satisfont l'\'equation d'onde \`a deux dimensions, comme dans le cas bosonique, et l'\'equation de mouvement pour les champs $\psi^\mu$ est l'\'equation de Dirac de masse nulle :
 
 \be i \gamma^a \partial_a \psi^\mu=0.\ee
 
 Les \'equations de mouvement doivent \^etre supplement\'ees par les conditions aux bords et les contraintes, qui sont les \'equations de la m\'etrique et du gravitino. Les conditions aux bords pour les coordonn\'ees bosoniques sont les m\^emes que dans le cas de la corde bosonique.  Pour les champs fermioniques ces conditions r\'esultent de l'annulation du terme de surface :
 
 \be \int d\tau \bar{\psi}^\mu \gamma^1\delta\psi_\mu\mid_{\sigma=0}^{\sigma=l}.\ee
 
  Le tenseur \'energie impulsion prend la forme :
  
  \be T_{ab}=\partial_a X^\mu \partial_bX_\mu + {i\over 2}\bar{\psi}^\mu\gamma_a\partial_b\psi_\mu+ {i\over 2}\bar{\psi}^\mu\gamma_b\partial_a\psi_\mu-{1\over 2}\eta_{ab}(\partial^c X_\mu\partial_c X^\mu + i\bar{\psi}^\mu\gamma^c\partial_c\psi_\mu).\ee
  
  La variation du gravitino dans l'action produit le supercourant :
  
  \be T_{Fa}= {1\over 2} \gamma_a\gamma^b\psi^\mu\partial_b X_\mu .  \ee
  
   Les contraintes \`a imposer aux \'etats physiques sont alors :
   
   \be T_{ab}=0, \quad T_{Fa}=0.\ee

  Comme dans le cas bosonique on va utiliser les coordonn\'ees du c\^one de lumi\`ere sur la surface d'univers, $\sigma^+$ et $\sigma^-$. L'action et les \'equations de mouvement se r\'e\'ecrivent :
  
   \be S={1\over \pi \alpha'}\int d^2\sigma\{\partial_+X\cdot\partial_-X+i(\psi_+\cdot\partial_-\psi_+ + \psi_-\cdot\partial_+\psi_-)\},\ee
   \ba \partial_+\partial_-X^\mu=0,\nonumber \\
   \partial_-\psi_+^\mu=\partial_+\psi_-^\mu=0.\ea
     
     La solution pour $X^\mu$ est \`a nouveau la somme de deux fonctions arbitraires $X^\mu_L(\sigma^+)$ et $X^\mu_R(\sigma^-)$ et le d\'eveloppement en oscillateurs est le m\^eme que dans le cas bosonique,  alors que pour les coordonn\'ees fermioniques on obtient: $\psi^\mu_+=\psi^\mu_+ (\sigma^+) $ et $\psi^\mu_-=\psi^\mu_- (\sigma^-) $ . Pour les cordes ferm\'ees la condition de p\'eriodicit\'e s'\'ecrit dans les coordonn\'ees du c\^one de lumi\`ere $ (\psi_+\delta \psi_++\psi_-\delta\psi_-)(\sigma)=(\psi_+\delta \psi_++\psi_-\delta\psi_-)(\sigma +2\pi)$, ce qui revient aux conditions :
     
     \ba \psi_+(\sigma )=\pm\psi_+(\sigma+2\pi ),\nonumber\\
       \psi_-(\sigma )=\pm\psi_-(\sigma+2\pi ),\ea
        et de m\^eme pour $\delta\psi_\pm$. On appelle conditions de Ramond (R) celles de p\'eriodicit\'e et conditions de Neveu-Schwarz(NS) celles de antip\'eriodicit\'e. Plus bri\` evement les fermions sur la surface d'univers respectent la condition :
  
  \be \psi(\sigma+2\pi)=e^{2\pi i \phi}\psi(\sigma),\ee
  avec $\phi=0$ pour le secteur de Ramond et $\phi={1\over 2}$ pour le secteur de Neveu-Schwarz. Avec cette condition 
 la solution g\'en\'erale de l'\'equation de Dirac s'\'ecrit :
 
 \ba \psi_+^\mu(\sigma,\tau)=\sum_{r\in Z+\phi}\bar{b}^\mu_r e^{-ir(\tau+\sigma)},\nonumber\\
\psi_-^\mu(\sigma,\tau)=\sum_{r\in Z+\phi} b^\mu_r e^{-ir(\tau-\sigma)},\ea
  avec $(b_r^\mu)^\dag=b_{-r}^\mu$ et $(\bar{b}_r^\mu)^\dag=\bar{b}_{-r}^\mu$  pour assurer la r\'ealit\'e des spineurs de Majorana $\psi_\pm^\mu$.        

 Dans les coordonn\'ees du c\^one de lumi\`ere les contraintes $T_{ab}=0, \ T_{Fa}=0$ prennent la forme :

\ba T_{++}&=&{1\over 2}\partial_+X\cdot\partial_+X+{i\over 2}\psi_+\cdot\partial_+\psi_+ =0,\nonumber\\
T_{--}&=&{1\over 2}\partial_-X\cdot\partial_-X+{i\over 2}\psi_-\cdot\partial_-\psi_- =0,\nonumber\\
T_{+-}&=&T_{-+}=0,\nonumber\\
T_{F+}&=&{1\over 2}\psi_+\cdot\partial_+X =0,\nonumber\\
T_{F-}&=&{1\over 2}\psi_-\cdot\partial_-X =0.\ea

et leurs modes de Fourier sont donn\'es par :

\ba L_m&=&L_m^{(\alpha)}+L_m^{(b)}={1\over 2}\sum_{n\in Z}\alpha_{m-n}\cdot\alpha_n+{1\over 2}\sum_{r\in Z+\phi}(r-{m\over 2})b_{m-r}\cdot b_{r},\nonumber\\
G_r&=&\sum_{n\in Z}b_{r-n}\cdot\alpha_{n}. \ea


\subsection{Quantification de la corde fermionique}
 Les r\'elations de commutation (\ref{comm1}) doivent \^etre compl\'et\'ees par les anticommutateurs :
 
 \ba &&\{\psi_+^\mu(\sigma,\tau),\psi_+^\nu(\sigma',\tau)\}=2\pi \eta^{\mu\nu}\delta(\sigma-\sigma'),\nonumber\\
 &&\{\psi_-^\mu(\sigma,\tau),\psi_-^\nu(\sigma',\tau)\}=2\pi \eta^{\mu\nu}\delta(\sigma-\sigma'),\nonumber\\
 &&\{\psi_+^\mu(\sigma,\tau),\psi_-^\nu(\sigma',\tau)\}=0,\ea
  ce qui implique pour les oscillateurs fermioniques :
 
 \be\{b_r^\mu,b_s^\nu\}=\eta^{\mu\nu}\delta_{r+s}.\ee
 
 Remarquons que les modes zero satisfont l'alg\`ebre de Clifford :
  \be\{b_0^\mu,b_0^\nu\}=\eta^{\mu\nu}.\ee
 
 L'op\'erateur nombre d'oscillateurs est donn\'e par :
 
 \be N=N^{(\alpha)}+N^{(b)}=\sum_{n=1}^\infty \alpha_{-n}\cdot\alpha_n+\sum_{r\in Z+\phi>0}^\infty r b_{-r}\cdot b_r.\ee
 
 Les op\'erateurs de Virasoro doivent \^etre red\'efinis en utilisant l'ordre normal et une constante, $a$, doit \^etre rajout\'ee dans les formules contenant $L_0$. Ils satisfont l'alg\`ebre de super-Virasoro :
 
 \ba &&[L_m,L_n]=(m-n)L_{m+n}+{d\over 8}m(m^2-2\phi)\delta_{m+n},\nonumber \\ &&[L_m,G_r]=( {m\over 2} -r) G_{m+r}, \nonumber\\
  &&\{G_r,G_s\}=2L_{r+s}+{d\over 2}(r^2-{\phi\over 2})\delta_{r+s}.\ea

L'\'etat fondamental est d\'efini par :
\be \alpha_m^\mu\mid\! 0\!\!>=b_r^\mu\mid\! 0\!\!>=0, \quad m,r>0.\ee

Dans le secteur de Neveu-Schwarz l'\'etat fondamental est un scalaire comme pour la corde bosonique. Ceci n'est plus le cas dans le secteur de Ramond \`a cause de l'existence du mode $b_0^\mu$ dans ce secteur. Il faut remarquer que $b_0^\mu$ commute avec l'op\'erateur de masse, $M^2={2\over \alpha'}(N+\bar{N}-2a)$, ce qui implique que les \'etats $\mid\! 0\!\!>$ et $b_0^\mu\mid\! 0\!\!>$ ont la m\^eme masse. En se rappellant que $b_0^\mu$ sont les g\'en\'erateurs de l'alg\`ebre de Clifford on peut conclure que l'\'etat fondamental dans le secteur de Ramond est un spineur de l'espace-temps. Comme les oscillateurs $\alpha_{-m}^\mu$ et $b_{-r}^\mu$ sont des vecteurs de l'espace-temps il en r\'esulte que tous les \'etats construits \`a partir de l'\'etat fondamental de Neveu-Schwarz sont des bosons de l'espace-temps et que tous les \'etats construits \`a partir de l'\'etat fondamental de Ramond sont des fermions de l'espace-temps. 

Des \'etats de norme n\'egative sont \`a nouveau pr\'esents dans le spectre et il est instructif d'\'etudier la quantification dans la jauge du c\^one de lumi\`ere qui est donn\'ee par les conditions:

\ba &&X^+=\alpha'p^+\tau,\nonumber\\ 
&&\psi^+=0.\ea
 
 Les contraintes permettent d'exprimer les oscillateurs $\alpha_m^-$ et $b_r^-$ en fonctions des modes transverses $\alpha_m^i$ et $b_r^ i, \ i=1...D-2$ et les \'etats de la corde sont construits en utilisant seulement les oscillateurs transverses.
 
 Le spectre de la corde ouverte (qui \'equivaut \` a celui des modes de gauche ou de droite de la corde ferm\'ee \`a un facteur $4$ pr\`es dans la formule de la masse) est donn\'e par:

 1.) Le secteur de Neveu-Schwarz: l'\'etat fondamental, $\mid\! 0\!\!>$, est un scalaire de masse $M^2=-{a\over \alpha'}$. Le premier \'etat excit\'e, $N=1/2$,  est donn\'e par $b_{-1/2}^i\mid\! 0\!\!>$ avec la masse $M^2={1\over \alpha'}({1\over 2}-a)$. Ceci est un vecteur de $SO(D-2)$ et l'invariance de Lorentz impose qu'il soit de masse nulle, c.a.d. $a_{\tt{NS}}={1\over 2}$. L'\'etat fondamental est encore une fois tachyonique. D'autre part la constante de l'ordre normale vaut $a=-{D-2\over 2}(\sum_{n=0}^\infty n -\sum_{r=1/2}^\infty r)$, qui, en utilisant la r\'egularisation de la fonction $\zeta$, donne le r\'esultat $a={D-2\over 2}({1\over 12}+{1\over 24})= {D-2\over 16}$. On en d\'eduit que la dimension de l'espace-temps vaut $D=10$.
  
 Les \'etats avec $N=1$ sont $\alpha_{-1}^i \mid\! 0\!\!>$, qui est un vecteur de $SO(8)$, et $b_{-1/2}^i b_{-1/2}^j\mid\! 0\!\!>$, tenseur antisym\'etrique de $SO(8)$, avec la masse $M^2={1\over 2\alpha'}$. Ces $8+28$ d\'egr\'es de libert\'e bosoniques forment un tenseur antisym\'etrique de $SO(9)$, qui est le petit groupe des \'etats massifs en dix dimensions. 
 
 Au niveau $N=3/2$ on obtient des \'etats de masse $M^2=1/\alpha'$ :\\ 
 $b_{-1/2}^i b_{-1/2}^jb_{-1/2}^k\mid\! 0\!\!>$, $\alpha_{-1}^ib_{-1/2}^j\mid\! 0\!\!>$ et $b_{-3/2}^i\mid\! 0\!\!>$, soit un total de 128 \'etats.\\

 2.) Le secteur de Ramond: l'\'etat fondamental est, comme on l'a vu, un spineur. En 10 dimensions on peut imposer simultan\'ement les conditions de Weyl et Majorana, ce qui conduit \`a 8 d\'egr\'es de libert\'e fermioniques ind\'ependants "on-shell", qui repr\'esentent les composantes d'un spineur de Majorana-Weyl, de masse nulle, du groupe $SO(8)$. La masse de l'\'etat fondamental \'etant donn\'ee par $M^2=-a/\alpha'$ il r\'esulte que $a_{\tt{R}}=0$. Il y a deux possibilit\'es pour la chiralit\'e de ce spineur conduisant \`a deux \'etats possibles $\mid\!a\!\!>$ et $\mid\!\bar{a}\!\!>$. Le niveau $N=1$ est form\'e des \'etats $\alpha^i _{-1}\mid\!a\!\!>$, $\alpha^i _{-1}\mid\!\bar{a}\!\!>$, $b^i _{-1}\mid\!a\!\!>$ et $b^i _{-1}\mid\!\bar{a}\!\!>$ de masse $M^2=1/ \alpha'$, soit $2\times 128$ \'etats. 
 
 \`A ce niveau le spectre obtenu n'est pas supersym\'etrique, mais la projection $GSO$ permet d'obtenir la supersym\'etrie. D'abord il faut \'eliminer le tachyon du secteur NS, ainsi que tous les \'etats de masse $(n+1/2)/\alpha', n\geq 0$ qui n'ont pas d'\'equivalent dans le secteur de Ramond. Il s'agit, donc, d'\'eliminer tous les \'etats avec un nombre pair d'oscillateurs de type $b^ i_{-r}$, c.a.d. tous les bosons ( par rapport \`a la surface d'univers) du secteur NS. Le projecteur $GSO$ dans le secteur NS est d\'efini par :
 
 \be P_{GSO}^{NS}={1-(-1)^{F_{NS}}\over 2},\ee
 
    avec $F_{NS}=\sum_{r=1/2}^\infty b^i_{-r}\cdot b_r^i$ le nombre fermionique sur la surface d'univers.
    
    Ensuite on remarque que pour obtenir l'\'equilibre entre le nombre des d\'egr\'es de libert\'e fermioniques (R) et bosonique (NS) pour les \'etats de masse nulle, par exemple, il est n\'ecessaire d`\'eliminer un des \'etats $\mid\!a\!\!>$ ou $\mid\!\bar{a}\!\!>$. L'op\'erateur qui permet d'effectuer cette op\'eration est 
    \be\Gamma=b_0^1...b_0^8(-1)^{\sum_{n>0}b_{-n}^ib_n^i}= \Gamma_9(-1)^{F_{R}},\ee
    o\`u $\Gamma_9=b_0^1...b_0^8$ est l'op\'erateur  de chiralit\'e  dans les dimensions transverses et $F_{R}$ le nombre fermionique sur la surface d'univers dans le secteur de Ramond. En demandant $\Gamma=1$ pour tous les \'etats (ou $\Gamma=-1$ pour tous les \'etats) on \'elimine la moiti\'e des d\'egr\'es de libert\'e en r\'etablissant l'\'equilibre avec les bosons du secteur NS.   
    
    Pour obtenir le spectre des cordes ferm\'ees il faut combiner les modes de droite avec celles de gauche en tenant compte de la contrainte $L_0-\bar{L}_0=0$, ou autrement $m_R^2=m_L^2$. Quatre secteurs apparaissent : (NS,NS) et (R,R), qui conduisent \`a des bosons de l'espace-temps, et (NS,R), (R,NS) qui contiennent les fermions de l'espace-temps. La projection $GSO$ peut \^etre effectu\'ee s\'epar\'ement pour les modes de droite et de gauche ce qui donne la possibilit\'e de choisir ind\'ependemment la projection dans le secteur R: $\Gamma_L=\Gamma_R=1$ ou $\Gamma_L=1,\ \Gamma_R=-1$.   
    Avec le premier choix on obtient la th\'eorie appell\'ee  IIB dont le spectre non massif est donn\'e par :\\
    
  secteur (NS,NS): un scalaire (le dilaton), le graviton, $g_{ij}$, et un tenseur antisym\'etrique, $B_{ij}$.
   
   secteur (R,R): une zero forme, $C_0$, une 2-forme, $C_2$, et une 4-forme auto-duale, $C_4^+$.
   
   secteurs (R,NS) et (NS,R): deux gravitinos de spin $3/2$ avec la m\^eme chiralit\'e et deux fermions de spin $1/2$. \\
   
   Le comptage des degr\'es de libert\'e fermioniques et bosoniques 
(128) indique qu'il s'agit d'une th\'eorie supersym\'etrique. La pr\'esence de deux\\ gravitions avec la m\^eme chiralit\'e indique que la th\'eorie IIB  est une th\'eorie chirale avec supersym\'etrie $N=(2,0)$.\\

   Le choix  $\Gamma_L=1,\ \Gamma_R=-1$ conduit \`a la th\'eorie IIA qui poss\`ede au niveau non massif les \'etats :\\
   
   secteur (NS,NS): un scalaire (le dilaton), le graviton, $g_{ij}$, et un tenseur antisym\'etrique, $B_{ij}$.
   
   secteur (R,R): une 1-forme, $C_1$ et une 3-forme, $C_3$.
      
   secteurs (R,NS) et (NS,R): deux gravitinos de spin $3/2$ avec des chiralit\'es oppos\'ees et deux dilatinos de spin $1/2$, un pour chaque chiralit\'e. \\
   
    La th\'eorie IIA est une th\'eorie non chirale avec supersym\'etrie $N=(1,1)$.\\   
   
               
\section{D\'eveloppement de Polyakov}

L'integrale de chemin de Feynman est une m\'ethode naturelle pour d\'ecrire les interactions en th\'eorie des cordes. 
 En m\'ecanique quantique les amplitudes sont obtenues en sommant sur toutes les trajectoires possibles qui relient les \'etats initials et finals, chaque trajectoire \'etant pond\'er\'ee par un facteur $\mbox{exp}(iS_{cl}/\hbar)$, o\`u $S_{cl}$ est l'action classique.
 
  Polyakov a g\'en\'eralis\'e ce proced\'e \`a la th\'eorie des cordes\cite{polyakov}: une amplitude des cordes est obtenue en sommant sur toutes les surfaces d'univers qui relient les courbes initiales et finales. 
 
\begin{figure}[!h]
\centering
\includegraphics[width=75mm]{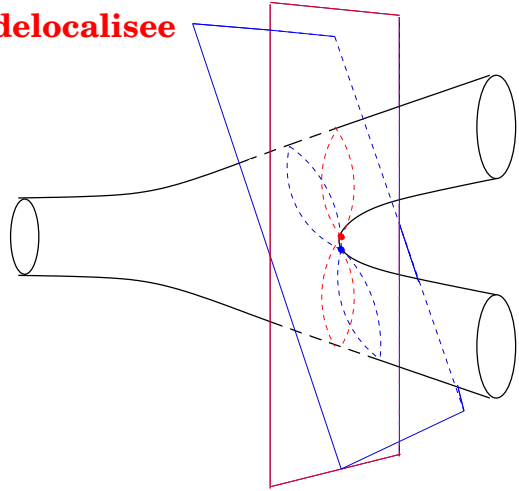}
\caption{Une corde ferm\'ee qui se d\'esintegre en deux cordes ferm\'ees}
\label{desinteg}
\end{figure}  
 Les interactions des cordes sont implicites dans la somme sur les surfaces d'univers. Par exemple la figure (\ref{desinteg}) montre la d\'esintegration d'une corde ferm\'ee en deux cordes. Les particules sont obtenues comme des diff\'erentes \'etats d'excitation des cordes et toutes les interactions du Mod\`ele Standard apparaissent \`a partir de ce type d'int\'eraction.

 Comme on peut le voir sur cette figure (\ref{desinteg}) la d\'esintegration  semble avoit lieu \`a des diff\'erents points de l'espace selon le rep\`ere de Lorentz choisi.  
 L'interaction est "\'etendue" dans l'espace, ce qui pourrait r\'esoudre les divergences de la gravit\'e quantique.
 
 \begin{figure}[!h]
\centering
\includegraphics[width=65mm]{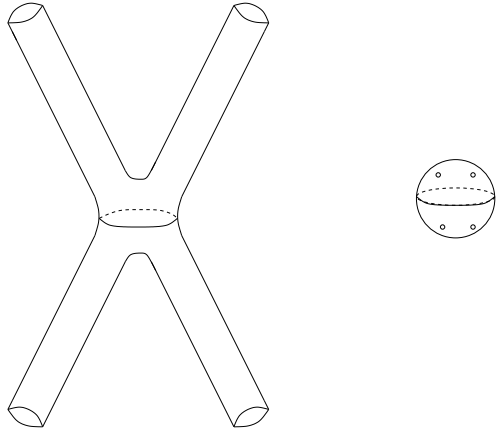}
\caption{Interaction avec quatre cordes ferm\'ees externes et la repr\'esentation \'equivalente de cette amplitude comme une sph\`ere avec quatre insertions ponctuelles}
\label{conformepdf}
\end{figure} 
 
 La somme sur les surfaces d'univers qui relient des courbes initiales et finales conduit \`a des amplitudes difficiles \`a calculer. Le calcul se simplifie lorsque les sources sont envoy\'ees \`a l'infini. Dans ce cas l'invariance conforme de la th\'eorie sur la surface d'univers rend \'equivalentes une surface d'univers avec des cordes externes et une surface compacte avec des insertions ponctuelles, comme la figure (\ref{conformepdf}) le montre.
 
   Pour calculer une amplitude il faut int\'egrer sur toutes les m\'etriques possibles de la surface d'univers, $g_{ab}(\tau,\sigma)$, et toutes les possibilit\'es de plonger la surface d'univers dans l'espace-temps, $X^\mu(\tau,\sigma)$ $\footnote{Nous avons remplac\'e la m\'etrique minkowskienne sur la surface d'univers avec une m\'etrique euclidienne.}$ :

  \be \int [dX\ dg] exp(-S),\label{chemin}\ee
 avec
  
   \be S={1\over 4\pi \alpha'}\int d^2\sigma g^{1/2}g^{ab}\partial_a X^\mu \partial_b X_\mu +\lambda \ \chi \label{lambda} ,\ee
     o\`u :
    
    \be\chi={1\over 4\pi}\int d^2\sigma g^{1/2}R, \ee
 est le nombre d'Euler de la surface d'univers qui pour une surface avec $h$ poign\'ees, $b$ fronti\`eres et $c$ crosscaps$\footnote{Un crosscap est une fronti\`ere avec les points diam\'etralement opppos\'es identifi\'es.}$ vaut $\chi=2-2h-b-c$.
 \begin{figure}[!h]
\centering
   \includegraphics[width=45mm]{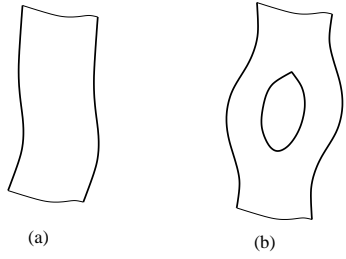}
\caption{(a) Une corde ouverte qui se propage librement. (b) Deuxi\`eme ordre en th\'eorie des perturbations}
\label{boucle}
\end{figure} 
 
  On peut remarquer que si on ajoute une fronti\`ere sur une surface d'univers (fig \ref{boucle}), ce qui \'equivaut \`a l'\'emission suivie de l'absorption d'une corde ouverte, le nombre d'Euler diminue d'une unit\'e et dans l'int\'egrale de chemin il y aura un facteur $e^\lambda$  de plus. Il en r\'esulte que l'amplitude pour \'emettre une corde ouverte est proportionnelle \`a $e^{\lambda/2}$.
 De la m\^eme fa\c{c}on on peut voir que le fait d'ajouter une poign\'ee sur la surface d'univers \'equivaut \`a l'\'emission suivi de l'absorption d'une corde ferm\'ee et, en m\^eme temps, diminue le nombre d'Euler de 2 unit\'es. L'amplitude d'\'emission pour une corde ferm\'ee est, donc, proportionnelle \`a $e^\lambda$.
 
 La constante de couplage en th\'eorie des cordes est reli\'ee \`a $\lambda$, qui semble \^etre un param\`etre libre de la th\'eorie. Mais si on veut g\'en\'eraliser l'action des cordes dans l'espace de Minkowski \`a une action d\'ecrivant la propagation des cordes dans un espace-temps courbe on se rend compte qu'en plus du remplacement de la m\'etrique de Minkowski $\eta_{\mu\nu}$ avec une m\'etrique g\'en\'erale $G_{\mu\nu}$, il faut inclure des couplages aux autres \'etats de masse nulle des cordes, car le graviton n'est qu'un \'etats de cordes parmi d'autres. Autrement dit il faut inclure des couplages au tenseur antisym\'etrique $B_{\mu\nu}$ et au dilaton $\Phi$ :
 
 \be S={1\over 4\pi \alpha'}\int d^2\sigma g^{1/2}[(g^{ab}G_{\mu\nu}(X)+i\epsilon^{ab}B_{\mu\nu}(X))\partial_aX^\mu \partial_b X^\nu+\alpha'R\Phi(X)].\ee
 
  On peut voir \`a partir de l'action (\ref{lambda}) que les diff\'erentes valeurs du param\`etre $\lambda$ correspondent , en fait,  \`a des diff\'erents fonds d'une seule  th\'eorie et non pas \`a des th\'eories diff\'erentes. La constante de couplage des cordes est donn\'ee par la valeur moyenne dans le vide du dilaton :
  
  \be g_s=e^{<\Phi>}.\ee

L'int\'egrale de chemin (\ref{chemin}) comporte un comptage erron\'e car les configurations $(X,g)$ reli\'es par les diff\'eomorphismes  et par la transformation de Weyl d\'ecrivent la m\^eme physique et devraient \^etre compt\'ees une seule fois. Pour cela il faut diviser la mesure d'int\'egration  par le volume des groupes de sym\'etrie:

\be Z=\int {[dX\ dg]\over V_{{\rm diff}\times {\rm Weyl}}} exp(-S).  \ee
  La mesure d'int\'egration correcte s'obtient par la m\'ethode de Faddeev-Popov. 
  
  Les reparam\'etrisations et l'invariance de Weyl permettent de fixer compl\`etement la m\'etrique \`a une valeur $\hat{g}_{ab}$.  La mesure  de Faddeev-Popov, $\Delta_{FP}$, est d\'efinie par:
  
  \be 1=\Delta_{FD}(g)\int [d\zeta]\delta(g-\hat{g}^\zeta),\label{FP}\ee
  o\`u $\zeta$ d\'enote une transformation sous les reparam\'etrisations combin\'ee avec une transformation de Weyl et $[d\zeta]$ est une mesure invariante de jauge. En ins\'erant (\ref{FP}) dans l'int\'egrale de chemin on obtient:
  
  \be Z[\hat{g}]=\int {[d\zeta \ dX\ dg]\over V_{{\rm diff}\times {\rm Weyl}}}  \Delta_{FD}(g)  \delta(g-\hat{g}^\zeta)\ exp(-S[X,g]).\ee
  
  Apr\`es int\'egration sur $g$ et une transformation de jauge l'int\'egrale de chemin devient:
  
  \be Z[\hat{g}]=\int {[d\zeta \ dX]\over V_{{\rm diff}\times {\rm Weyl}}}  \Delta_{FD}(\hat{g}) \ exp(-S[X,\hat{g}]).\ee
  
  On peut effectuer l'int\'egrale sur $\zeta$ qui se simplifie avec le volume du groupe de jauge, le r\'esultat \'etant:
  
  \be  Z[\hat{g}]=\int [dX]  \Delta_{FD}(\hat{g}) \ exp(-S[X,\hat{g}]).\label{chem}\ee  
  
  Le d\'eterminant de Faddeev-Popov vaut\cite{string}:
  
  \be \Delta_{FD}(\hat{g})=\int[db\ dc] exp(-S_g),\ee  
  o\`u $S_g=(1/2\pi) \int d^2\sigma\ \hat{g}^{1/2}b_{ab}\nabla^a\ c^b$, avec $b$ et $c$ des fant\^omes 
  de Grassmann. 
  
  En fait l'in\'egrale de chemin (\ref{chem}) d\'epend du facteur conforme, $\omega$, avec $\hat{g}_{ab}=e^{2\omega}\delta_{ab}$, sauf si la dimensions de l'espace-temps est 26. Dans ce cas il n'y a pas d'anomalie conforme et l'expression (\ref{chem}) est correcte.  
  

\section{Compactification toroidale, T-dualit\'e, D-branes}
\label{toroidal}

 Nous avons vu que la corde bosonique et la supercorde pr\'edisent une dimension de l'espace-temps plus grande que $D=4$, qui est la dimension de notre espace-temps, telle qu'elle est per\c{c}ue \`a l'heure actuelle. Ceci pourrait \^etre expliqu\'e par le fait que les dimensions suppl\'ementaires sont compactifi\'ees \`a une \'echelle tr\`es r\'eduite qui rend impossible leur d\'etection aux \'energies actuelles.  

Ici on va consid\'erer le cas le plus simple, \`a savoir la compactification d'une coordonn\'ee sur un cercle pour la corde bosonique :

\be  X\equiv X+2\pi R.   \ee

 La p\'eriodicit\'e implique que l'impulsion du centre de masse de la corde est quantifi\'ee, comme pour la compactification de Kaluza-Klein en th\'eorie des champs :
 
 \be p={m\over R}, \ m\in Z.\ee

 En th\'eorie des cordes la compactification a un deuxi\`eme effet. Une corde ferm\'ee peut s'enrouler autour de la dimension compacte:
 
 \be X(\sigma + 2\pi)=X(\sigma)+2\pi w,\  w=nR, n\in Z.\ee
$w$ s'appelle nombre d'enroulement (winding number) et repr\'esente le nombre de tours effectu\'es par la corde autour de la dimension $X$. 
 
  Le d\'evelopppement en oscillateurs pour la coordonn\'ee compacte est donn\'e par :
  
   \be X(\tau,\sigma)=x+2\alpha'{m\over R}\tau + 2nR\sigma+i\sqrt{\alpha'\over 2}\sum_{k\neq 0}\{{\alpha_k\over k}e^{-2ik(\tau+\sigma)}+{\bar{\alpha}_k\over k}e^{-2ik(\tau-\sigma)}\},\ee
qui, en utilisant la formule (\ref{xlxr}) permet de determiner :

\ba &&p_L=(2/\alpha')^{1/2}\alpha_0={m\over R}+{nR\over \alpha'},\nonumber\\
&&p_R=(2/\alpha')^{1/2}\bar{\alpha}_0={m\over R}-{nR\over \alpha'}.\ea

Les op\'erateurs de Virasoro $L_0$ et $\bar{L_0}$ sont donn\'es par$\footnote{A noter que $L_0-\bar{L}_0=0$ n'implique plus $N=\bar{N}$, mais $N-\bar{N}=mn$.}$ :

\ba L_0={\alpha'p_L^2\over 4}+\sum_{n=1}^\infty\alpha_{-n}\alpha_n,\nonumber\\
\bar{L}_0={\alpha'p_R^2\over 4}+\sum_{n=1}^\infty\bar{\alpha}_{-n}\bar{\alpha}_n\ea
et l'op\'erateur de masse devient :

\ba M^2&=& -p^\mu p_\mu={1\over 2}(p_L^2+p_R^2)+{2\over \alpha'}(N+\bar{N}-2)=   \nonumber\\
&=&{m^2\over R^2}+{n^2R^2\over \alpha'^2}+{2\over \alpha'}(N+\bar{N}-2).
 \ea

On peut voir \`a partir de cette formule que lorsque $R\rightarrow \infty$ les modes d'enroulement deviennent infiniment massifs et les valeurs de l'impulsion compacte tendent vers un spectre continu, comme pour une dimension non-compacte. Dans la limite $R \rightarrow 0$ ce sont les valeurs de l'impulsion qui deviennent infinies et les modes d'enroulement qui tendent vers un continuum, ce qui ressemble \`a nouveau au spectre d'une dimension non-compacte. En fait les deux limites sont \'equivalentes, ce qui n'est pas le cas en th\'eorie des champs o\`u il n'y a pas l'\'equivalent du nombre d'enroulement. L'invariance du spectre sous la transformation :

\be R\rightarrow R'={\alpha'\over R}\ , \quad n\leftrightarrow m \ee
s'appelle T-dualit\'e. L'\'equivalence des limites $R \rightarrow 0$ et $R\rightarrow \infty$ en th\'eorie des cordes montre que la g\'eom\'etrie de l'espace-temps \`a courte distance est vue diff\'eremment par les cordes et par les particules ponctuelles. Les th\'eories in\'equivalentes sont celles pour lesquelles $R\geq R_{\tt{self-dual}}=\alpha'^{1/2}$.

\'Echanger $n$ et $m$ revient \`a une parit\'e sur l'impulsion $p_R$ et, implicitement, sur la coordonn\'ee $X_R$. Apr\`es la T-dualit\'e la th\'eorie est d\'ecrite en termes de la coordonn\'ee $X'=X_L(\sigma^+)-X_R(\sigma^-)$.

La T-dualit\'e agit de mani\`ere non-trivialle sur le dilaton. Le couplage de la th\'eorie compactifi\'ee doit \^etre invariant sous la T-dulait\'e. Ce couplage est obtenu en integrant 
l'action de la th\'eorie \`a  26 dimensions sur la coordonn\'ee compacte, il sera, donc, proportionnel au p\'erim\`etre du cercle et, donc, au rayon $R$. Il en r\'esulte que $Re^{-2\Phi}$ doit \^etre invariant sous la T-dualit\'e, ce qui implique la transformation suivante pour $\Phi$ :





\be e^{\phi'}=   {\alpha'^{1/2}\over R} e^\Phi.\ee

Pour les cordes ouvertes il n'y a pas de nombre d'enroulement car les conditions aux bords de Neumann indiquent que les bouts des cordes ouvertes se d\'epalcent librement dans l'espace ce qui permet toujours de d\'erouler la corde de la dimension compacte. Dans la limite $R\rightarrow 0$ on n'obtient pas un continuum d'\'etats comme pour la corde ferm\'ee. Apr\`es la T-dualit\'e les cordes ouvertes se d\'eplacent en 25 dimensions. L'interpretation de ce fait est que les cordes ouvertes vibrent toujours \`a 26 dimensions, mais que leurs bouts sont contraints \`a appartenir \`a une hypersurface 25-dimensionnelle. En effet, la th\'eorie T-duale \'etant d\'ecrite en utilisant la coordonn\'ee $ X'^{(25)}=X_L^{(25)}(\tau+\sigma)-X_R^{(25)}(\tau-\sigma)$, la condition de Neumann dans cette direction devient, apr\`es la T-dualit\'e, une condition de Dirichlet :

\be \partial_\sigma X^{(25)}=\partial_\tau X'^{(25)}=0.\ee

Les hypersurfaces auxquelles les bouts des cordes ouvertes sont restreintes sont des nouveaux objets dynamiques appel\'es $D$-branes. En effectuant des T-dualit\'es sur $25-p$ coordonn\'ees on obteint une $Dp$-brane, un objet avec $p$ dimensions spatiales. Comme la T-dualit\'e \'echange les conditions aux bords Neumann et Dirichlet il en r\'esulte qu'en effectuant une T-dualit\'e le long d'une coordonn\'ee parall\`ele \`a une $Dp$-brane on la transforme en une $D(p-1)$-brane, alors qu'une T-dualit\'e dans une direction perpendiculaire \`a la brane la transforme dans une $D(p+1)$-brane. \\

 Par analogie avec le cas de la corde l'action effective pour une $Dp$-brane pourrait s'\'ecrire :
 
 \be S_p=-T_p\int d^{p+1}\xi\  e^{-\Phi}(-det\ G_{ab})^{1/2},\label{ess}\ee   
 o\`u $\xi^a, \ a=0,...,p$ sont les coordonn\'ees qui param\'etrisent la brane et $G_{ab}$ est la m\'etrique induite sur la brane :
 
 \be G_{ab}(\xi)={\partial X^\mu\over \partial\xi ^a}{\partial X^\nu\over \partial\xi^b}G_{\mu\nu}(X(\xi)).\ee 
 $T_p$ est la tension de la $Dp$-brane, donn\'ee par :
\be T_p=
{M_P\over 32\sqrt{2}}e^{-\Phi_0}(4\pi \alpha')^{(11-p)/2},\ee
avec $M_P$ la masse de Planck et $\Phi_0$ la valeur moyenne du dilaton dans le vide. 
 
 La dependance dans le dilaton, $e^{-\Phi}$, intervient puisqu'il s'agit d'une action des cordes ouvertes au niveau des arbres, la premi\`ere contribution venant du disque($\chi=1$).  
   
   La m\'etrique (\ref{ess}) fait intervenir les champs  $X^\mu(\xi)$ qui d\'ecrivent le plongement de la brane dans l'espace-temps et leur couplage \`a  la m\'etrique induite. Il n'y a aucune raison pour ne pas inclure des couplages aux autres \'etats de masse nulle des cordes ferm\'ees et ouvertes: le tenseur antisym\'etrique, $B_{\mu\nu}$, qui doit appara\^itre dans l'action comme le tenseur induit, $B_{ab}$ :
   
 \be  B_{ab}(\xi)={\partial X^\mu\over \partial\xi ^a}{\partial X^\nu\over \partial\xi^b}B_{\mu\nu}(X(\xi))\ee  
 et le champ de jauge $A_a(\xi)$. Ces champs apparaissent dans l'action des cordes sur la surface d'univers comme :
 
 \be {1\over 4\pi\alpha'}\int_M d^2\sigma g^{1/2}\epsilon^{ab}\partial_a X^\mu\partial_b X^\nu B_{\mu\nu}+\int_{\partial M}d\tau A_\mu \dot{X}^\mu={1\over 2\pi\alpha'}\int_M B+\int_{\partial M} A\ee   
 Cette action est invariante sous la transformation de jauge de l'espace-temps $\delta A=d\lambda$. Par contre la transformation de jauge $\delta B=d\zeta$ conduit \`a un terme de surface non nul qui peut \^etre annul\'e si $A$ se transforme sous la sym\'etrie de jauge de $B$ comme: $\delta A=-\zeta/2\pi\alpha'$. La combinaison $B_{ab}+2\pi\alpha'F_{ab}$, avec $F=dA$, est invariante sous les deux transformations de jauge. C'est cette combinaison qui doit appara\^itre dans l'action. Par cons\'equent l'action de la brane s'\'ecrit :
 
 \be S_p=-T_p\int d^{p+1}\xi\  e^{-\Phi}[-det(G_{ab}+B_{ab}+2\pi \alpha' F_{ab})]^{1/2}.\label{bi}\ee  
(\ref{bi}) s'appelle l'action de Born-Infeld.

 Une $Dp$-brane couple naturellement \`a une $p+1$ forme :
 
 \be S=-q\int d^{p+1}\xi\  A_{p+1},\label{electric}\ee
o\`u $q$ est la charge de la brane, appel\'ee charge de Ramond-Ramond (RR), car les formes apparaissent dans le secteur (R,R) du spectre des cordes fermioniques. L'action (\ref{electric}) repr\'esente le couplage "\'electrique". La duale d'une $(p+1)$ forme \'etant  
 une $(7-p)$ forme il existe \'egalement un couplage "magn\'etique" \`a une $(6-p)$ brane. 

Une antibrane, $\bar{D}p$ , a la m\^eme trension qu'une brane et la charge oppos\'ee.

Les branes et les antibranes sont des objets BPS, c'est \`a dire qu'elles pr\'eservent une partie de la supersym\'etrie. Plus pr\'ecisement les branes brisent la moiti\'e des supersym\'etries et les antibranes brisent l'autre moiti\'e. La combinaison brane-antibrane brise compl\`etement la supersym\'etrie et forme un syst\`eme non-BPS. 

La th\'eorie de type IIB contient des branes BPS avec dimension impaire et la th\'eorie de type IIA conteint les branes BPS de dimension paire. En plus la th\'eorie de type IIB contient des branes non-BPS de dimension paire et la th\'eorie IIA contient des branes non-BPS de dimension impaire\cite{sen}. Une $D2p$ -brane non-BPS en th\'eorie IIB est d\'efinie comme une combinaison d'une $D2p$ brane BPS et une $\bar{D}2p$ brane BPS de la th\'eorie IIA  soumise \`a l'op\'eration $(-1)^{F_L}$, avec $F_L$ le nombre fermionique gauche de l'espace-temps. Par cette transformation la th\'eorie IIA est transform\'ee dans la th\'eorie IIB. Comme cette op\'eration change le signe des champs du secteur (R,R) et que les branes sont charg\'ees sous un champ RR, il en r\'esulte qu'une $D$ - brane se transforme en une antibrane et inversement. Par cons\'equent  le syst\`eme brane-antibrane est invariant sous cette transformation. De la m\^eme mani\`ere on peut d\'efinir des branes impaires non-BPS en th\'eorie IIA. Etant construites \`a partir des paires brane-antibrane les branes non-BPS ne sont pas charg\'ees sous les formes de RR. Leur tension est reli\'ee \`a celle des branes BPS par

 \be T_{\rm non-BPS}=\sqrt{2}T_{BPS}\label{tension}.\ee

Les combinaisons brane-antibrane sont d\'ecrites en th\'eorie IIB par l'amplitude de l'anneau
$\footnote{Pour les notations voir le chapitre 2.}$\cite{nonbps}:

\ba \tilde{\mathcal{A}}_{pp}&=&2^{-(p+1)/2}[\al m+n\ar^2 V_8-\al m-n\ar^2 S_8]\nonumber\\
\mathcal{A}_{pp}&=&(m\bar{m}+n\bar{n})(V_8-S_8)+(m\bar{n}+n\bar{m})(O_8-C_8),\ea
avec $m$ le nombre de branes et $n$ le nombre d'antibranes. Les caract\`eres $so(8)$ se d\'ecomposent selon:

\ba V_8=V_{p-1}O_{9-p}+O_{p-1}V_{9-p}, \quad O_8=O_{p-1}O_{9-p}+V_{p-1}V_{9-p}\nonumber\\
S_8=S_{p-1}S_{9-p}+C_{p-1}C_{9-p}, \quad C_8=S_{p-1}C_{9-p}+C_{p-1}S_{9-p},\ea
o\`u le deuxi\`eme caract\`ere de chaque produit d\'ecrit des degr\'es  de libert\'e internes. Par exemple $V_8$ se d\'ecompose en un vecteur avec $p-1$ composantes  et $9-p$ scalaires.

 Apr\`es l'op\'eration $(-1)^{F_L}$ les branes et antibranes sont interchang\'ees, ce qui implique en particulier une identification des facteurs de Chan-Paton $n=m=N$:
 
 \ba \tilde{\mathcal{A}}_{pp}&=&2\times 2^{-(p+1)/2} N\bar{N}V_8\nonumber\\
 \mathcal{A}_{pp}&=&N\bar{N}[O_8+V_8-S_8-C_8]\ea 
  
  Le facteur 2 dans l'amplitude $\tilde{\mathcal{A}}_{pp}$ confirme la relation (\ref{tension}).  Le spectre contient un vecteur, un tachyon, $(9-p)$ scalaires et un fermion non-chiral dans l'adjointe du groupe de jauge $U(N)$.
   
   Les th\'eories de type I $\footnote{Voir le chapitre 2 pour leur d\'efinitions.}$, $SO(32)$ et $USp(32)$, contiennent des $D9$, $D5$ et $D1$ branes BPS et des branes non-BPS pour les autres valeurs de $p$. Pour $p$ pair les branes non-BPS sont obtenues \`a partir des branes non-BPS de la th\'eorie IIB en agissant avec la projection d'orientifold. Les branes non-BPS $D(-1)$, $D3$ et $D7$ sont obtenues \`a partir des paires brane-antibrane BPS de la th\'eorie IIB.  La $D(-1)$ brane non-BPS dans la th\'eorie $SO(32)$ et la $D3$ brane non-BPS dans la th\'eorie $USp(32)$ sont stables, car leurs spectres ne contiennent pas des tachyons. 
  


\section{Dualit\'es des cordes}

 Il existe 5 th\'eories des cordes supersym\'etriques \`a 10 dimensions. Deux d'entre elles ont \'et\'e discut\'ees dans les sections pr\'ec\'edentes: les th\'eories de type IIA et type IIB. Ce sont des th\'eories des cordes ferm\'ees orient\'ees. Une troisi\`eme th\'eorie est obtenue en jaugeant la parit\'e sur la surface d'univers, $\Omega$, en th\'eorie IIB. Cette nouvelle th\'eorie, dite de type I, contient des cordes ferm\'ees et ouvertes non-orient\'ees. 
Il existe aussi deux th\'eories des cordes ferm\'ees, avec supersym\'etrie $N=1$, dites les cordes h\'et\'erotiques\cite{heterotique}. Nous avons vu que pour les cordes ferm\'ees les secteurs gauche et droit sont pratiquement ind\'ependants, n'\'etant reli\'es que par la condition de raccordement des niveaux. Ceci est le point de d\'epart pour la construction de la corde h\'et\'erotique, qui est obtenue en mettant ensemble le secteur gauche de la corde  bosonique en 26 dimensions avec le secteur droit de la corde fermionique \`a 10 dimensions. 16  des champs bosoniques de gauche, $X^a_L(\tau+\sigma), \ a=1,...,16$ sont, en fait, des degr\'es de libert\'e internes et la th\'eorie r\'esultante est une th\'eorie \`a 10 dimensions. Les champs $X^a_L$ vivent sur un tore 16-dimensionnel. Leurs impulsions prennent des valeurs discr\`etes et sont des vecteurs d'un r\'eseau  de dimension seize, $\Gamma$ :

\be p^a_L\in \Gamma , \quad p_L^a =p_i e_i^a, \ a=1,...,16, \ p_i\in \mathbb{Z},\ee
o\`u $e_i^a$ forment une base de vecteurs de $\Gamma$. L'invariance modulaire \`a une boucle, qui sera discut\'ee dans le chapitre suivant, impose des restrictions sur les r\'eseaux possibles. $\Gamma$ doit \^etre un r\'eseau pair et auto-dual. \`A 16 dimensions il n'y a que deux choix possibles pour  $\Gamma$ conduisant \`a deux th\'eories diff\'erentes: la corde h\'et\'erotique $E_8\times E_8$ et la corde h\'et\'erotique $SO(32)$. Le spectre de masse nulle de ces th\'eories contient les m\^emes \'etats que les secteurs (NS,NS) (graviton, tenseur antisym\'etrique et dilaton) et (NS,R) (gravitino et dilatino) des cordes de type I, plus les bosons de jauge de $E_8\times E_8$ ou $SO(32)$, obtenus par le produit tensoriel des modes bosoniques internes de gauche avec le secteur NS droit, et leurs partenaires supersym\'etriques, les jauginos(produit tensoriel des modes bosoniques internes de gauche avec le secteur R droit).

 \begin{figure}[!h]
\centering
\includegraphics[width=75mm]{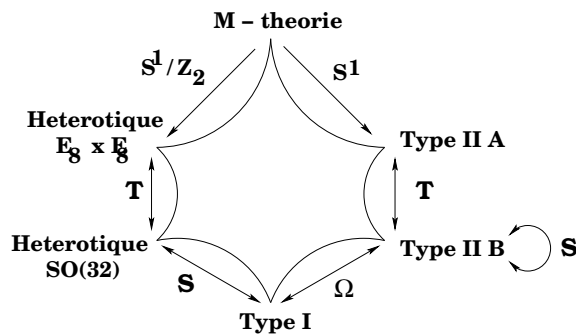}
\caption{Les dualit\'es entre les diff\'erentes th\'eories des cordes}
\label{mtheory}
\end{figure}

 L'existence de cinq th\'eories diff\'erentes mettait en danger l'unicit\'e de la th\'eorie des cordes. Mais en 1995 il a \'et\'e conjectur\'e \cite{mtheory} qu'en fait ces cinq th\'eories sont reli\'ees par des dualit\'es et pourraient \^etre des diff\'erentes facettes d'une seule th\'eorie, pas encore \'elucid\'ee, la M-th\'eorie. Cette id\'ee a \'et\'e sugg\'er\'ee par le fait que la limite de basse \'energie de la th\'eorie de type IIA peut \^etre obtenue par r\'eduction dimensionnelle de la th\'eorie de supergravit\'e \`a 11 dimensions. La M-th\'eorie correspondrait \`a une th\'eorie fondamentale qui est r\'eli\'e \`a la limite de couplage fort de la th\'eorie IIA et dont la limite de basse \'energie serait la supergravit\'e onze-dimensionnelle. La compactification de la M-th\'eorie sur un cercle serait \'equivalente au couplage fort de la th\'eorie IIA, alors que la compactification sur un intervalle avec deux bords sur lequels vivent des champs de jauge $E_8$ correspondrait \`a la corde h\'et\'erotique $E_8\times E_8$.
 
 Les dualit\'es qui relient les diff\'erentes th\'eories de cordes supersym\'etriques en 10 dimensions sont :
 
 \begin{itemize}
\item[a)] La T-dualit\'e, discut\'ee pr\'ec\'edemment, qui rend \'equivalentes une th\'eorie compactifi\'ee sur un cercle de rayon $R$ et une th\'eorie compactifi\'ee sur un cercle de rayon $R'=\alpha'/R$. Sont reli\'ees par cette dualit\'e les th\'eories de type IIA et IIB et \'egalement les deux th\'eories de la corde h\'et\'erotique. La T-dualit\'e est une sym\'etrie  perturbative, satisfaite \`a tous les ordres du d\'eveloppement en puissance de la constante de couplage $g_S$. La T-dualit\'e entre deux th\'eories peut \^etre verifi\'ee ordre par ordre dans le d\'eveloppement perturbatif.
\item[b)] La S-dualit\'e, qui change le signe de la valeur moyenne du dilaton $<\Phi>\rightarrow -<\Phi>$, rendant \'equivalents le r\'egime de couplage faible d'une th\'eorie avec le r\'egime de couplage fort d'une autre th\'eorie. La th\'eorie de type IIB est auto-duale sous la S-dualit\'e, et la th\'eorie de type I est reli\'ee par S-dualit\'e \`a 
la th\'eorie de la corde h\'et\'erotique $SO(32)$. La S-dualit\'e est de nature non-perturbative. La conjecture de  S-dualit\'e entre deux th\'eories est \'etablie en comparant l'action effective des modes de masse nulle qui est fix\'ee par la supersym\'etrie et ne peut pas \^etre modifi\'ee par les corrections des boucles. Un autre outil est offert par le spectre des \'etats BPS, comme les branes, par exemple, qui reste inchang\'e lorsque la constante de couplage varie.
\end{itemize}


\section{Anomalies}

\begin{center}
  \begin{figure}[!h]
\centering
\includegraphics[width=60mm]{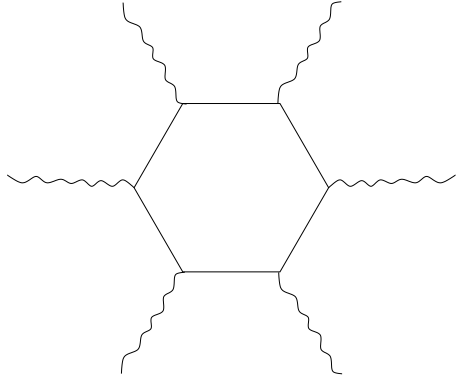}
\caption{Diagramme hexagonal g\'en\'erateur d'anomalies en 10 dimensions.}
\label{hexagone}
\end{figure}    
 \end{center}

 Les sym\'etries des th\'eories classiques peuvent \^etre bris\'ees par des effets quantiques qu'on appelle des anomalies. Les anomalies proviennent des diagrammes de Feynman qui n'admettent pas des r\'egulateurs compatibles avec la conservation simultan\'ee de tous les courants attach\'es. L'interpr\'etation des anomalies diff\`ere selon s'il s'agit d'une sym\'etrie globale ou locale qui est bris\'ee. La brisure des sym\'etries globales peut \^etre utile dans certaines th\'eories pour des raisons ph\'enom\'enologiques. En revanche la brisure des sym\'etries locales, comme l'invariance de jauge ou la covariance de la th\'eorie , conduit \`a des inconsistances. Les anomalies apparaissent dans les th\'eories chirales. En cons\'equece les anomalies existent seulement en dimension paire$\footnote{En dimension impaire on ne peut pas avoir des spineurs de Weyl.}$$\footnote{Des anomalies de parit\'e peuvent exister en dimension impaire.}$. En dimension $D=2k$, le plus simple diagramme de Feynman, qui est potentiellement source d'anomalies, est une boucle avec $k+1$ bras externes. Pour les supercordes, $D=10$, cela correspond \`a un diagramme hexagonal(fig \ref{hexagone}). Les lignes externes peuvent \^etre des bosons de jauge ou des gravitons et les lignes internes sont des fermions ou des bosons auto-duaux(ou anti-auto-duaux).
 En dimension $D$ les anomalies sont repr\'esent\'ees par des $(D+2)$-formes, $I_{D+2}$, qui sont des polyn\^omes de $trF^m$ et $trR^m$, o\`u $F$ est le tenseur du champ de jauge et $R=d\omega+\omega^2$, avec $\omega$ la connexion de spin, est d'une certaine mani\`ere l'\'equivalent de $F$ pour la gravitation$\footnote{Pour une revue compl\`ete du formalisme des anomalies voir le volume 2 de "Superstring Theory", M.B.~Green, J.M.~Schwarz et E.~Witten, Cambridge University Press, 1987.}$. $I_{D+2}$ est, par construction, invariante de jauge et exacte. Elle d\'etermine localement une $(D+1)$-forme, $I_{D+1}$ :
 
 \be I_{D+2}=dI_{D+1},\ee
 dont la variation sous les transformation de jauge est exacte, en cons\'equence de l'invariance de jauge de $I_{D+2}$ :
 
 \be \delta I_{D+1}=dI_D.\ee
 
  L'anomalie est d\'etermin\'ee comme l'int\'egrale de $I_D$. La condition d'annulation des anomalies est donn\'ee par $I_{D+2}=0$.
  
  Rappelons nous que la th\'eorie de type IIB est une th\'eorie chirale. Elle peut donc pr\'esenter des anomalies. La th\'eorie IIB ne contient pas de bosons de jauge, les anomalies possibles sont, donc, de nature purement gravitationnelle.

   En d\'efinissant :
   
   \be Y_{2m}=\sum_{i=1}^{2k+1} ({1\over 2}x_i)^{2m}= {1\over 2}\left(-{1\over 4}\right)^m tr R^{2m},\ee
   les polyn\^omes pour les anomalies gravitationelles sont donn\'ees par$\footnote{Un facteur global $-i(2\pi)^{-D/2}$ a \'et\'e enlev\'e dans les formules suivantes.}$ :

  \begin{itemize}
\item[i) pour un fermion chiral de spin 1/2:] \ba I_{1/2}=\prod_{i=1}^{2k+1} \left({{1\over 2}x_i\over \mbox{sinh}{1\over 2}x_i}\right)&=&1-{1\over 6}Y_2+{1\over 180}Y_4+{1\over 72}Y_2^2-{1\over 2835}Y_6-\nonumber \\ &-&{1\over 1080}Y_2Y_4-{1\over 1296}Y_2^3+...,\label{i}\ea

\item[ii) pour un fermion chiral de spin 3/2:] $$I_{3/2}=I_{1/2}(-1+2\prod_{i=1}^{2k+1} \mbox{cosh} x_i)=I_{1/2}(D-1+4Y_2+{4\over 3}Y_4+{8\over 45}Y_6...) ,$$

\item[iii) pour une tenseur auto-dual:]\ba I_A&=&-{1\over 8}\prod_{i=1}^{2k+1} \left({x_i\over \mbox{tanh}x_i}\right)=-{1\over 8}-{1\over 6}Y_2+({7\over 45}Y_4-{1\over 9}Y_2^2)+\nonumber\\ &+&{1\over 2835}(-496Y_6+588Y_2Y_4-140Y_2^3)+...,\ea

\end{itemize}

 Ces expressions contiennent les contributions des anomalies pour les dimensions de la forme $4k+2$ pour tout $k$. Pour obtenir l'anomalie pour la dimensions $D=4k+2$ il faut extraire 
de ces expressions les termes d'ordre $2(k+1)$. En revenant \`a la th\'eorie IIB le spectre de la th\'eorie de basse \'energie contient un gravitino de Weyl, un spineur de Weyl de chiralit\'e opos\'ee et une 4-forme auto-duale. Leurs contributions :

\ba (I_{1/2})_6&=& -{1\over 2835}Y_6-{1\over 1080}Y_2Y_4-{1\over 1296}Y_2^3 , \nonumber\\
(I_{3/2})_6&=&-{495\over 2835}Y_6-{225\over 1080}Y_2Y_4+{63\over 1296}Y_2^3,\nonumber\\
(I_A)_6&=&-{496\over 2835}Y_6+{224\over1080}Y_2Y_4-{64\over1296}Y_2^3,\ea
  au polyn\^ome total d'anomalie: $I_{IIB}=(I_{1/2})_6-(I_{3/2})_6 +(I_{A})_6$ s'annulent compl\`etement: $I_{IIB}=0$. La th\'eorie IIB est consistante. 
  
\begin{figure}[!h]
\centering
\includegraphics[width=75mm]{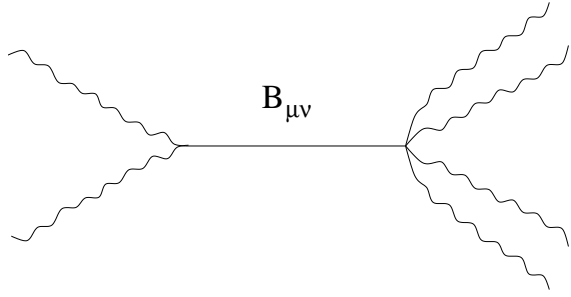}
\caption{Diagramme qui contribue aux anomalies. Une 2-forme de masse nulle est \'echang\'ee entre deux bosons de jauge(ou gravitons) et quatre bosons de jauge(ou quatre gravitons ou deux bosons de jauge et quatre gravitons).}
\label{anomalb}
\end{figure}

  Pour les th\'eories qui contiennent des bosons de jauge il faut prendre en compte les anomalies de jauge ou mixtes. La formule utile dans ce cas est :
  
  \be I_{1/2}(F,R)=tr(e^{iF})I_{1/2}(R),\ee  
   avec $I_{1/2}(R)$ donn\'e par (\ref{i}). Par exemple la th\'eorie de type I, qui sera pr\'esent\'ee dans le chapitre suivant, contient des bosons de jauge et les anomalies de jauge et mixtes re\c{c}oivent une contribution venant des jauginos. Cette contribution est annul\'ee par le m\'ecanisme de Green-Schwarz \cite{gs}. Ce m\'ecanisme consiste \`a introduire un nouveau couplage, qui n'est pas pr\'esent dans la supergravit\'e minimale, du tenseur antisym\'etrique, $B_{\mu\nu}$, $\footnote{En r\'ealit\'e, comme on le verra dans la chapitre suivant, le tenseur antisym\'etrique, $B_{\mu\nu}$, du secteur (NS,NS) est \'elimin\'e dans la th\'eorie de type I, mais une 2-forme est toujours pr\'esente, la $C_2$ du secteur (R,R).}$ aux bosons de jauge et/ou au graviton :
  
   \be S_2=\int d^{10}x B\wedge tr F^4,\label{couplge}\ee
  o\`u $tr F^4$ est un expression g\'en\'erique pour les termes d'ordre 4, $(tr F^2)^2$, $tr R^4$,  $tr F^2tr R^2$ etc. Dans la supergravit\'e minimale  de type I il y a d\'ej\`a un terme de couplage du tenseur $B$ aux bosons de jauge :

\be S_1=\int d^{10}x {}^*(dB)\wedge tr(A\wedge dA),\ee
qui provient du terme $H^2$. En effet dans la supergravit\'e minimale N=1 en 10 dimensions, coupl\'ee \`a un syst\`eme Yang-Mills, la supersym\'etrie impose que le tenseur de la 2-forme, $B_{\mu\nu}$, soit g\'en\'eralis\'e de $H=dB$ \`a :

\be H=dB- \omega_{3Y}, \ee
avec $\omega_{3Y}$ la 3-forme  Chern-Simons, donn\'ee par:
\be \omega_{3Y}=tr(AdA+{2\over 3}A^2).\label{chern}\ee
Le tenseur de $B$, $H$, est invariant sous une transformation de jauge:

\be \delta A=d\Lambda +[A,\Lambda],\ee
 ce qui implique la transformation de jauge suivante pour $B$: $\delta B=\omega_{2Y}$, avec $\omega_{2Y}=tr(\Lambda dA)$, $\delta\omega_{3Y}=d\omega_{2Y}$. $S_1$ est invariante de jauge si $H$ est invariant de jauge, c'est \`a dire si $B$ n'est pas invariant de jauge. En revanche, $S_2$ est invariante de jauge si $B$ est invariant de jauge.
$S_1$ et $S_2$ ne peuvent, donc,  pas \^etre invariants de jauge en m\^eme temps. Le diagramme de Feynman de la figure \ref{anomalb} contient les deux couplages et g\'en\'ere une contribution aux anomalies qui compense les anomalies de jauge et gravitationnelles venant du diagramme hexagonal. 

 Les anomalies peuvent \^etre compens\'ees par le m\'ecanisme de Green-Schwarz si le polyn\^ome d'anomalies peut \^etre factoris\'e sous la forme:
 
 \be I_{12}=(trR^2+k \ trF^2)X_8,\label{factor}\ee
 avec $k$ une constante et $X_8$ un polyn\^ome d'ordre quatre en $F$ et $R$.  Comme $I_{12}$ est exacte il en r\'esulte que $X_8$ est exact aussi. Il y a deux choix possibles pour $I_{11}$. On a $trF^2=d\omega_{3Y}$ et, \'egalement, $tr R^2=d\omega_{3L}$ , o\`u $\omega_{3L}$ est donn\'ee par une expression analogue \`a (\ref{chern}) en rempla\c{c}ant $A$ par la connexion de spin, $\omega$. 
 Si on d\'efinit $X_7$ comme $X_8=d X_7$ la forme la plus g\'en\'erale pour $I_{11}$ est donn\'ee par:
 
 \be I_{11}={1\over 3}(\omega_{3L}+k\ \omega_{3Y})X_8+{2\over 3}(tr R^2+k\ tr F^2)X_7+\alpha d[((\omega_{3L}+k\omega_{3Y})X_7],\ee 
 o\`u le dernier terme repr\'esente l'ambiguit\'e dans la d\'efinition de $I_{11}$ avec $\alpha$ un param\`etre arbitraire. $I_{10}$ est alors obtenue, modulo une forme exacte, par $\delta I_{11}=d I_{10}$, donc 
 
 \be I_{10}=({2\over 3}+\alpha)(tr R^2+k\ tr F^2) X_6+({1\over 3}-\alpha)(\omega_{2L}+k\ \omega_{2Y})X_8,\ee
 avec $\delta X_7=dX_6$ et $\delta \omega_{3L}=d\omega_{2L}$. L'anomalie, $G=\int I_{10}$ , est donn\'ee par:
 
 \be G=({2\over 3}+\alpha)\int (\omega_{3L}+k\ \omega_{3Y})dX_6+ ({1\over 3}-\alpha)\int (\omega_{2L}+k\ \omega_{2Y})X_8. \ee
 
 Cette anomalie peut \^etre annul\'ee par l'ajout dans le Lagrangian des termes:
 
 \be S=({1\over 3}-\alpha)\int B\ X_8-({2\over 3}+\alpha)\int (\omega_{3L}+k\ \omega_{3Y})X_7\ee
 
 et si la transformation de jauge pour le champ $B$ est g\'en\'eralis\'e \`a :
 
 \be \delta B=\omega_{2Y}-\omega_{2L},\ee 
 c'est \`a dire si le tenseur du $B$ est g\'en\'eralis\'e \`a :
 
 \be H=dB-\omega_{3Y}+\omega_{3L}.\ee
 
 Pour la th\'eorie de Type I (et la corde h\'et\'erotique $SO(32)$) l'anomalie totale donn\'ee par les diagrammes hexagonales est:
 \be I_{\rm type\  I}= I_{3/2}(R)-I_{1/2}(R)-I_{1/2}(F,R)\ee
  et explicitement pour $m$ jauginos:
  
  \ba I_{12}=&&-{1\over 720}Tr F^6+{1\over 24\cdot 48}TrF^4tr R^2\nonumber\\ &&-{1\over 256}Tr F^2[{1\over 45}tr R^4+{1\over 36}(tr R^2)^2]\nonumber\\ &&+{m-496\over 64}[{1\over 2\cdot 2835}tr R^6+{1\over 4\cdot 1080}tr R^2 tr R^4+{1\over 8\cdot 1296}(tr R^2)^3]\nonumber\\  &&+{1\over 384}tr R^2 tr R^4+{1\over 1536}(tr R^2)^3,\label{12}\ea
 o\`u $Tr$ designe la trace dans la repr\'esentation adjointe(les traces dans la repr\'esentation fondamentale sont not\'ees $tr$). Pour que l'anomalie (\ref{12}) puisse \^etre factoris\'ee il faut que le coefficient de $tr R^6$ s'annule, c'est \`a dire que le groupe de jauge doit avoir 496 g\'en\'erateurs, $m=496$. Ensuite en exprimant les traces dans la repr\'esentation adjointe en fonction de celles dans la r\'epr\'esentation fondamentale, pour un groupe $SO(n)$,
 
 \ba Tr F^2&=&(n-2)tr F^2\nonumber\\
 Tr F^4&=&(n-8)tr F^4+3(tr F^2)^2\nonumber\\
 Tr F^6&=& (n-32)tr F^6+15 tr F^2tr F^4,\ea
 on voit que pour le groupe $SO(32)$ il n'y a pas de contribution $tr F^6$ dans les anomalies. En plus ce groupe a $32\cdot(32-1)/2=496$ g\'en\'erateurs. L'anomalie se factorise dans la forme (\ref{factor}) avec $k=-1/30$ et 
 
 \be X_8={1\over 24} Tr F^4-{1\over 7200}(Tr F^2)^2-{1\over 240}Tr F^2 tr R^2+{1\over 8}tr R^4+{1\over 32}(tr R^2)^2.\ee
 
 L'anomalie peut, donc, \^etre annul\'ee par le m\'ecanisme de Green-Schwarz.
 
  Un autre groupe avec 496 g\'en\'erateurs est $E_8\times E_8$ ($E_8$ poss\`ede 248 g\'en\'erateurs), qui correspond \`a la corde h\'et\'etorique. Pour ce groupe $Tr F^6$ est proportionnelle \`a $(TrF^2)^3$ et $Tr F^4$ \`a $(TrF^2)^2$. Le factorisation dans la forme (\ref{factor}) peut avoir lieu.

  D'autres possibilit\'es pour le groupe de jauge sont $[U(1)]^{496}$ et $E_8\times [U(1)]^{248}$, mais on ne conna\^it pas des th\'eories des cordes qui poss\`edent ces groupes de jauge.



\chapter{Orbifolds et Orientifolds}
\section{Fonction de partition - le tore}
À la difference de la théorie des champs, en théorie des cordes
 les amplitudes du vide
 jouent un role important, car elles determinent le spectre perturbtif. Int\'eressons-nous \`a l'amplitude du vide \`a une boucle pour les cordes ferm\'ees bosoniques.
 
 \begin{figure}[!h]
\centering
\includegraphics[width=55mm]{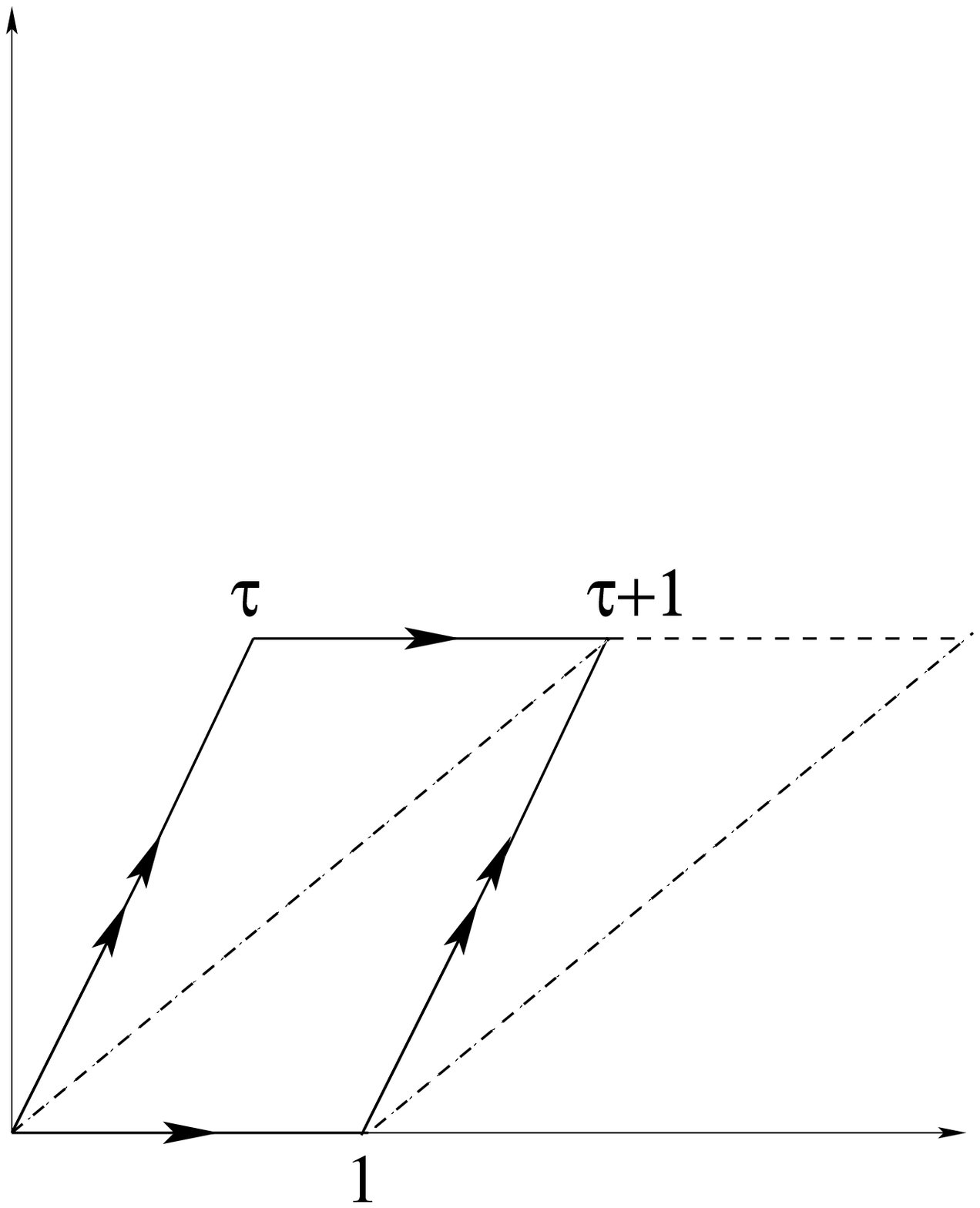}
\caption{Repr\'esentation du tore dans le plan complexe}
\label{tore}
\end{figure} 
  Une corde fermée qui effectue une boucle dessine un tore. Le tore peut \^etre  
  repr\'esent\'e comme un parall\'elogramme dans le plan complexe, avec les c\^ot\'es oppos\'ees identifi\'ees (fig \ref{tore}). La structure complexe du tore est d\'ecrite par un param\`etre complexe, $\tau=\tau_1+i\tau_2$, appel\'e param\`etre de Teichm\"uller. Mais toutes les valeurs de $\tau$ ne correspondent pas \`a des tores in\'equivalents. Plus pr\'ecisement toutes les valeurs reli\'ees par la transformation : 
  
  \be \tau\rightarrow {a\tau+b\over c\tau+d},\quad\mbox{avec}\  ad-bc=1,\quad  a,b,c,d\in \mathbb{Z}\label{modular}\ee
  d\'ecrivent le m\^eme tore. La transformation (\ref{modular}) d\'ecrit le groupe modulaire $PSL(2,\mathbb{Z})=SL(2,\mathbb{Z})/\mathbb{Z}_2$, qui est engendr\'e par les deux transformations :
  
  \be T: \tau\rightarrow\tau+1, \quad S: \tau\rightarrow -{1\over \tau}.\ee
  On peut noter que la transformation $S$ \'echange les c\^ot\'es horizontal et oblique, alors que la transformation $T$ red\'efinit le c\^ot\'e oblique. 
  
  Les valeurs ind\'ependantes de $\tau$ sont contenues \`a l'int\'erieur de la r\'egion fondamentale :
  
  \be \mathcal{F}= \{-1/2<\tau_1\le 1/2, \mid \tau\mid \ge 1\}.\ee
   
  L'expression de l'amplitude du vide à une boucle pour un champ de masse $M$ est donn\'ee par (voir \cite{os}) :
 
 \be \Gamma=-{V\over 2(4\pi)^{D/2}}\int_\epsilon^\infty {dt\over t^{D/2+1}}\mbox{Str}(e^{-tM^2}),\label{str}\ee
 o\`u la trace Str prend en compte la multiplicit\'e et les signe "-" pour les \'etats fermioniques, $V$ est le volume de l'espace-temps  et $\epsilon$ est un cut-off ultraviolet.

 Appliquons cette formule aux cordes ferm\'ees bosoniques. On rappelle que la dimension critique dans ce cas vaut $D=26$ et que l'op\'erateur de masse est donn\'e par :
 
 \be M^2={2\over \alpha'}(N+\bar{N}-2),\ee
 avec la contrainte $N-\bar{N} =0$. Cette contrainte peut \^etre impos\'ee en introduisant une fonction $\delta$ :
 
 \be \delta(N-\bar{N})=\int_{-1/2}^{1/2}ds\  e^{2\pi i s (N-\bar{N})}.\ee 
   
   On obtient alors :
   
   \be \Gamma= -{V\over 2(4\pi)^{13}}\int_{-1/2}^{1/2}ds\int_\epsilon^\infty {dt\over t^{14}}tr\left(e^{-{2\over \alpha'}(N+\bar{N}-2)t}  e^{2\pi i s(N-\bar{N} )}\right)\label{tr}.\ee  
   
   L'expression (\ref{tr}) peut \^etre mise sous une forme plus concise en d\'efinissant le param\`etre de Schwinger :

   \be \tau=\tau_1+i\tau_2=s+i{t\over \alpha'\pi}\ee
  et en d\'efinissant :
   
   \be q=e^{2\pi i \tau}, \quad \bar{q}=e^{-2\pi i \bar{\tau}},\ee
  pour obtenir :
   
\be \Gamma = -{V\over
  2(4\pi^2\alpha')^{13}}\int_{-1/2}^{1/2}d\tau_1\int_\epsilon^\infty
  {d\tau_2\over \tau_2^{14}}tr\ q^{N-1}\ \bar{q}^{\bar{N}-1}.\ee

 Le param\`etre de Schwinger, $\tau=\tau_1+i\tau_2$, est identifi\'e au param\`etre Teichm\"uller  du tore. Par cons\'equent il faut restreindre l'int\'egration \`a la r\'egion fondamentale $\mathcal{F}$. Cette restriction introduit un cutoff UV. Apr\'es un changement de normalisation l'amplitude sur le tore prend la forme :

\be \mathcal{T}=\int_{\mathcal{F}} {d^2\tau\over \tau_2^2}{1\over \tau_2^{12}}tr\  q^{N-1}\bar{q}^{\bar{N}-1}\label{tor}.\ee 
 Cette expression d\'efinit la fonction de partition pour la corde ferm\'ee bosonique et peut \^etre g\'en\'eralis\'ee pour tout mod\`ele de cordes ferm\'ees orient\'ees si on conna\^it  les op\'erateurs de Virasoro. 
  
  Dans la suite on va calculer explicitement l'amplitude (\ref{tor}) pour la corde bosonique. On rapelle l'expression des op\'erateurs de Virasoro dans ce cas : 
  
  \be N=\sum_{n+1}^\infty n\ a_n^{\dag i}\ a_{n,i}, \quad \bar{N}=\sum_{n=1}^\infty n\ \bar{a}_n^{\dag i}\ \bar{a}_{n,i}, \ee
  o\`u nous avons utilis\'e les op\'erateurs d'annihilation et de cr\'eation avec la normalisation habituelle pour l'oscillateur harmonique. On obtient alors :
  
  \ba &&tr\ q^{N-1}={1\over q}\prod_{i=1}^{24}\prod_{n=1}^{\infty} tr\ q^{ n\ a_n^{\dag i}\ a_{n,i}}={1\over q}\prod_{i=1}^{24}\prod_{n=1}^{\infty}\sum_{k=0}^{\infty}<k\mid q^{ n\ a_n^{\dag i}\ a_{n,i}}\mid k>=\nonumber\\
 && ={1\over q}\prod_{i=1}^{24}\prod_{n=1}^{\infty}\sum_{k=0}^{\infty} (q^n)^k ={1\over q}\prod_{i=1}^{24}\prod_{n=1}^{\infty}(1+q^n+q^{2n}+...)={1\over q}\prod_{i=1}^{24}\prod_{n=1}^{\infty}{1\over 1-q^n}=\nonumber\\ 
&&  =\left({1\over q^{1/24} \prod_{n=1}^{\infty}(1-q^n)}\right)^{24}
  \ea
   et la fonction de partition vaut :
  
  \be \mathcal{T}=\int_{\mathcal{F}} {d^2\tau\over \tau_2^2}{1\over \tau_2^{12}}{1  \over \mid \!\!\eta(\tau)\!\!\mid^{48}},\ee
   o\`u $\eta$ est la fonction de Dedekind :
  
  \be \eta(\tau)=q^{1/24} \prod_{n=1}^{\infty}(1-q^n).\ee
  
   Les transformations modulaires de la fonction $\eta$ :
  
  \be T:\ \eta(\tau+1)=e^{i\pi\over 12}\eta(\tau), \quad S:\ \eta(-1/\tau)=\sqrt{-i\tau}\eta(\tau),\ee
   impliquent l'invariance de la combinaison $\tau_2^{1\over 2}\mid\!\eta\!\mid^2$. La mesure $
d^2\tau/ \tau_2^2$ \'etant \'egalement invariante sous $S$ et $T$ il r\'esulte que la fonction de partition est invariante sous le groupe modulaire.
    


\section{Orientifold}
 
 Les mod\`eles d'orientifold\cite{orientifold} sont des th\'eories de cordes ferm\'ees non-orient\'ees ou des th\'eories des cordes ferm\'ees et ouvertes non-orient\'ees.

\subsection{Cordes fermées non-orientées}

 Jusqu'\`a pr\'esent nous avons discut\'e des th\'eories des cordes orient\'ees. Mais  
 la surface d'univers poss\`ede une sym\'etrie que nous n'avons pas prise en compte. La transformation de coordonn\'ees :
 
 \be \sigma'=l-\sigma, \quad \tau'=\tau,\label{omega}\ee
 avec $l=\pi$ pour la corde ouverte et $l=2\pi$ pour la corde ferm\'ee, change l'orientation 
 de la surface d'univers. Cette sym\'etrie est g\'en\'er\'ee par la parit\'e sur la surface d'univers, $\Omega$. A partir des d\'eveloppements en oscillateurs (\ref{xlxr}) et (\ref{open}) on peut en d\'eduire l'action de $\Omega$ sur les oscillateurs :
 
 \be \Omega \alpha_n^\mu\Omega^{-1}=\bar{\alpha}_n^\mu,\quad \Omega \bar{\alpha}_n^\mu\Omega^{-1}=\alpha_n^\mu, \ee 
  pour la corde ferm\'ee, et
 
 \be \Omega \alpha_n^\mu\Omega^{-1}=(-1)^n\alpha_n^\mu \ee
 pour la corde ouverte. A partir de (\ref{omega}) on voit que $\Omega^2=1$. Le spectre est, donc, partag\'e en \'etats pairs et impairs sous $\Omega$. \'Etant donn\'e que deux \'etats impairs qui interagissent donnent naissance \`a un \'etat pair il r\'esulte que que la seule mani\`ere consistente de tronquer le spectre est de garder les \'etats pairs. Par exemple pour  les cordes ferm\'ees bosoniques au niveau non-massif on gardera le dilaton et le graviton qui sont pairs sous $\Omega$, alors que le tenseur antisym\'etrique $B_{\mu\nu}$ sera \'elimin\'e. Au niveau non-massif  le mod\`ele des cordes ferm\'ees orient\'ees bosoniques contenait $(24)^2$ degr\'es de libert\'e, dans  le mod\`ele des cordes non-orient\'ees il restent $24(24+1)/2$ degr\'es de libert\'e. A noter que le tachyon reste pr\'esent dans les th\'eories non-orient\'ees.
 
 \begin{figure}[!h]
 \centering
\includegraphics[width=85mm]{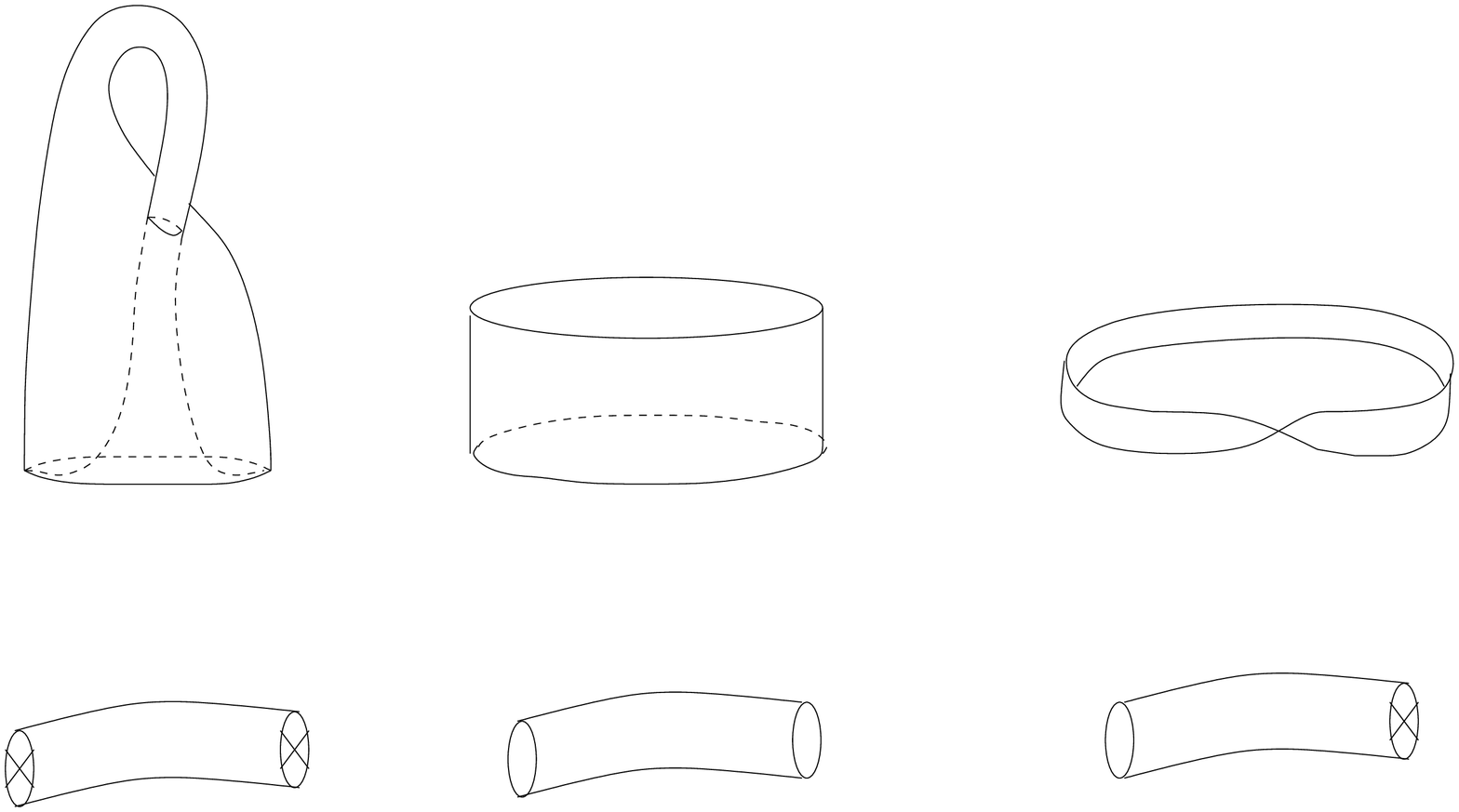}
\caption{La bouteille de Klein. L'anneau. La bande de M\"obius (en haut: la canal direct; en bas: le canal transverse)}
\label{klanmo}
\end{figure}

 Dans la section pr\'ecedente nous avons calcul\'e l'amplitude du vide \`a une boucle pour les cordes ferm\'ees orient\'ees qui est donn\'ee par l'amplitude sur le tore, qui est la seule surface ferm\'ee orient\'ee avec nombre d'Euler z\'ero. En revanche les cordes non-orient\'ees peuvent parcourir aussi des surfaces non-orient\'ees, ce qui laisse encore une possibilit\'e, la bouteille de Klein$(h=0,\ b=0, \ c=2)$.
 
 \begin{figure}[!h]
\centering
\includegraphics[width=65mm]{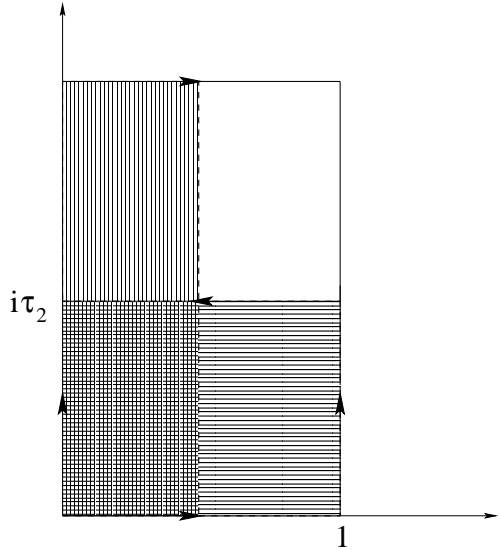}
\caption{Les polygones fondamentaux pour la bouteille de Klein}
\label{klein}
\end{figure} 
 
  Comme pour le tore on peut repr\'esenter la bouteille de Klein dans le plan complexe (fig \ref{klein}) et on a le choix entre deux polygones. Pour le premier, de c\^ot\'es 1 et $i\tau_2$, les c\^ot\'es verticaux sont identifi\'es comme dans le cas du tore, mais les c\^ot\'es horizontaux ont des orientations oppos\'ees. Cette repr\'esentation d\'ecrit la propagation dans une boucle d'une corde ferm\'ee qui subit une inversion de son orientation, $\tau_2$ \'etant le temps propre sur la surface d'univers. Le deuxi\`eme polygone est obtenu en divisant par deux le c\^ot\'e horizontal et en doublant le  c\^ot\'e vertical. 
  On obtient alors un tube termin\'e par deux crosscaps et le temps s'\'ecoule "horizontalement", cette repr\'esentation d\'ecrivant la propagation d'une corde ferm\'ee \`a l'ordre des arbres entre deux crosscaps(fig \ref{klanmo}). Ces deux choix de temps propre sont reli\'ees par une transformation $S$.
  
   La bouteille de Klein est obtenue \`a partir du tore de double recouvrement, avec param\`etre de Teichm\"uller $2i\tau_2$, par l'involution $z\rightarrow 1-\bar{z}+i\tau_2$. 
 
 Pour d\'ecrire le spectre des cordes ferm\'ees non-orient\'ees il faut ins\'erer dans la fonction de partition le projecteur ${1+\Omega\over 2}$. Ceci revient \`a diviser par deux la contribution du tore et \`a rajouter l'amplitude provenant de la bouteille de Klein :

 \be \Gamma=\int {d^2\tau\over \tau_2^2}{1\over \tau_2^{12}}tr \
 {1+\Omega\over 2}q^{N-1}\ \bar{q}^{\bar{N}-1}= {1\over 2}{\cal
   T}+{\cal K},\ee
où: 
\be {\cal K}= {1\over 2}\int {d^2\tau\over \tau_2^2}{1\over \tau_2^{12}}tr \left(
  \ q^{N-1}\ \bar{q}^{\bar{N}-1}\Omega\right).\ee

En utilisant le fait que $\Omega\mid L,R>=\mid R,L>$ on obtient:
\ba &&tr \left(q^{N-1}\ \bar{q}^{\bar{N}-1}\Omega\right)= \sum_{L,R} <L,R\mid q^{N-1}\ \bar{q}^{\bar{N}-1}\Omega\mid L,R>= \nonumber \\
&&=\sum_{L,R} <\!\!L,R\!\mid q^{N-1}\bar{q}^{\bar{N}-1}\mid\!\!R,L\!\!>=\sum_L<\!\!L,L\!\!\mid (q\bar{q})^{N-1}\mid\!\!L,L\!\!>.\ea

Le calcul de la trace est ensuite analogue \`a celui effectu\'e dans le cas du tore et la r\'esultat est :

\be \mathcal{K}={1\over 2}\int_0^\infty {d\tau_2\over \tau_2^{14}}{1\over \eta^{24}(2i\tau_2)}.\ee

Ceci repr\'esente l'amplitude de la bouteille de Klein dans la canal direct (ou \`a une boucle) et elle d\'epend du temps propre "vertical" $\tau_2$. L'amplitude dans le canal transverse ( ou \`a l'ordre des arbres) est obtenue par une transformation $S$ et elle d\'epend du temps propre "horizontal" $\tilde{\tau}=-1/\tau_{\tt{Klein}}=-1/(2i\tau_2)=il$ :

\be \tilde{\mathcal{K}}={2^{13}\over 2}\int_0^\infty dl{1\over \eta^{24}(il)}.\ee

Il est int\'eressant de regarder le d\'eveloppement en puissances de $q$ des int\'egrands de $\mathcal{T}$ et $\mathcal{K}$, en faisant attention \`a ne garder pour le tore que les termes avec puissances \'egales de $q$ et $\bar{q}$ qui respectent la condition de raccordement des niveaux :

\ba {1\over 2}\mathcal{T}\rightarrow {1\over 2}((q\bar{q})^{-1}+(24)^2+...),\nonumber \\
\mathcal{K}\rightarrow {1\over 2}((q\bar{q})^{-1}+(24)+...).\ea

 On voit donc que la combinaison ${1\over 2}\mathcal{T}+\mathcal{K}$ reproduit bien le spectre projet\'e, en particulier au niveau non-massif on retouve  $24(24+1)/2$ degr\'es de libert\'e et le tachyon est toujours pr\'esent.

L'amplitude de la bouteille de Klein n'est pas proteg\'ee par l'invariance modulaire, comme le tore, et elle pr\'esente dans le canal direct une divergence ultraviolete, $\tau_2\rightarrow 0$, qui, dans le canal transverse, devient une divergence infrarouge, $l\rightarrow \infty$. Il est plus simple d'examiner cette divergence dans le canal transverse. Un \'etat  de masse $M$ ayant une contribution proportionnelle \`a:

\be \int_0^\infty dl\ e^{-M^2l}={1\over M^2},\ee
on peut voir que la divergence provient des \'etats de masse nule avec impulsion nulle. Dans le cas des cordes ferm\'ees bosoniques il s'agit du dilaton. Comme on va le voir dans la suite, l'introduction d'un secteur des cordes ouvertes dans la th\'eorie permet d'\'eliminer la divergence.


\subsection{Cordes ouvertes - facteur de Chan-Paton - annulation des tadpoles}

Les cordes ouvertes admettent une g\'en\'eralisation gr\^ace au fait qu'elles poss\`edent deux points sp\'eciaux, leurs bouts. On peut attacher \`a chaque bout une charge, nomm\'ee charge de Chan-Paton\cite{chanpaton}. Ces nouveaux degr\'es de libert\'e ne sont pas dynamiques, mais permettent l'introduction des groupes de jauge non-abeliens en th\'eorie des cordes\cite{groupes}. 

\begin{figure}[!h]
\centering
\includegraphics[width=65mm]{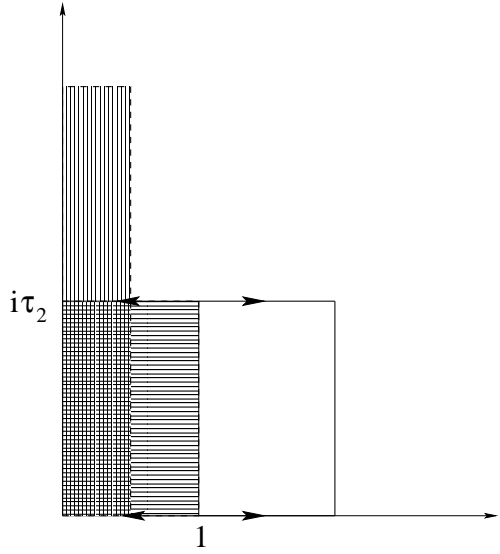}
\caption{Les polygones fondamentaux pour l'anneau}
\label{anneau}
\end{figure}

Une corde ouverte qui effectue une boucle d\'ecrit un anneau. Comme pour le tore et la bouteille de Klein l'anneau peut \^etre repr\'esent\'e dans le plan complex par un polygone(fig \ref{anneau}) et il y a deux choix pour ce polygone. Pour la premi\`ere, de c\^ot\'es 1 et $i\tau_2$, les c\^ot\'es horizontaux sont identifi\'es et les c\^ot\'es verticaux correspondent aux deux fronti\`eres. $\tau_2$ est le temps propre pour une corde ouverte qui se propage dans une boucle. Dans la deuxi\`eme repr\'esentation le temps s'\'ecoule horizontalement et c'est une corde ferm\'ee qui se propage entre deux fronti\`eres(fig \ref{klanmo}). Ces deux repr\'esentations sont reli\'ees par une transformation $S$. L'anneau peut \^etre obtenu \`a partir de son tore de double recouvrement de param\`etre de Teichm\"uller $i\tau_2/2$ par les involutions: $z\rightarrow -\bar{z}$ et $z\rightarrow 2-\bar{z}$. 

 Calculons l'amplitude de l'anneau pour la corde bosonique. On associe une multiplicit\'e $N$ \`a chaque bout de la corde ouverte. En utilisant l'expression de l'op\'erateur de masse : $M^2= {1\over \alpha'}(N-1)$ dans la formule (\ref{str}) on obtient :
 
 \be \mathcal{A}={N^2\over 2}\int_0^\infty {d\tau_2\over \tau_2^{14}} tr(q^{{1\over 2}(N-1)})={N^2\over 2}\int_0^\infty {d\tau_2\over \tau_2^{14}} {1\over \eta^{24}(i\tau_2/2)}.\label{annulus}  \ee

L'amplitude dans le canal transverse est obtenue par la transformation $S$ \`a partir de  (\ref{annulus}) et d\'epend du param\`etre $\tilde{\tau}=-1/\tau=2/(i\tau_2)=il$ :

\be \tilde{\mathcal{A}}={N^22^{-13}\over 2}\int_0^\infty dl {1\over \eta^{24}(il)}.\ee

\begin{figure}[!h]
\centering
\includegraphics[width=65mm]{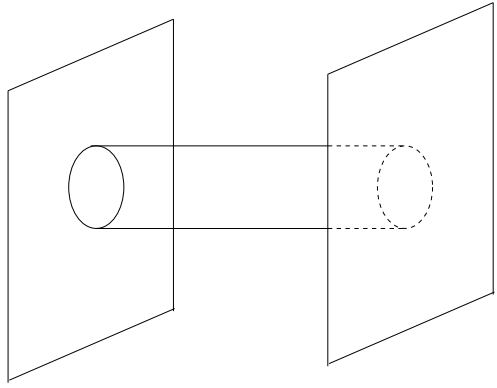}
\caption{Deux $Dp$ - branes interagissant par l'\'echange d'une corde ferm\'ee.}
\label{branes}
\end{figure}

 Dans le chapitre pr\'ecedant nous avons introduit la notion de $Dp$-brane. Deux $Dp$-branes 
 interagissent en \'echangeant des cordes ferm\'ees comme dans la figure \ref{branes}. L'amplitude est un anneau et peut \^etre vue \'egalement comme la propagation \`a une boucle d'une corde ouverte tendue entre les $Dp$-branes. Elle est donn\'e par une expression de la forme (\ref{annulus}). Si on r\'e\'ecrit l'amplitude (\ref{annulus}) en reintroduisant tous les facteurs initiaux,
 
 \be \mathcal{A}={N^2V_{26}\over 2(4\pi^2\alpha')^{13}}\int_0^\infty {d\tau_2\over \tau_2^{14}}{1\over \eta^{24}(i\tau_2/2)},\ee 
 alors l'interaction de deux $Dp$-branes s'obtient en rempla\c{c}ant le nombre de dimensions 26 avec $p+1$ et en rajoutant un terme de masse additionnel li\'e \`a la tension de la corde ouverte tendue entre les deux $Dp$-branes\cite{branes} :
 
 \be \mathcal{A}={V_{p+1}\over (4\pi^2\alpha')^{(p+1)/2}}\int_0^\infty {d\tau_2\over \tau_2^{(p+1)/2+1}} e^{-\tau_2Y\cdot Y\over4\pi\alpha'} {1\over \eta^{24}(i\tau_2/2)},\label{intbranes}\ee 
 o\`u $Y^m=x_1^m-x^m_2$ est la s\'eparation des branes et le facteur de Chan-Paton $N^2$ est devenu 2, car il y a deux orientations possibles de la corde ouverte.
 
\begin{figure}[!h]
\centering
\includegraphics[width=65mm]{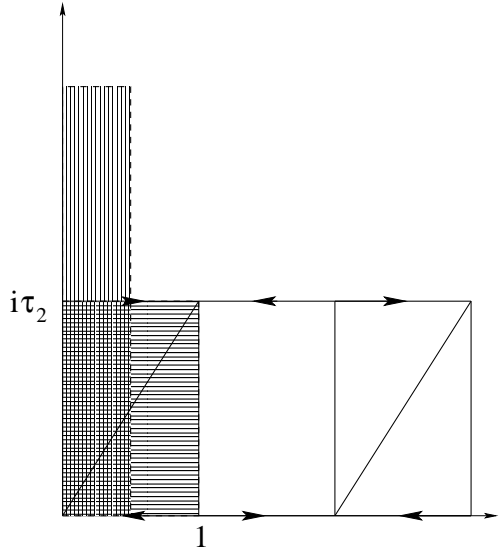}
\caption{Les polygones fondamentaux pour la bande de M\"obius}
\label{mobius}
\end{figure}

 Une corde ouverte qui effectue une boucle et qui subit une inversion d'orientation d\'ecrit une bande de M\"obius. Dans le polygone de la figure \ref{mobius} de c\^ot\'es 1 et $i\tau_2$ les c\^ot\'es horizontaux sont identifi\'es, mais avec des orientations oppos\'ees et les c\^ot\'es verticaux d\'ecrivent deux portions de la m\^eme fronti\`ere. $\tau_2$ est le temps propre qui s'\'ecoule lorsqu'une corde ouverte parcourt la bande de M\"obius. On peut \'egalement chosir un temps "horizontal", dans ce cas la bande de M\"obius est repr\'esent\'ee comme un tube termin\'e par un crosscap et une fronti\`ere et c'est une corde ferm\'ee qui se propage. Le param\`etre de Teichm\"uller pour le tore de double recouvrement est $\tau=1/2+i\tau_2/2$ et on obtient la bande de M\"obius par l'involution: $z\rightarrow 1-\bar{z}+i\tau_2$. Le canal direct et la canal transverse sont reli\'es dans le cas de la bande de M\"obius par la transformation $P=TST^2S$ :
 
 \be P:{1\over 2}+{i\tau_2\over 2}\xrightarrow{S}-{2\over 1+i\tau_2}\xrightarrow{T}{-1+i\tau_2\over 1+i\tau_2}\xrightarrow{T}{2i\tau_2\over 1+i\tau_2}\xrightarrow{S}-{1+i\tau_2\over 2i\tau_2}\xrightarrow{T} {1\over 2}+{i\over 2\tau_2}.\ee

 L'amplitude sur la bouteille de M\"obius est donn\'ee par :
 
 \be \mathcal{M}={N\over 2}\int_0^\infty {d\tau_2\over \tau_2^{14}}tr(\Omega q^{{1\over 2}(N-1)})={\epsilon N\over 2}\int_0^\infty{d\tau_2\over \tau_2^{14}} {1\over \hat{\eta}^{24}(i\tau_2/2+1/2)},
 \ee 
 avec $\hat{\eta}(1/2+\tau)=q^{1/24}\prod_{n=1}^\infty (1-(-1)^nq^n)$. Le $1/2$ dans l'argument de la fonction $\hat{\eta}$ est une cons\'equence de la r\'elation: $\Omega\alpha_n^\mu\Omega^{-1} =(-1)^n \alpha_n^\mu$ et 
$\epsilon=\pm 1$ est un signe qui provient de l'action de $\Omega$ sur les facteurs de Chan-Paton. Si on d\'esigne par $| \psi, ij>$ un \'etat de la corde ouverte, avec $\psi$ l'\'etat des champs de la surface d'univers et $i$ et $j$ les facteurs de Chan-Paton associ\'es aux extr\'emit\'es gauche et droite, l'action de $\Omega$ sur $| \psi, ij>$ s'\'ecrit\cite{gimonpol}:

\be \Omega : \ | \psi, ij> \rightarrow (\gamma_{\Omega})_{ii'} | \psi, j'i'>(\gamma_{\Omega}^{-1})_{j'j}.\label{actomeg}\ee
 En agissant deux fois avec $\Omega$:
 
 \be \Omega^2 : \ | \psi, ij> \rightarrow (\gamma_{\Omega}(\gamma_{\Omega}^T)^{-1})_{ii'} | \psi, j'i'>(\gamma_{\Omega}^T\gamma_{\Omega}^{-1})_{j'j}.\ee
 et en imposant $\Omega^2=1$ il r\'esulte que : $\gamma_{\Omega}^T=\pm \gamma_\Omega$. Si $\gamma_{\Omega}$ est sym\'etrique on peut choisir une base telle que $\gamma_{\Omega}=I$. Dans le cas o\`u  $\gamma_{\Omega}$ est antisym\'etrique on peut choisir une base telle que:
 
 \be  \gamma_{\Omega}=\rl\begin{array}{cc}0& iI\\ -iI&0\end{array}\rr.\ee
 Le signe $\epsilon$ de l'amplitude de la bande de M\"obius est donn\'e par le facteur $Tr(\gamma_{\Omega}\gamma_{\Omega}^{-1})$ qui provient de l'action (\ref{actomeg}) de $\Omega$ sur les facteurs de Chan-Paton.
   
La transformation $P$ permet d'obtenir l'amplitude de la bande de M\"obius dans le canal transverse :

\be \tilde{\mathcal{M}}=2\cdot {\epsilon N\over 2}\int_0^\infty dl {1\over \hat{\eta}^{24}(il+1/2)},
 \ee 
 avec $l=1/(2\tau_2)$. Le facteur $2$ refl\`ete la possibilit\'e d'interpreter $\tilde{\mathcal{M}}$ comme la propagation d'une corde ferm\'ee d'une fronti\`ere \`a un crosscap ou d'un crosscap \`a une fronti\`ere. Ceci montre que $ \tilde{\mathcal{M}}$ peut \^etre aussi calcul\'e directement \`a partir de $\tilde{\mathcal{K}}$ et $\tilde{\mathcal{A}}$ comme leur moyenne g\'eom\'etrique.
 
 Le d\'eveloppment en puissance de $\sqrt{q}$ de $\mathcal{A}+\mathcal{M}$ :
 
 \ba \mathcal{A}\rightarrow {N^2\over 2}((\sqrt{q})^{-1}+(24)+...),\nonumber \\
 \mathcal{M}\rightarrow {\epsilon N\over 2}((\sqrt{q})^{-1}-(24)+...),\label{dev}\ea 
 montre que pour $\epsilon=1$ au niveau de masse nulle il y a $N(N-1)/2$ vecteurs ce qui correspond \`a un groupe de jauge orthogonal. Le cas $\epsilon=-1$ conduit \`a un groupe de jauge symplectique.
 
 L'anneau et la bande de M\"obius pr\'esent aussi une divergence qui provient dans la canal transverse de la propagation des \'etats de masse nulle avec un impulsion z\'ero. Les termes divergents de $\mathcal{K}$, $\mathcal{A}$ et $\mathcal{M}$ ont une contribution proportionnelle \`a$\footnote{La contribution de la bande de M\"obius comporte un signe "-" car l'\'etat de masse nulle appara\^it avec un signe "-" dans le d\'eveloppement (\ref{dev}).}$ :
 
 \be \mathcal{K} + \mathcal{A} + \mathcal{M} \sim {1\over 2}(2^{13}+2^{-13}N^2-2\epsilon N)={2^{-13}\over 2}(N-\epsilon 2^{13})^2,\label{tadpole}\ee
  qui s'annule pour $N=2^{13}=8192$ et $\epsilon=1$. Le groupe de jauge de Chan-Paton est donc $SO(8192)$\cite{dilatontadpole}. La condition que la somme des divergences (\ref{tadpole}) s'annule s'appelle une condition de tadpole, car les divergences apparaissent \`a partir des fonctions \`a un point pour le dilaton devant une fronti\`ere ou un crosscap. A noter que la divergence infrarouge due \`a la propagation du dilaton dans le canal transverse pour $l\rightarrow \infty$ ne refl\`ete pas une inconsistance de la th\'eorie, mais simplement le fait que le vide a \'et\'e mal d\'efini. Ceci ne sera pas le cas pour d'autres \'etats qui apparaissent, par exemple, dans les mod\`eles supersym\'etriques et dans ces cas la condition d'annulation des tadpoles sera obligatoire pour la consistence de la th\'eorie, ce qui justifiera l'introduction d'un secteur des cordes ouvertes. 
  
  
\subsection{Amplitudes du vide pour les cordes supersym\'etriques}

  Pour calculer les amplitudes du vide pour la corde fermionique nous avons besoin de nous rappeler que dans ce cas la dimension critique vaut $D=10$ et que le spectre contient deux secteurs, le secteur de Neveu-Schwarz (NS) et le secteur de Ramond(R) et l'op\'erateur de masse est donn\'e par :
  
  \be M^2={2\over \alpha'}(N+\bar{N}-2a),\ee  
o\`u $N=N^{(b)}+N^{(f)}$ englobe les contributions des fermions et des bosons et $a$ vaut $1/2$ dans le secteur NS et 0 dans le secteur R. Dans les sections pr\'ec\'edentes nous avons d\'ej\`a calcul\'e la contribution des bosons. Pour les oscillateurs fermioniques on obtient :

\be tr(q^{\sum_r r b_{-r}^i b_{r,i}})=\prod_{i=1}^8 \prod_r tr(q^{r b_{-r}^i b_{r,i}})=\prod_r(1+q^r)^8,\ee
 o\`u $r$ est un entier allant de 1 \`a $\infty$ pour le secteur R et un demi-entier dans le secteur NS et o\`u nous avons tenu compte du principe de Pauli.
 
 En mettant ensemble les contributions des bosons et des fermions on obtient pour le secteur NS:
 
\be tr_{\tt{NS}}(q^{N-1/2})={\prod_{r=1/2}^\infty (1+q^r)^8\over q^{1/2}\prod_{m=1}^\infty(1-q^m)^8}={\prod_{m=1}^\infty (1+q^{m-1/2})^8\over q^{1/2}\prod_{m=1}^\infty(1-q^m)^8}\ee
et pour le secteur R :

\be tr_{\tt{R}}(q^N)=16{\prod_{m=1}^\infty (1+q^m)^8\over \prod_{m=1}^\infty(1-q^m)^8},\ee
o\`u le coefficient 16 refl\`ete la d\'eg\'en\'erescence de l'\'etat fondamental dans le secteur R. 

 Dans la suite on va s'int\'eresser aux amplitudes du vide pour la th\'eorie IIB. Il faut donc  inclure dans la trace la pojection GSO :
 
 \be tr_{\tt{NS}}\left({1-(-1)^{F_{\tt{NS}}}\over 2}q^{N-1/2}\right)={\prod_{m=1}^\infty(1+q^{m-1/2})^8-\prod_{m=1}^\infty(1-q^{m-1/2})^8\over 2q^{1/2}\prod_{m=1}^\infty(1-q^m)^8}\label{ns},\ee

 \be tr_{\tt{R}}\left({1+\Gamma_{9}(-1)^{F_{\tt{R}}}\over 2}q^{N}\right) = {16\over 2} {\prod_{m=1}^\infty (1+q^m)^8\over \prod_{m=1}^\infty(1-q^m)^8}, \ee  
 o\`u $F_{\tt{NS,R}}$ sont les op\'erateurs nombre fermionique sur la surface d'univers et l'insertion de  $(-1)^{F_{\tt{NS,R}}}$ change le signe des contributions avec un nombre impair d'oscillateurs fermioniques et $\Gamma_9$ est l'op\'erateur de chiralit\'e dans les dimensions transverses.
 
 L'amplitude r\'esultante s'\'ecrit de mani\`ere compacte en utilisant les fonctions de Jacobi, d\'efinies par:
 
 \ba&&\theta{\scriptsize\left[\!\!\begin{array}{c}\alpha\\\beta\end{array}\!\!\right]}(z\!\mid\!\tau)=\sum_n q^{{1\over 2}(n+\alpha)^2}e^{2\pi i (n+\alpha)(z+\beta)}=\\
&& =e^{2\pi iÊ\alpha(z+\beta)}q^{\alpha^2\over 2}\prod_{n=1}^{\infty}(1-q^n)(1+q^{n+\alpha-1/2}e^{2\pi i (z+\beta)})(1+q^{n-\alpha-1/2}e^{-2\pi i(z+\beta)}).\nonumber
 \ea
 
 Sous les transformations modulaires les fonctions $\theta$ se transforment de la mani\`ere suivante :

 \ba&& T: \theta{\scriptsize\left[\!\!\begin{array}{c}\alpha\\\beta\end{array}\!\!\right]}(z\!\mid\! \tau+1)=e^{-i\pi \alpha (\alpha -1) }\theta{\scriptsize\left[\!\!\begin{array}{c}\alpha\\\beta+\alpha-1/2\end{array}\!\!\right]}(z\!\mid\! \tau),\nonumber\\
&& S: \theta{\scriptsize\left[\!\!\begin{array}{c}\alpha\\\beta\end{array}\!\!\right]}(z\!\mid\!-{1\over\tau})=(-i\tau)^{1/2}e^{2\pi i \alpha \beta+i\pi z^2/\tau}\theta{\scriptsize\left[\!\!\begin{array}{c}\beta\\-\alpha\end{array}\!\!\right]}(z\!\mid\!\tau).\label{TS}\ea

En notant que$\footnote{A noter \'egalement le fait que $\theta{\scriptsize\left[\!\!\begin{array}{c}1/2\\1/2\end{array}\!\!\right]}(0\!\mid\!\tau)=0$.}$ :

\ba&& {\theta^4{\scriptsize\left[\!\!\begin{array}{c}1/2\\0\end{array}\!\!\right]}(0\!\mid\!\tau)\over\eta^{12}(\tau)}=
  {\theta_2^4(0\!\mid\!\tau)\over \eta^{12}(\tau)}=16{\prod_{m=1}^\infty(1+q^m)^8\over \prod_{m=1}^\infty(1-q^m)^8},\nonumber\\
  &&{\theta^4{\scriptsize\left[\!\!\begin{array}{c}0\\0\end{array}\!\!\right]}(0\!\mid\!\tau)\over\eta^{12}(\tau)}=
  {\theta_3^4(0\!\mid\!\tau)\over \eta^{12}(\tau)}={\prod_{m=1}^\infty(1+q^{m-1/2})^8\over q^{1/2} \prod_{m=1}^\infty(1-q^m)^8},\nonumber\\
  &&{\theta^4{\scriptsize\left[\!\!\begin{array}{c}0\\1/2\end{array}\!\!\right]}(0\!\mid\!\tau)\over\eta^{12}(\tau)}=
  {\theta_4^4(0\!\mid\!\tau)\over \eta^{12}(\tau)}={\prod_{m=1}^\infty(1+q^{m-1/2})^8\over q^{1/2} \prod_{m=1}^\infty(1-q^m)^8},\nonumber\\
\ea
on peut voir que l'amplitude du tore pour la th\'eorie IIB s'\'ecrit :
 
 \be \mathcal{T}=\int{d^2\tau\over\tau_2^6}\al{\theta_3^4(0\!\mid\!\tau)-\theta_4^4(0\!\mid\!\tau)-\theta_2^4(0\!\mid\!\tau)\over 2\eta^{12}(\tau)}\ar^2.\label{2b}\ee
 
 On peut facilement v\'erifier \`a l'aide des transformations (\ref{TS}) que (\ref{2b}) est invariante modulaire. \'Egalement la fameuse {\it aequatio identica satis abstrusa} de Jacobi :
 
 \be \theta_3^4-\theta_4^4-\theta_2^4=0,\ee
montre que l'amplitude (\ref{2b}) est nulle, ce qu'il fallait attendre puisque la th\'eorie IIB est supersym\'etrique et contient, donc, un nombre \'egal de degr\'es de libert\'e fermioniques et bosoniques.
 
 L'amplitude (\ref{2b}) peut s'\'ecrire de mani\`ere encore plus compacte en utilisant les caract\`eres du groupe $so(8)$ :
 
 \ba O_8={\theta_3^4+\theta_4^4\over 2\eta^4}, \quad V_8= {\theta_3^4-\theta_4^4\over 2\eta^4},\nonumber\\ 
  S_8={\theta_2^4+\theta_1^4\over 2\eta^4},\quad C_8={\theta_2^4-\theta_1^4\over 2\eta^4}.\ea

  $V_8$ rep\'esente la trace sur les \'etats du secteur NS qui subsitent apr\`es la projection GSO (\ref{ns}) et $O_8$ repr\'esente la trace sur toutes les autres \'etats:
    
  \be O_8=tr_{\tt{NS}}\left({1+(-1)^{F_{\tt{NS}}}\over 2}q^{N-1/2}\right).\ee
  
      Si on d\'eveloppe $O_8$ et $V_8$ en puissances de $q$ :
  
  \ba &&O_8=q^{-1/6}(1+28q+...),\nonumber\\
  &&V_8=q^{-1/6}(8q^{1/2}+64q^{3/2}+...),\ea
  et si l'on tient compte de l'\'energie du point z\'ero pour les bosons de la surface d'univers, $(q^{-1/(24)})8 =q^{-1/3}$, on voit que le d\'eveloppement de $O_8$ commence par $q^{-1/2}$, ce qui correspond \`a un scalaire de masse $M^2=-1/\alpha'$, le tachyon de la corde ouverte, et que $V_8$ commence avec un \'etat non-massif avec 8 degr\'es de libert\'e, qui est le vecteur de la corde ouverte fermionique dans le secteur NS apr\`es la projection GSO.
   
   Les caract\`eres $S_8$ et $C_8$ d\'ecrivent chacun une des deux parties du spectre du secteur R: $\Gamma=1$ et $\Gamma=-1$, avec $\Gamma=\Gamma_9(-1)^{F_{\tt{R}}}$.
    
    En g\'en\'eral pour une alg\`ebre $so(2n)$  les caracth\'eres sont d\'efinis par :
    
    \ba&& O_{2n}={\theta_3^n+\theta_4^n\over 2\eta^n},\quad V_{2n}={\theta_3^n-\theta_4^n\over 2\eta^n},\nonumber\\
 &&   S_{2n}={\theta_2^n+i^{-n}\theta_1^n\over 2\eta^n}, \quad C_{2n}={\theta_2^n-i^{-n}\theta_1^n\over 2\eta^n}.\ea
 
 Les trasformations modulaires pour le vecteur $\left(\begin{array}{c}O_{2n}\\ V_{2n} \\ S_{2n}\\ C_{2n}\end{array}\right)$ sont donn\'ees par les matrices:
 
 \be S={1\over 2}\left(\begin{array}{cccc}1&1&1&1\\1&1&-1&-1\\1&-1&i^{-n}&-i^{-n}\\1&-1&-i^{-n}&i^{-n}\end{array}\right),\ee
  \be T=e^{-in\pi/12}\ \mbox{diag}(1,-1,e^{in\pi/4},e^{in\pi/4}),\ee      
        \be P=\left(\begin{array}{cccc}c&s&0&0\\ s&-c&0&0\\0&0&\zeta c&i\zeta s\\0&0&i\zeta s&\zeta c\end{array}\right),\ee        
        o\`u $c=\mbox{cos}(n\pi/4)$, $s=\mbox{sin}(n\pi/4)$, $\zeta=e^{-in\pi/4}$ et la transformation $P$ agit sur les caract\'eres $\hat{\chi}_i=e^{-i\pi(h_i-c/24)}\chi_i$, o\`u $h_i$ sont les poids conformes $\{h_O,h_V,h_S,h_C\}=\{0,1/2,n/8,n/8\}$.
        
        Les amplitudes du vide pour les supercordes s'expriment en fonction des caract\'eres $so(8)$ divis\'es par $\eta^8(\tau)$ et les matrices des transformations modulaires $T$ et $P$ prennent une forme tr\`es simple pour les combinaisons :
        
        \be {O_8\over \eta^8}, \ \ {V_8\over \eta^8}, \ \ {S_8\over \eta^8}, \ \ {C_8\over \eta^8},      
        \ee   
        
        \be T=\mbox{diag}(-1,1,1,1), \ \ P=\mbox{diag}(-1,1,1,1).\ee     
        
        Avec les caract\'eres $so(8)$ l'amplitude du tore pour la th\'eorie IIB s'\'ecrit :
        
        \be \mathcal{T}=\int_{\mathcal{F}}{d^2\tau\over \tau_2^2} {1\over \tau_2^4\mid \! \eta\!\mid^{16}}\mid\!V_8-S_8\!\mid^2\ee
                et on peut facilement v\'erifier qu'elle est bien invariante sous les transformations modulaires. Dans la suite on va laisser implicites dans les expression l'int\'egration et la contributions des bosons de la surface d'univers, donc :
           
           \be\mathcal{T}_{\tt{IIB}}=\mid\!V_8-S_8\!\mid^2.\ee
                
                Pour la th\'eorie IIA l'amplitude du tore s'\'ecrit :
                
                \be \mathcal{T}_{\tt{IIA}}=   (\bar{V}_8-\bar{S}_8)(V_8-C_8).\ee

             A partir des caract\'eres $so(8)$ on peut construire encore deux amplitudes, invariantes sous les transformations modulaires, qui correspondent aux deux th\'eories nonsupersym\'etriques 0A et 0B\cite{0A0B} :
             
             \ba&&\mathcal{T}_{\tt{0A}}=\mid\!O_8\!\mid^2+\mid\!V_8\!\mid^2+\bar{S}_8C_8+\bar{C}_8S_8,\nonumber\\ 
&&\mathcal{T}_{\tt{0B}}=\mid\!O_8\!\mid^2+\mid\!V_8\!\mid^2+\mid\!S_8\!\mid^2+\mid\!C_8\!\mid^2.\ea

 A noter que les th\'eories 0A et 0B ne contiennent pas des fermions d'espace-temps et que le tachyon est pr\'esent dans le spectre. 
 La projection GSO pour la th\'eorie 0B est donn\'ee par $(1+(-1)^{F_L+F_R})/2$ dans les secteurs NS et R, alors que pour la th\'eorie 0A la projection  est  $(1+(-1)^{F^{(NS)}_L+F^{(NS)}_R})/2$ dans le secteur NS et $(1-(-1)^{F^{(R)}_L+F^{(R)}_R})/2$ dans le secteur R.

 Rapellons qu'il existe en 10 dimensions deux autres th\'eories supersym\'etriques, la corde h\'et\'erotique $SO(32)$ et la corde h\'et\'erotique $E_8\times E_8$. Leurs fonctions de partitions sont donn\'ees par:
 
\ba  \mathcal{T}_{SO(32)}&=&(V_8-S_8)(\bar{O}_{32}+\bar{S}_{32}) =\nonumber\\
&=&(V_8-S_8)(\bar{O}_{16}^2+\bar{V}_{16}^2+\bar{S}_{16}^2+\bar{C}_{16}^2) \ea

\be \mathcal{T}_{E_8\times E_8}=(V_8-S_8)(\bar{O}_{16}+\bar{S}_{16}) (\bar{O}_{16}+\bar{S}_{16}) \ee
 
 Les caract\`eres $so(32)$ et $so(16)$ d\'ecrivent les degr\'es de libert\'e internes.
 
 
 \subsection{La th\'eorie de type I}
 
  Dans cette s\'ection on va discuter l'orientifold de la th\'eorie IIB. Les amplitudes de la bouteille de Klein, de l'anneau et de la bande de M\"obius dans le canal direct sont obtenues de fa\c{c}on similaire que pour la corde bosonique :
  
  \ba \mathcal{K}&=&{1\over 2}\int_0^\infty {d\tau_2\over \tau_2^6}{(V_8-S_8)(2i\tau_2)\over \eta^8(2i\tau_2)},\nonumber\\
  \mathcal{A}&=&{N^2\over 2}\int_0^\infty {d\tau_2\over \tau_2^6}{(V_8-S_8)({1\over 2}i\tau_2)\over \eta^8({1\over 2}i\tau_2)},\nonumber\\ 
 \mathcal{M}&=&{\epsilon N\over 2}\int_0^\infty {d\tau_2\over \tau_2^6}{(\hat{V}_8-\hat{S}_8)({1\over 2}i\tau_2+{1\over 2})\over \eta^8({1\over 2}i\tau_2+{1\over 2})}\label{type1}\ea
 
  La bouteille de Klein sym\'etrise le secteur NS-NS en \'eliminant au niveau de masse nulle le tenseur antisym\'etrique, $B_{\mu\nu}$ et en gardant le graviton et le dilaton. Le secteur R-R est antisym\'etris\'e, ce qui \'elimine la 0-forme et la 4-forme et garde la 2-forme. Comme l'amplitude du tore doit \^etre multipli\'ee avec 1/2, la moiti\'e des \'etats des secteurs NS-R et R-NS subsistent: un gravitino et un dilatino(au niveau de masse nulle). Le spectre r\'esultant est supersym\'etrique avec $N=(1,0)$. 
  Les amplitudes dans les canaux transverses sont obtenues par les transformations $S$ et $P$ :
  
  \ba \tilde{\mathcal{K}}&=&{2^5\over 2}\int_0^\infty dl{(V_8-S_8)(il)\over \eta^8(il)},\nonumber\\
\tilde{\mathcal{A}}&=&{2^{-5}N^2\over 2}\int_0^\infty dl{(V_8-S_8)(il)\over \eta^8(il},\nonumber\\ 
\tilde{ \mathcal{M}}&=&2{\epsilon N\over 2}\int_0^\infty dl{(\hat{V}_8-\hat{S}_8)(il+{1\over 2})\over \eta^8(il+{1\over 2})}\ea  
 et la condition d'annulation des tadpoles,
 
 \be {2^5\over 2}+{2^{-5}N^2\over 2}+   2{\epsilon N\over 2}={2^{-5}\over 2}(N+32\epsilon)^2=0,\label{tad}\ee    
 implique $N=32$ et $\epsilon=-1$ ce qui correspond au groupe $SO(32)$. A noter que la condition (\ref{tad}) s'applique autant au dilaton qu'au secteur R-R. L'amplitude du cylindre $\tilde{\mathcal{A}}$ d\'ecrit l'interaction entre deux $D9$-branes, selon la formule (\ref{intbranes}). De la m\^eme mani\`ere l'amplitude de la bouteille de Klein $\tilde{\mathcal{K}}$ d\'ecrit l'interaction entre deux objets non-dynamiques, des orientifolds $O9$. Les orientifolds ou $Op$ - planes sont des hyperplanes avec $p$ dimensions spatiales qui sont invariantes sous la transformation $\Omega$.  Comme les $Dp$ - branes,  les $Op$ - planes  couplent  \`a une $p+1$ forme et ils poss\`edent  une tension et une charge, \`a la diff\'erence que leur tension peut \^etre n\'egative. Les orientifolds ne contiennent pas de la mati\`ere. 
Il y a plusieurs types d'orientifolds :\\

 \begin{tabular}{p{1.5cm}|c|c}
 & \quad \quad tension \quad \quad & \quad \quad charge \quad \quad \\
\hline 
& & \\
$O_+$ & - & -\\
\hline
& & \\
$\overline{O}_+$ & - & +\\
\hline 
& & \\
$O_-$ & + & +\\
\hline 
& & \\
$\overline{O}_-$ & + & -
\end{tabular}\\

 La th\'eorie de type I contient des orientifolds $O9_+$ et la condition d'annulation des tadpoles traduit la compensation de la charge et de la tension entre les $O9_+$ - planes et le $D9$ - branes.
 Les conditions d'annulation des tadpoles NS-NS et R-R sont tr\`es diff\'erentes. Comme nous l'avons expliqu\'e dans le cas de la corde bosonique, un tadpole NS-NS non nul indique que le vide de la th\'eorie doit \^etre red\'efini. En revanche la loi de Gauss interdit l'existence d'une charge non nulle dans un espace compact.  Ceci laisse une deuxi\`eme possibilit\'e \cite{sugimoto} pour construire une th\'eorie consistante, mais avec un tadpole r\'esiduel pour le dilaton.  Dans ce deuxi\`eme mod\`ele l'amplitude de la bande de M\"obius vaut :
 
 \ba  \mathcal{M}&=& {N\over 2} (\hat{V}_8+\hat{S}_8),\nonumber\\
  \tilde{ \mathcal{M}}&=& N (\hat{V}_8+\hat{S}_8)\label{usp}\ea
         
         La condition d'annulation des tadpoles de R est toujours satisfaite pour $N=32$, en revanche le tadpole du dilaton est non nul. Les orientifolds dans ce mod\`ele sont des $\overline{O}9_-$ -planes. La pr\'esence simultan\'ee des $D$-branes et antiorientifolds brise la supersym\'etrie. La formule de l'amplitude de l'anneau dans le canal direct (\ref{type1}) et l'\'equation (\ref{usp}) montrent qu'au niveau de masse nulle il y a $N(N+1)/2$ vecteurs ce qui correspond au groupe $USp(32)$. Les fermions de masse nulle sont en nombre de $N(N-1)/2$, la supersym\'etrie est, donc, bris\'ee. 
                                 

\section{Compactification toroidale}

Dans la section (\ref{toroidal}) nous avons discut\'e la compactification d'une coordonn\'ee sur un cercle. A pr\'esent nous allons exposer la mani\`ere dont cette compactification se traduit dans les amplitudes du vide \`a une boucle. Nous rappelons que les impulsions gauche et droite pour la coordonn\'ee compacte sont donn\'ees par :

\be p_{L,R}={m\over R}\pm{nR\over \alpha'}\ee
et que les op\'erateurs de Virasoro s'\'ecrivent :
\ba L_0={\alpha'p_L^2\over 4}+\sum_{n=1}^\infty\alpha_{-n}\alpha_n,\nonumber\\
\bar{L}_0={\alpha'p_R^2\over 4}+\sum_{n=1}^\infty\bar{\alpha}_{-n}\bar{\alpha}_n\ea

Dans le calcul des amplitudes l'integrale sur l'impulsion est remplac\'ee par une somme discrete :

\be {1\over \sqrt{\tau_2}\eta(\tau)\eta(\bar{\tau})}\rightarrow \sum_{m,n}{q^{\alpha'p_L^2\over4}\bar{q}^{\alpha'p_R^2\over4}\over \eta(\tau)\eta(\bar{\tau})}\ee

 Par exemple la fonction de partition pour la th\'eorie IIB devient:
 
 \be \mathcal{T}=\mid V_8-S_8\mid^2\sum_{m,n}{q^{\alpha'p_L^2\over4}\bar{q}^{\alpha'p_R^2\over4}\over \eta(\tau)\eta(\bar{\tau})}\ee
 
  Pour la bouteille de Klein la projection $\Omega$ ne permet que la propagation des \'etats avec $p_L=p_R$ ou autrement $n=0$. Par cons\'equent l'amplitude de la bouteille de Klein devient :
  
  \be\mathcal{K}={1\over 2}(V_8-S_8)(2i\tau_2)\sum_m {q^{\alpha'm^2/4R^2}\over \eta(2i\tau_2)},\ee
 avec $q=e^{-4\pi\tau_2}$.  Pour obtenir l'amplitude dans le canal transverse on utilise la formule de resommation de Poisson:
 
 \be \sum_{n\in \mathbb{Z}} e^{-\pi an^2+2\pi i bn}=a^{-1/2}\sum_{m \in \mathbb{Z}}e^{-{\pi(m-b)^2\over a}}.\ee
 Dans la canal transverse ne se propagent que les \'etats avec $m=0$ et nombre d'enroulement pair :
 
 \be \tilde{\mathcal{K}}={2^5\over 2}{R\over \sqrt{\alpha'}}(V_8-S_8)(il)\sum_n{q^{(2n)^2R^2/4\alpha'}\over \eta(il)},\ee
avec $q=e^{-2\pi l}$.

 Le tadpole de R-R pour $n=0$ impose l'introduction d'un secteur des cordes ouvertes.  Les cordes ouvertes avec conditions aux bords de Neumann ne poss\`edent pas de nombre d'enroulement.  Les amplitudes de l'anneau et de la bande de M\"obius s'\'ecrivent :
 
 \ba \mathcal{A}&=&{N^2\over 2}(V_8-S_8)({i\tau_2\over 2})\sum_m {q^{\alpha'm^2/2R^2}\over \eta(i\tau_2/2)},\nonumber\\
 \mathcal{M}&=&-{N\over 2}(\hat{V}_8-\hat{S}_8)({i\tau_2\over 2}+{1\over 2})\sum_m {q^{\alpha'm^2/2R^2}\over \hat{\eta}(i\tau_2/2+1/2)} , \ea
  avec $q=e^{-2\pi \tau_2}$.  Dans le canal transverse de l'anneau se propagent que les \'etats avec $m=0$, alors dans le canal transverse de la bande de M\"obius se propagent les \'etats avec $m=0$ et nombre d'enroulement pair :
  \ba \tilde{\mathcal{A}}&=&{2^{-5}N^2\over 2}{R\over \sqrt{\alpha'}}(V_8-S_8)(il)\sum_n{q^{n^2R^2/4\alpha'}\over \eta(il)},\nonumber\\
  \tilde{\mathcal{M}}&=&-{2N\over 2}{R\over \sqrt{\alpha'}}(\hat{V}_8-\hat{S}_8)(il+1/2)\sum_n{q^{(2n)^2R^2/4\alpha'}\over \eta(il+1/2)},\ea 
  
  
\section{Orbifolds}

Les orbifolds \cite{dixon} \cite{ibanez}sont des mod\`eles qui pr\'esentent des spectres int\'eressantes pour la ph\'enom\`enologie. Des alternatives ph\'enom\`enologiques des orbifolds sont les compactifications sur les vari\'et\'es de Calabi Yau\cite{calabiyau} et les constructions des cordes fermioniques en quatre dimensions\cite{fermionic}.

Du point de vue g\'om\'etrique l'orbifold \cite{ginsparg} g\'en\'eralise la
notion de vari\'et\'e 
en permettant qu'un ensemble discret des points, qui sont les points
fixes, de la vari\'et\'e soient
singuliers. Les points fixes de l'action $G:M\rightarrow M$ d'un
groupe discret $G$ sur une 
vari\'et\'e $M$ sont les points $x \in M$ pour lesquels $gx=x, \ \forall
g\in G $($g \neq $ identit\'e).
L'orbifold est d\'efini comme l'espace
quotient $M/G$ construit par l'identification des points de la vari\'et\'e
$M$ par la relation d'\'equivalence $x \sim gx ,\ \forall g \in G$.

{\it Exemple.} Le cercle $M=S^1$ param\'etris\'e par $x \equiv x+ 2\pi r$
avec l'action du groupe $G=\mathbb{Z}_2: S^1\rightarrow S^1$ qui agit
comme $g:x\rightarrow -x$ cr\'eent l'orbifold $S^1/\mathbb{Z}_2$ qui est
le segment $[0,\pi r]$ avec les points fixes $x=0$ et $x=\pi r$.

 En th\'eorie des cordes 
la notion d'orbifold  correspond \`a une th\'eorie invariante modulaire
obtenue en jaugeant  une sym\'etrie $G$ dans une th\'eorie invariante
modulaire $T$, o\`u $G$ est une sym\'etrie discr\`ete de la th\'eorie
initiale. Pour construire l'orbifold $T/G$ il faut d'abord projeter
l'espace de Hilbert de la th\'eorie $T$ dans le sous-espace invariant
sous $G$.  Ensuite puisque les points $x$ et $gx$ sont \'echivalents
 il faut consid\'erer les \'etats o\`u la corde ferme modulo une
 transformation de $G$: $X^{\mu}(\sigma+2\pi)=g X^{\mu}(\sigma)$. Ces
 nouveaux \'etats s'appellent des \'etats twist\'es et sont confin\'es aux points fixes. Le secteur des \'etats twist\'es doit \^etre \`a son tour projet\'e dans le sous-espace invariant sous l'action de $G$.
L'existence du secteur twist\'e, ainsi que la n\'ecessit\'e de le projeter dans le sous-espace invariant sous l'action de $G$ sont impos\'ees par l'invariance modulaire.  

 Du point de vue de l'amplitude sur le tore  la projection du spectre untwist\'e revient \`a consid\'erer des \'etats avec une p\'eriodicit\'e modulo une transformation de $G$ dans la direction du type "temps" : $X(z+\tau)=gX(z), \  X(z+1)=X(z)$.  Mais comme la transformation $S$ \'echange les deux cycles du tore, il faut \'egalement ajouter des \'etats avec une p\'eriodicit\'e : $X(z+\tau)=X(z), \ X(z+1)=gX(z)$. Ceci sont les \'etats twist\'es. Sous une transformation $T$ le secteur twist\'e se transforme dans un secteur avec 
  $X(z+\tau)=g X(z), \ X(z+1)=gX(z)$, qui repr\'esente la projection dans les \'etats invariants sous $G$. 
  
   Dans la suite on va d\'eduire la fonction de partition de l'orbifold $S^1/\mathbb{Z}_2$ pour la corde bosonique:
   
   \be \mathcal{T}=\int_{\mathcal{F}} {d^2\tau\over \tau_2^2}{1\over \tau_2^{12}}tr_{\tt{Untw+Tw}}\  {(1+g)\over 2}q^{N-1}\bar{q}^{\bar{N}-1},\ee  
   o\`u $g$ est le g\'en\'erateur du groupe $\mathbb{Z}_2$, $g^2=1$ et la trace porte sur les secteurs untwist\'e et twist\'e. Le premier terme correspond \`a l'amplitude du tore de la th\'eorie initiale divis\'ee par 2.  Ce terme est invariant modulaire. Le deuxi\`eme terme est la trace sur le secteur untwist\'e avec l'action de $g$.  La contribution \`a ce terme d'un boson complexe
   pour un \'element $e^{2\pi i\beta}$ d'un groupe g\'en\'erique est donn\'e par:
   
   \ba tr(e^{2\pi i \beta} q^{N-1/12})&=&q^{-1/12}\prod_n {1\over 1-q^ne^{2\pi i \beta}}{1\over 1-q^n e^{-2\pi i \beta}}=\nonumber\\
  & =&{\eta \over \theta{\scriptsize\left[\!\!\begin{array}{c}1/2\\ 1/2+\beta\end{array}\!\!\right]}(0\!\mid\!\tau)  } \ 2\mbox{sin}(\pi \beta)
   \ea

   Pour le groupe $\mathbb{Z}_2$ on a $\beta=1/2$.   En prenant en compte aussi les modes de droite on obteint la contribution totale pour un boson complexe  $\al    {2\eta \over \theta_2 }\ar   ^2$. Dans notre cas une seule coordonn\'ee bosonique est affect\'ee par l'op\'eration $\mathbb{Z}_2$ et
   la contribution du secteur untwist\'e \`a la fonction de partition s'\'ecrit :
   
   \be \mathcal{T}={1\over 2}\sum_{m,n}{q^{\alpha'p_L^2\over4}\bar{q}^{\alpha'p_R^2\over4}\over \eta(\tau)\eta(\bar{\tau})}+{1\over 2}\al{2\eta\over \theta_2}\ar.\ee  
   
   La transform\'ee $S$ du dernier terme g\'en\'ere la contribution du secteur twist\'e $2\cdot {1\over 2}\mid{\eta\over \theta}_4\mid$. Ensuite par la transformation $T$ de ce terme on obtient la contribution $tr_{\tt{Tw}} \ {g\over2} \ q^{N-1}\bar{q}^{\bar{N}-1}=2\cdot {1\over 2}\mid{\eta\over \theta_3}\mid$. Au final la fonction de partition est donn\'ee par:
   
   \be    \mathcal{T}={1\over 2}\sum_{m,n}{q^{\alpha'p_L^2\over4}\bar{q}^{\alpha'p_R^2\over4}\over \eta(\tau)\eta(\bar{\tau})}+{1\over 2}\al{2\eta\over \theta_2}\ar+{1\over 2}\left\{2\al{\eta\over \theta_4}\ar+2\al{\eta\over \theta_3}\ar\right\}.\ee
   
   Le facteur 2 devant les contributions du secteur twist\'e repr\'esente le nombre de points fixes. Les \'etats twist\'es sont localis\'es aux points fixes et il y a autant de secteurs twist\'es que de points fixes.



\chapter{Brisure de supersym\'etrie}
\section{Brisure de supersym\'etrie en th\'eorie des cordes}
 La brisure de la supersym\'etrie est un point central de la ph\'enom\'enologie des cordes. Dans les mod\`eles d'orientifolds quatre m\'ecanismes de brisure de supersym\'etrie sont connues : 
 
 \begin{description} 
 
\item[ $\bullet$] La supsersym\'etrie peut \^etre bris\'ee d\'es  le d\'epart. Il s'agit des orientifolds de la th\'eorie de type 0A et 0B. Ces mod\`eles contiennent en g\'en\'eral des tachyons, mais la projection $\Omega'=\Omega \cdot (-1)^{f_L}$ en 0B, avec $(-1)^{f_L}$ le nombre fermionique de la surface d'univers,  conduit \`a une th\'eorie sans tachyons, appel\'ee 0'B\cite{0primb}.

\item[ $\bullet$] Le m\'ecanisme de Scherk-Schwarz, qui consite \`a d\'eformer les modes de Kaluza-Klein ou les modes d'enroulement, brise la supersym\'etrie \`a l'\'echelle de la compactification\cite{sssusybreak}.  Ce m\'ecanisme sera pr\'esent\'e dans la suite.

\item[ $\bullet$]La supersym\'etrie peut \^etre bris\'ee  \`a l'\'echelle des cordes en utilisant des configurations non-BPS de branes et antibranes, ainsi que des orientifolds et anti-orientifolds\cite{branesusybreak}.  Nous avons d\'ej\`a vu un exemple avec la th\'eorie $USp(32)$\cite{sugimoto}. Ce m\'ecanisme s'appelle "brane supersymmetry breaking". 
\item[ $\bullet$]Une autre possibilit\'e consiste \`a briser la supersym\'etrie dans le secteur des cordes ouvertes par l'introduction des champs magn\'etiques internes\cite{mag}. La supersym\'etrie est bris\'ee \`a l'\'echelle de compactification et la T-dualit\'e relie ces mod\`eles \`a des configurations avec des branes inclin\'ees (branes at angles)\cite{angle}. 

\end{description}

\section{Brisure de supersym\'etrie par compactification \`a la Scherk-Schwarz }
 Le m\'ecanisme de brisure de supersym\'etrie de Scherk-Schwarz est une g\'en\'eralisation de la compactification de Kaluza-Klein. Consid\'erons un champ scalaire $\Phi$ en D dimensions. Si une des coordonn\'ees est compacte $y\sim y+2\pi R$, la p\'eriodicit\'e du champ $\Phi$, $\Phi(x^{\mu},y)=\Phi(x^{\mu},y+2\pi R)$ implique que l'impulsion dans la direction $y$ est quantifi\'ee, $p=n/R$, et que la champ $\Phi$ peut \^etre d\'evelopp\'e en modes de Fourier :
 
 \be \Phi(x^{\mu},y)=\sum_{n\in Z}e^{iny\over R}\phi_n(x^{\mu}).\ee
 
 L'\'equation de Klein Gordon en D dimensions,
 
 \be \partial_M\partial^M \Phi=0,\ee
 devient:
 
 \be \partial_\mu\partial^\mu \phi_n(x^\mu)={n^2\over R^2} \phi_n(x^\mu).\ee
 Le champ $\Phi$ se d\'ecompose dans un nombre infini des champs D-1 dimensionels avec des masses donn\'ees par $m_n^2=n^2/R^2$. Ceci est le m\'ecanisme de Kaluza-Klein. 
 
 Une g\'en\'eralisation possible est de permettre une p\'eriodicit\'e des champs modulo une transformation $M$:
 
 \be \Phi(x^{\mu},y+2\pi R)= M \Phi(x^{\mu},y).\ee 
 
 Si la transformation $M$ est, par exemple, le nombre fermionique, $M=(-1)^F$, on obtient une brisure de supersym\'etrie. En effet, les bosons restent p\'eriodiques, mais les champs fermioniques satisfont une condition d'antip\'eriodicit\'e:
 
 \be \Psi(y+2\pi R)=-\Psi(y),\ee
  qui a comme effet une modification des modes de Kaluza-Klein:
 
 \be \Psi(y)=\sum e^{i(n+1/2)y\over R} \Psi_n .\ee
 
 La masse des fermions est d\'ecal\'ee par rapport \`a celle des bosons et on obtient une brisure de supersym\'etrie.
 
 En th\'eorie des cordes on a \'egalement la possibilit\'e de d\'ecaler les modes d'enroulement. Ce deuxi\`eme  m\'ecanisme s'appelle "M-theory breaking" car il peut \^etre reli\'e par des dualit\'es \`a un m\'ecanisme Scherk-Schwarz conventionnel le long de la onzi\`eme coordonn\'ee\cite{eg}. Ces mod\`eles pr\'esentent le ph\'enom\`ene de "brane supersymmetry": les excitations des branes plong\'ees dans un bulk nonsupersym\'etrique peuvent \^etre supersym\'etriques, au premier ordre, car les branes sont orthogonales \`a la direction de la d\'eformation et la brisure n'affecte pas le secteur de masse z\'ero des cordes ouvertes. 
 
 \subsection{Scherk-Schwarz parall\`ele} 
 
 Le point de d\'epart est l'orbifold de la  th\'eorie IIB, compactifi\'ee sur un cercle $x^9=x^9+2\pi R$, avec le g\'en\'erateur $(-1)^F\delta$, avec $F=F_L+F_R$ le nombre fermionique de l'espace-temps et $\delta$ le shift: $x^9\rightarrow x^9+\pi R$.  La fonction de partition r\'esultante est donn\'ee par :
 
 \ba \mathcal{T}&=&{1\over 2}\left[|V_8-S_8 |^2\Lambda_{m,n}+\al V_8+S_8\ar^2(-1)^m\Lambda_{m,n}\right]+\nonumber\\
 &+&{1\over 2}\left[\al O_8-C_8\ar^2\Lambda_{m,n+1/2}+\al O_8+C_8\ar^2(-1)^m\Lambda_{m,n+1/2}\right] ,\ea
 avec 
 \be \Lambda_{m+a,n+b}=\sum_{m,n}{q^{{\alpha'\over 4}\rl {(m+a)\over R}+{(n+b)R\over \alpha'}\rr^2}\bar{q}^{{\alpha'\over 4}\rl {(m+a)\over R}-{(n+b)R\over \alpha'}\rr^2}\over \eta(q)\eta(\bar{q})}=\sum_{m,n}Z_{m+a,n+b} .\ee
 
 La fonction de partition peut \^etre mise sous la forme :
 
 \ba \mathcal{T}&=&(V_8\bar{V}_8+S_8\bar{S}_8 )\Lambda_{2m,n}+(O_8\bar{O}_8+C_8\bar{C}_8 )\Lambda_{2m,n+1/2} \nonumber\\
 &-&(V_8\bar{S}_8+S_8\bar{V}_8 )\Lambda_{2m+1,n}- (O_8\bar{C}_8+C_8\bar{O}_8 )\Lambda_{2m+1,n+1/2},  \ea
 avec 
 
 \ba \Lambda_{2m,n+a}=\sum_{m,n}{1+(-1)^m\over 2}Z_{m,n+a}, \quad \Lambda_{2m+1,n+a}=\sum_{m,n}{1-(-1)^m\over 2}Z_{m,n+a},\nonumber\\ 
 \Lambda_{m+a,2n}=\sum_{m,n}{1+(-1)^n\over 2}Z_{m+a,n}, \quad \Lambda_{m+a,2n+1}=\sum_{m,n}{1-(-1)^n\over 2}Z_{m+a,n}.\nonumber\\\ea 
 Il est utile d'effectuer la transformation $R\rightarrow R/2$ qui permet de mieux visualiser la connection avec le m\'ecanisme de Scherk-Schwarz en th\'eorie des champs. La fonction de partition se r\'e\'ecrit :

 \ba \mathcal{T}&=&(V_8\bar{V}_8+S_8\bar{S}_8 )\Lambda_{m,2n}+(O_8\bar{O}_8+C_8\bar{C}_8 )\Lambda_{m,2n+1} \nonumber\\
 &-&(V_8\bar{S}_8+S_8\bar{V}_8 )\Lambda_{m+1/2,2n}- (O_8\bar{C}_8+C_8\bar{O}_8 )\Lambda_{m+1/2,2n+1}.\label{basess} \ea  
 
 Dans cette nouvelle base on voit que les fermions ont des modes de Kaluza-Klein d\'ecal\'es avec 1/2 par rapport aux bosons. Pour $R<2\sqrt{\alpha'}$ le spectre des cordes ferm\'ees contient un tachyon.  Dans la limite $R\rightarrow \infty$ on retrouve la th\'eorie IIB, tandis que la limite $R\rightarrow0$ conduit \`a la th\'eorie 0B. 
  
 Dans la bouteille de Klein se propagent les \'etats avec $p_L=p_R \Leftrightarrow n=0$ :
 
 \be \mathcal{K}={1\over 2}(V_8-S_8)P_{m},\ee
 avec $P_{am+b}=\sum_m q^{\alpha'(am+b)^2\over 4R^2}$. La transformation $S$ fournit l'amplitude dans la canal transverse :
 \be  \tilde{\mathcal{K}}={2^5\over 4}{R\over \sqrt{\alpha'}}(V_8-S_8) W_{2n},\ee
 o\`u $W_{an+b}=\sum_n q^{(an+b)^2R^2\over \alpha'}$. 
 
  Le tadpole de R-R impose l'introduction des $D$ - branes. L'amplitude de l'anneau s'\'ecrit :
 
 \be \tilde{\mathcal{A}}={2^{-5}N^2\over 4}{R\over \sqrt{\alpha'}}\pl(V_8-S_8) W_{2n}+(O_8-C_8)W_{2n+1}\pr\ee
 
  L'amplitude de la bande de M\"obius dans la canal transverse est obtenue comme la moyenne g\'eom\'etrique de $\tilde{\mathcal{K}}$ et $\tilde{\mathcal{A}}$ :
  
  \be \tilde{\mathcal{M}}=-{N\over 2}{R\over \sqrt{\alpha'}}[\pm(V_8-(-1)^nS_8) W_{2n} ]\ee
  L'annulation des tadpoles de R-R impose $N=32$. 
  
   Les amplitudes des cordes ouvertes sont donn\'ees par :
   
   \ba \mathcal{A}&=&{N^2\over 4}\pl (V_8-S_8)(P_m+P_{m+1/2})+(V_8+S_8)(P_m-P_{m+1/2})\pr=\nonumber\\
   &=&{N^2\over 2}[V_8P_{m}-S_8P_{m+1/2}],\nonumber\\
  \mathcal{M}&=&-{N\over 2}[\pm V_8 P_{m}-S_8 P_{m+1/2}]\ea
  
  Le m\'ecanisme de brisure de supersym\'etrie \`a la Scherk-Schwarz correspond au signe "+" dans $\mathcal{M}$. Le groupe de jauge dans ce cas est $SO(32)$. Le signe "-" correspond au groupe $USp(32)$.

  
  \subsection{M-theory breaking}
  \label{modele}
 
 On va consid\'erer le m\^eme mod\`ele que dans la section pr\'ecedante, mais avec un shift antisym\'etrique $\delta x^9_L=x^9_L+\pi R/2, \  \delta x^9_R=x^9_R-\pi R/2 $.  La fonction de partition est alors donn\'ee par :
 
 \ba \mathcal{T}&=&{1\over 2}\left[|V_8-S_8 |^2\Lambda_{m,n}+\al V_8+S_8\ar^2(-1)^n\Lambda_{m,n}\right]+\nonumber\\
 &+&{1\over 2}\left[\al O_8-C_8\ar^2\Lambda_{m+1/2,n}+\al O_8+C_8\ar^2(-1)^n\Lambda_{m+1/2,n}\right] \ea
 et devient apr\`es le rescaling $R\rightarrow R/2$  :
 
 \ba \mathcal{T}&=&(V_8\bar{V}_8+S_8\bar{S}_8 )\Lambda_{2m,n}+(O_8\bar{O}_8+C_8\bar{C}_8 )\Lambda_{2m+1,n} \nonumber\\
 &-&(V_8\bar{S}_8+S_8\bar{V}_8 )\Lambda_{2m,n+1/2}- (O_8\bar{C}_8+C_8\bar{O}_8 )\Lambda_{2m+1,n+1/2}. \ea  
 
 Pour $R> 2\sqrt{\alpha'}$ le spectre des cordes ferm\'ees contient un tachyon.  Dans la limite $R\rightarrow 0$ on retrouve la th\'eorie IIB, tandis que la limite $R\rightarrow\infty$ conduit \`a la th\'eorie 0B. 
 
 La projection $\Omega$ impose $p_L=p_R$, donc seulement les \'etats avec nombre d'enroulement z\'ero se propagent dans la bouteille de Klein :
 
 \be \mathcal{K}={1\over 2}\pl (V_8-S_8)P_{2m}+(O_8-C_8)P_{2m+1}\pr.\ee
 
 La transformation $S$ permet d'obtenir l'amplitude dans le canal transverse :
 
 \ba \tilde{\mathcal{K}}&=& {2^5\over 2}{R\over \sqrt{\alpha'}} \left\{ (V_8-S_8){W_n\over 2}+(V_8+S_8){(-1)^nW_n\over 2}\right\}=\nonumber\\
 &=& {2^5\over 2}{R\over \sqrt{\alpha'}} \left\{ V_8 W_{2n} -S_8W_{2n+1}\right\}.\ea
  
  Les \'etats de R-R qui se propagent dans le canal transverse sont tous massifs, il n'y a, donc,  pas de tadpole de R-R. Le mod\`ele contient des $O9$ et $\bar{O}9$ planes et leurs charges de R-R se compensent. Pour que le secteur des cordes ouvertes n'introduise pas de tadpoles de R-R, il faut qu'il comporte des $D$ et $\bar{D}$ branes. On va appeler $N$ le nombre des branes et $M$ le nombre de antibranes. L'amplitude de l'anneau dans le canal transverse s'\'ecrit :
  
  \ba \tilde{\mathcal{A}}&=&{2^{-5}\over 2}{R\over \sqrt{\alpha'}}\pl (N^2+M^2)(V_8-S_8)+2NM(-1)^n(V_8+S_8)\pr W_n=\nonumber\\
  &=&{2^{-5}\over 2}{R\over \sqrt{\alpha'}}\pl[N+(-1)^n M]^2V_8-[N-(-1)^nM]^2S_8\pr W_n.\ea
  
  Les amplitudes de la bouteille de Klein et de l'anneau dans le canal transverse permettent d'obtenir l'amplitude transverse de M\"obius :
  \ba   \tilde{\mathcal{M}}&=& {R\over \sqrt{\alpha'}}\pl\pm(N+  (-1)^n M)V_8W_{2n}-(N-(-1)^nM)S_8W_{2n+1}\pr=\nonumber\\
  &=&{R\over \sqrt{\alpha'}}(N+M)(\pm V_8W_{2n}+S_8W_{2n+1}),\ea
  qui devient dans le canal direct :
  \be \mathcal{M}= {(N+M)\over 2}(\pm V_8 P_m+S_8(-1)^mP_m)\label{mob}\ee
  
   L'annulation des tadpoles de R-R implique que pour $n$ pair $N-(-1)^nM=0$, c'est \`a dire $N=M$.   
  L'amplitude de l'anneau dans le canal direct :
  
  \be \mathcal{A}={1\over 2}\pl (N^2+M^2)(V_8-S_8)P_m+2NM(O_8-C_8)P_{m+1/2}\pr\ee
  montre qu'au niveau de masse nulle, pour le signe "-" dans l'amplitude (\ref{mob}) on a effectivement un spectre supersym\'etrique avec le groupe de jauge $SO(M)\times SO(N)$. Si on impose \'egalement l'annulation des tadpoles de NS on a :
  
  \be {2^5\over 2}+{2^{-5}\over 2}(N+M)^2-(N+M)={2^5\over 2}[(N+M)-2^5]^2=0\ \Rightarrow N+M=32,\ee  
  donc le groupe de jauge est $SO(16)\times SO(16)$.
   
  Une T-dualit\'e sur la coordonn\'ee $x^9$ transforme les modes d'enroulement en modes de Kaluza-Klein et on retrouve le m\'ecanisme de Scherk-Schwarz habituel. La T-dualit\'e transforme \'egalement les $D9$ - branes et $O9$ - planes en $D8$ et $O8$ et elle g\'en\'ere une parit\'e $\Pi_{x^9}$. Les $D8$ - branes et $O8$ - planes sont plac\'es en $x^9=0$, tandis que les $\bar{D}8$ - branes et $\bar{O}8$ - planes se trouvent en $x^9=\pi R$.  D\'eformer les modes d'enroulement est, donc, \'equivalent, \`a la d\'eformation des modes de Kaluza-Klein dans une coordonn\'ee orthogonale aux branes. 
  
  En \cite{ej} un mod\`ele ressemblant \`a \'et\'e propos\'e. Ce mod\`ele utilise une parit\'e  particuli\`ere sur la surface d'univers: $\Omega'=\Omega (-1)^{f_L}$, avec $f_L$ le nombre fermionique sur la surface d'univers. Cette projection a le m\'erite d'\'eliminer le tachyon pr\'esent dans le spectre des cordes ferm\'ees.  Le mod\`ele r\'esultant interpole entre la th\'eorie de type I supersym\'etrique et la th\'eorie de type I avec groupe de jauge $USp(32)$.  
  
La bouteille de  Klein dans ce mod\`ele vaut:

\ba 
K &=& \frac{1}{2}\left\{ (V_8-S_8)P_{2m}-
(O_8-C_8)P_{2m+1}\right \}. \nonumber
\ea
 Le tachyon est antisym\'etris\'e et \'elimin\'e du spectre. Dans le canal transverse l'amplitude de la bouteille de Klein devient:
  
  \ba 
\widetilde{K}
= \frac{2^5}{2}[(V_8-S_8)W_n-(V_8+S_8)(-1)^n W_n]
= \frac{2^5}{2}(V_8W_{2n+1}-S_8W_{2n}) \ .  
\ea
Comme on peut le voir sur cette amplitude il n'y a pas de tadpole de NS-NS, mais il y a un tadpole 
de R-R. Pour annuler ce tadpole il est n\'ecessaire d'introduire des $D9$-branes. La mod\`ele T-dual conteint des $O8_+$ - planes \`a l'origine et des $\bar{O}8_-$ - planes en $\pi R$. Les $D8$ - branes sont distribu\'ees entre ces deux points. Lorsque des $D8$ - branes sont plac\'ees sur les $\bar{O}8_-$ - planes la supersym\'etrie est bris\'ee.

 Si les branes sont plac\'ees au m\^eme point l'amplitude de l'anneau dans le canal transverse s'\'ecrit:
 
 \be \tilde{\mathcal{A}}={2^{-5}\over 2}N^2(V_8-S_8)W_n\ .\ee

 L'amplitude de la bouteille de M\"obius est alors donn\'ee par:
 
 \be \tilde{\mathcal{M}}=\pm NV_8W_{2n+1}+NS_8W_{2n}.\ee
   Le signe "-" correspond au cas o\`u dans le mod\`ele T-dual les $D8$ - branes sont plac\'ees sur les $O8$ - planes, alors que le signe "+" d\'ecrit la configuration avec les $D8$ - branes sont situ\'ees en $\pi R$ avec les $\bar{O}8_-$.   L'annulation des tadpoles implique $N=32$.
   
   Les amplitudes des cordes ouvertes :
   
   \ba \mathcal{A}&=&{N^2\over 2}(V_8-S_8)P_m,\nonumber\\
   \mathcal{M}&=&\pm{N\over 2}V_8(-1)^mP_m+{N\over 2}S_8P_m,\ea
  
   determinent le groupe de jauge qui est $SO(32)$ pour le signe "-" et $USp(32)$ pour le signe "+".  Si $N$ branes sont plac\'ees \`a l'origine et $32-N$ en $\pi R$ le groupe de jauge est $SO(N)\times USp(32-N)$.

\chapter{Introduction \`a la cosmologie}

\section{Le mod\`ele du Big Bang}

La cosmologie est bas\'ee sur l'hypoth\`ese que, dans une premi\`ere approximation, l'univers est isotrope et homog\`ene dans les coordonn\'ees spatiales $\footnote{Les observations montrent que l'univers n'est pas statique, donc on ne peut pas consid\`erer l'homog\'en\'eit\'e et l'isotropie dans le temps.}$. Cela veut dire que l'univers peut \^etre "coup\'e" en tranches $t=\mbox{const}.$ qui sont des hypersurfaces du genre temps isotropes et homog\`enes.  La m\'etrique de l'espace-temps peut donc s'\'ecrire:

\be ds^2=-dt^2+g_{ij}(t,x^k)dx^idx^j,\ee
o\`u $g_{ij}$ est une m\'etrique avec sym\'etrie maximale pour $t$ fix\'e. Les m\'etriques avec sym\'etrie maximale sont les m\'etriques avec sym\'etrie sph\'erique et de courbure constante. Ceci conduit aux solutions:

\be ds^2=-dt^2+a^2(t)\left[{dr^2\over 1-\kappa r^2}+r^2(d\theta^2+sin^2\theta d\phi^2)\right],\ee
dites m\'etriques de Robertson-Walker, avec $\kappa=-1,0,1$ selon la courbure des hypersurfaces spatiales. Le cas avec courbure n\'egative, $\kappa=-1$, correspond \`a l'espace hyperbolique, dit \'egalement ouvert, car son volume est infini. Le cas $\kappa=0$ correspond \`a l'espace euclidien, appell\'e espace plat. L'espace avec courbure positive est la sph\`ere, qui est un espace ferm\'e.

 La mati\`ere dans l'univers peut \^etre d\'ecrite par un fluide parfait dont le tenseur \'energie-impulsion peut \^etre mis sous la forme:
 
 \be T^\mu_\nu=\mbox{diag}(-\rho(t),p(t),p(t),p(t)),\ee 
o\`u $\rho$ est la densit\'e d'\'energie et $p$ la pression. Les \'equations d'Einstein s'\'ecrivent alors:

\ba (00)\ &:&\left({\dot{a}\over a}\right)^2={8\pi G \rho\over 3}-{\kappa\over a^2},\label{first}\\
(ij)\ &:&\quad\ {\ddot{a}\over a}=-{1\over 2}\left({\dot{a}\over a}\right)^2-{\kappa\over 2a^2}-4\pi G p.\ea
La derni\`ere \'equation peut \^etre simplifi\'ee \`a l'aide de la premi\`ere:

\be  {\ddot{a}\over a}=-{4\pi G\over 3}(\rho+3p).\label{friedmann2}\ee
Mises sous cette forme ces \'equations s'appellent les \'equations de Friedmann-Lema\^itre. Une cons\'equence directe est l'equation de continuit\'e:
 
 \be \dot{\rho}+3H(\rho+p)=0,\label{cont}\ee  
 o\`u $H=\dot{a}/a$ est le param\`etre de Hubble qui d\'ecrit le taux d'expansion de l'univers. Sa valeur \`a l'\'epoque pr\'esente est la constante de Hubble $H_0$, qui vaut environ 76 km/s/Mpc \ $\footnote{1Mpc(Megaparsec.)=$3\cdot 10^{24}$cm}$.
 
 Pour determiner l'\'evolution cosmologique il faut compl\'eter les \'equations obtenues jusqu'\`a maintenant avec une \'equation d'\'etat, qui pour les fluides parfaits les plus populaires en cosmologie prend une forme tr\`es simple:

\be p=w\ \rho,\label{etat}\ee
avec $w$ une constante ind\'ependante de temps qui vaut, par exemple, $1/3$ pour un gas des particules relativistes ou pour la radiation \'electromagn\'etique et $0$ pour la mati\`ere non-relativiste$\footnote{Par exemple les \'etoiles et les galaxies, pour lesquelles la pression est n\'egligeable par rapport \`a la densit\'e d'\'energie.}$.

 \`A l'aide de l'\'equation d'\'etat (\ref{etat}) on peut int\'egrer l'\'equation de continuit\'e (\ref{cont})  pour obtenir:
 
 \be \rho=\rho_0 \left({a\over a_0}\right)^{-3(1+w)},\ee
o\`u l'indice $0$ indique les quantit\'es \'evalu\'ees \`a pr\'esent. Pour un univers domin\'e par la mati\`ere la densit\'e d'\'energie varie comme $\rho\propto a^{-3}$, ce qui refl\` ete la d\'ecroissance de la densit\'e volumique des particules dans un univers en expansion. Dans un univers domin\'e par la radiation la densit\'e d'\'energie varie plus rapidement, $\rho\propto a^{-4}$, \`a cause de la perte d'\'energie par effet Doppler.

Pour $\kappa=0$, l'\'equation (\ref{first}) permet alors d'obtenir rapidement la d\'ependance dans le temps du facteur d'\'echelle:  $\dot{a}^2\propto a^{-(1+3w)}\Rightarrow a(t)\propto t^{2/3}$, pour un univers domin\'e par la mati\`ere et $a(t)\propto t^{1/2}$, pour un univers domin\'e par la radiation.

On peut consid\'erer aussi le cas d'une constante cosmologique$\footnote{Ou \'energie du vide}$, $\Lambda$. Les \'equations d'Einstein avec une constante cosmologique s'\'ecrivent :

 \be G_{\mu\nu}=8\pi G T_{\mu\nu}-\Lambda g_{\mu\nu}\ee
 et peuvent \^etre vues comme les \'equations d'Einstein habituelles, mais avec un tenseur \'energie-impulsion pour le vide de la forme:
 
 \be T_{\mu\nu}^{(\mbox{vac})}=-{\Lambda\over 8 \pi G }g_{\mu\nu}.\ee
 Ceci a la forme d'un tenseur \'energie-impulsion pour un fluide parfait avec:
 
 \be \rho=-p= {\Lambda\over 8 \pi G }.\ee
 La constante cosmologique satisfait, donc,  une \'equation d'\'etat de la forme (\ref{etat}) avec $w=-1$. Ceci implique que la densit\'e d'\'energie, $\rho$, est ind\'ependante de $a$ et qu'un univers en expansion qui pr\'esente une \'energie du vide non nulle sera, apr\'es un certain temps, domin\'e par l'\'energie du vide, car les densit\'es d'\'energie pour la mati\`ere et la radiation d\'ecroissent avec l'expansion. Pour un univers domin\'e par l'\'energie du vide le facteur d'\'echelle a une \'evolution exponentielle:
 
 \be a(t)\propto \mbox{exp}(Ht).\ee
  
Il est utile d'introduire le param\`etre "densit\'e":

\be \Omega={8\pi G\over 3H^2}\rho={\rho\over\rho_{\tt{crit}}},\ee
avec \be\rho_{\tt{crit}}={3H^2\over 8\pi G},\ee appel\'ee densit\'e critique. La premi\`ere \'equation de Friedmann-Lema\^itre (\ref{first}) se r\'e\'ecrit alors:

\be \Omega-1={\kappa\over H^2a^2}.\label{plat}\ee

 La valeur de la courbure $\kappa$ est, donc, d\'etermin\'ee par la densit\'e d'\'energie:
 
 \ba &&\rho<\rho_{\tt{crit}} \longleftrightarrow \Omega<1\longleftrightarrow \kappa=-1 \longleftrightarrow \mbox{espace\ hyperbolique} \nonumber\\
 &&\rho=\rho_{\tt{crit}} \longleftrightarrow \Omega=1\longleftrightarrow \kappa=0\longleftrightarrow \mbox{espace\  plat} \nonumber\\
&& \rho>\rho_{\tt{crit}
} \longleftrightarrow \Omega>1\longleftrightarrow \kappa=1 \longleftrightarrow \mbox{sph\`ere}  \ea

 Les observations cosmologiques donnent les valeurs suivantes pour les param\`etres de densit\'e:\\
 
 Baryons: $\Omega_b\simeq 0.05$,\\
 
 Mati\`ere noire $\Omega_x\simeq 0.25$,\\
 
 Constante cosmologique $\Omega_\Lambda\simeq 0.7$,\\
 
 Photons: $\Omega_\gamma\simeq 5\cdot 10^{-5}$.\\

 Notre univers est domin\'e \`a pr\'esent par l'\'energie du vide, ce qui implique qu'il acc\`el\'ere, $\ddot{a}>0$. Mais en extrapolant dans le pass\'e on trouve que la radiation a \'et\'e dominante \`a une certaine \'epoque, car $\rho_\gamma\propto a^{-4}$ d\'ecro\^it plus vite que $\rho_{\mbox{\scriptsize mati\`ere}}$. \`A cette \'epoque l'univers d\'ecellerait, ce qui implique que, plus on remonte dans le temps, plus l'univers \'etait en expansion de plus en plus rapide et au moment $t=0$ on trouve une singularit\'e. Cette singularit\'e s'appelle le {\bf Big Bang}. 
 A l'approche de cette singularit\'e la relativite g\'en\'erale ne permet plus une description valide de la nature, car les effets de la gravit\'e quantique deviennent importantes. La th\'eorie des cordes pourrait offrir les outils necessaires pour comprendre les ph\'enom\`enes  qui se sont d\'eroul\'es dans cette r\'egion de l'espace-temps.

 
 \section{Les probl\`emes du mod\`ele du Big Bang}

 Le mod\`ele du Big Bang a remport\'e beaucoup des succe\`s, en particulier ses pr\'edictions sont en excellent accord avec les observations sur la temp\'erature du CMB(cosmic microwave background)$\footnote{fond diffus cosmologique}$ et le sc\'enario de la nucl\'eosynth\`ese est en accord avec les abondances des \'elements D, $^3\mbox{He}$, $^4\mbox{He}$, $^7\mbox{Li}$\cite{nucleosinthese}. Mais en plus du probl\`eme th\'eorique mentionn\'e plus haut, concernant la non validit\'e du traitement classique de la gravit\'e dans les r\'egions proches de la singularit\'e, le mod\`ele cosmologique standard \'echoue dans l'explication de plusieurs probl\`emes observationnels comme les anisotropies du CMB, la formation des structures et l'origine de la mati\`ere. Egalement le mod\`ele cosmologique standard est bas\'e sur des hypoth\`eses qu'il ne justifie pas, comme l'homog\'en\'eit\'e et l'isotropie de l'univers. Voici un apper\c{c}u des principaux probl\`emes laiss\'es sans r\'eponse par le mod\`ele cosmologique standard:
 
 \begin{enumerate}
\item Le probl\`eme de l'homog\'en\'eit\'e et de l'isotropie. \\
 
 Les observations confirment que l'univers est homog\`ene et isotrope avec une tr\`es bonne pr\'ecision. Comme les inhomog\'en\'eit\'es ont tendance \`a s'accentuer dans le temps, \`a cause de la gravitation, cela veut dire que dans le pass\'e elles \'etaient encore plus petites et rien n'explique pourquoi l'univers \'etait si homog\`ene dans le pass\'e.
 
 \item Le probl\`eme de la platitude.\\
 
 Selon les observations, $\Omega$ est tr\`es proche de 1($\kappa=0$). Dans ce cas la premi\`ere \'equation de Friedmann implique $H\sim 1/t$ et $a\sim t^{2\over 3(1+w)}$, donc $(aH)^{-2}\sim t^{2-{4\over 3(1+w)}}$. Pour la mati\`ere non relativiste($w=0$) et la radiation($w=1/3$) cela implique que $(aH)^{-2}$ cro\^it dans le temps. En m\^eme temps l'\'equation (\ref{plat}):
 
 \be \Omega-1={\kappa\over H^2a^2}\nonumber\ee

 implique que la valeur de $\Omega$ s'\'eloigne rapidement de 1(sauf si elle est exactement 1). Le fait que $\Omega$ soit si proche de 1 de nos jours demande un enorme ajustement fin de $\Omega$ pr\`es de 1 au d\'ebut de l'univers.
 
  \item Le probl\`eme de l'horizon. \\
    
   Les observations du CMB montrent qu' \`a l'\'epoque de la derni\`ere diffusion (last scattering) l'univers  \'etait quasi-homog\`ene. Ce qui pose probl\`eme est le fait que, selon le mod\`ele cosmologique standard, toutes les r\'egions de l'univers n'avait pas eu le temps d'entrer  en contact causal jusqu'\`a cette \'epoque, il n'y a, donc, aucune raison qu'elles aient, par exemple,  la m\^ eme temp\'erature.
   
   \item L'absence des d\'efauts topologiques.\\
   
   La densit\'e d'\'energie dans l'univers primordial \'etait assez \'elev\'ee pour produire en abondance des d\'efauts topologiques, pr\'edites par les th\'eories de grande unification, comme les monopoles, les murs des domaines (domain walls) ou les cordes cosmiques. Mais ces d\'efauts n'ont pas \'et\'e observ\'es jusqu'\`a pr\'esent.  En fait, la densit\'e pr\'edite est si grande qu'elle entra\^inerait le collapse de l'univers. 
   
   \item L'origine des anisotropies du CMB ne trouve pas d'explication dans la th\'eorie du Big Bang.

\end{enumerate}


 \section{Inflation}

   Certains des probl\`emes du mod\`ele du Big Bang trouvent une r\'eponse dans le sc\'enario  de l'inflation. L'inflation correspond \`a une p\'eriode pendant laquelle $(aH)^{-2}$ d\'ecro\^it, c'est \`a dire que le facteur d'\'echelle $a$ cro\^it plus vite que le rayon de l'horizon $H^{-1}$. Ceci permet de r\'esoudre le probl\`eme de la platitude. 
   
   Pendant l'inflation l'univers subit une phase d'expansion acc\'el\'er\'ee et, d'une certaine mani\`ere, on peut dire que l'expansion est plus rapide que la vitesse de la lumi\`ere. Les objets qui ont \'et\'e en contact causal peuvent \^etre separ\'es, \`a cause de l'inflation, par des distances plus grandes que le rayon de Hubble(plus grandes, donc, que leurs horizons respectifs). Ceci veut dire que des objets, qui, apr\`es l'inflation, semblent ne pas avoir \'et\'e en contact causal, ont pu \^etre en contact avant l'inflation, ce qui explique pourquoi elles ont les m\^emes propri\'et\'es et r\'esout le probl\`eme de l'horizon.
   
   Egalement l'expansion tr\`es rapide de l'univers pendant l'inflation implique une forte dilution des d\'efauts topologiques expliquant le fait qu'ils n'ont pas \'et\'e observ\'es. Ceci est \'egalement le cas pour les inhomog\'en\'eit\'es, ce qui explique la quasi homog\'en\'eit\'e de l'univers.   

 De la m\^eme mani\`ere la mati\`ere existant dans l'univers primordial serait dilu\'ee par l'inflation. La cr\'eation de la mati\`ere observ\'ee aujourd'hui et la thermalisation trouvent une explication dans les m\'echanismes de r\'echauffement (reheating) et pr\'echauffement (preheating) \cite{reheating}. 
 Ces m\'echanismes doivent \^etre contr\^ol\'es pour ne pas produire \`a leur tour des d\'efauts topologiques.  

 L'inflation pr\'edit des anisotropies dans le CMB avec les bonnes propri\'et\'es(adiabatiques et invariantes d'\'echelle\cite{lindebook}), mais pour obtenir la bonne amplitude des anisotropies des ajustements fins sont n\'ecessaires.  
 
 La condition pour avoir un r\'egime d'inflation est donn\'ee par:
 
 \be {d\over dt}{1\over a^2H^2}={d\over dt}{1\over \dot{a}^2}=-{2\ddot{a}\over \dot{a}^3}<0,\ee
 c'est \`a dire $\ddot{a}>0$ pour un univers en expansion($\dot{a}>0$), donc une expansion acc\'el\'er\'ee. L'\'equation de Friedmann (\ref{friedmann2}):
 
\be  {\ddot{a}\over a}=-{4\pi G\over 3}(\rho+3p),\ee
 montre qu'on peut obtenir un tel r\'egime dans un univers qui contient un fluide parfait avec pression n\'egative. La constante cosmologique peut \^etre alors un bon candidat, mais un univers domin\'e par une constante cosmologique sera domin\'e dans le futur par une constante cosmologique et ne pourra pas donner lieu \`a une \'epoque domin\'ee par la mati\`ere ou la radiation. L'univers sera en expansion exponentielle \'eternelle. 
 
 Le bon candidat s'av\`ere \^etre un champ scalaire, appel\'e {\it inflaton}, qui poss\`ede une \'energie potentielle:
 
 \be S=\int d^4x \sqrt{-g}\left[{1\over 2}\partial^\mu\phi\partial_\mu\phi -V(\phi)\right].\ee

  Le tenseur \'energie impulsion:
  
  \be T_{\mu\nu}=\partial_\mu\phi\partial_\nu\phi-g_{\mu\nu}\left({1\over 2}\partial^\mu\phi\partial_\mu\phi -V(\phi)\right),\ee 
  dans le cas d'un univers et mati\`ere homog\`enes, d\'ecrit un fluide parfait avec la densit\'e d'\'energie:
  
  \be \rho={1\over 2}\dot{\phi}^2+V(\phi)\ee
   et la pression:
   
   \be p={1\over 2}\dot{\phi}^2-V(\phi).\ee

    L'\'equation du mouvement de l'inflaton dans un univers FLRW$\footnote{Friedmann-Lama\^itre-Robertson-Walker}$ est donn\'ee par:
    
    \be \ddot{\phi}+3H\dot{\phi}+V'=0\label{infl1}\ee
    et les \'equations de Friedmann prennent la forme :
   
   \ba &&H^2={8\pi G\over 3}\left( {1\over 2}\dot{\phi}^2+V(\phi)\right),\nonumber\\   
   &&\dot{H}=-4\pi G\dot{\phi}^2, \label{infl2}\ea
  la derni\`ere \'equation \'etant redondante. 
   Ce syst\`eme conduit \`a une expansion acc\'el\'er\'ee dans l'approximation de {\it slow-roll}(descente lente), qui consiste \`a n\'egliger l'\'energie cin\'etique de l'inflaton devant son \'energie potentielle:
   
   \be \dot{\phi}^2\ll V(\phi),\ee
   ce qui implique $p\sim-\rho$. Une deuxi\`eme condition est : $\ddot{\phi}\ll V'$. 
   Les \'equations (\ref{infl1}) et (\ref{infl2}) deviennent alors:
   
   \ba H^2\simeq {8\pi G\over 3} V,\nonumber\\   
   3H\dot{\phi}+V'=0.\ea

   Les conditions de slow-roll sont g\'en\'eralement d\'ecrites par les param\`etres:

   \ba &&\epsilon={M_P^2\over 2}\left({V'\over V}\right)^2\ll1,\nonumber\\
   &&\eta=M_P^2{V''\over V}\ll1,\ea
    avec $M_P=1/\sqrt{8\pi G}$, la masse de Planck quadri-dimensionnelle. 
   
   L'expansion pendant l'inflation peut se m\'esurer en nombre de e-foldings:
   
   \be N=\mbox{ln}{a(t_f)\over a(t_i)},\ee
   qui est reli\'e au potentiel de l'inflaton par le fait que $\mbox{ln} a\sim \int H dt\sim \int H{d\phi/\dot{\phi}}$:
   
  \be N= {1\over M_P^2}\int {V\over V'}d\phi.\ee
   Pour r\'esoudre, par exemple, le probl\`eme de l'ajustement fin de $\Omega$, 70 e-foldings seraient n\'ecessaires.

   Le premier mod\`ele de type inflation a \'et\'e propos\'e par Alexei Starobinsky \cite{staro}, mais la motivation du mod\`ele n'\'etait pas la r\'esolution des probl\`emes de la th\'eorie du Big Bang. Alan Guth a propos\'e en  1981, un mod\`ele plus simple, qu'on appelle "old inflation"\cite{guth}. Dans ce mod\`ele l'univers subit une inflation exponentielle dans un \'etat de "faux" vide avec une large densit\'e d'\'energie. Le vrai vide aparra\^it sous forme de bulles qui entrent en collision. Le probl\`eme de ce sc\'enario est que pour r\'esoudre les difficult\'es du mod\`ele standard cosmologique  le taux de cr\'eation des bulles doit \^etre plut\^ot  faible, mais dans ce cas, comme l'espace est en expansion, les bulles ne s'unissent pas. Un sc\'enario avec plus de chances de reussite a \'et\'e propos\'e par A. Linde\cite{new}. Ce mod\`ele, appel\'e "new inflation", demande un potentiel tr\`es plat pr\`es de $\phi=0$, ce qui est assez artificiel, et n'arrive pas \`a expliquer tous les probl\`emes du Big Bang. Le mod\`ele de l'inflation chaotique\cite{chaotique} a montr\'e que l'inflation peut \^etre obtenue avec des potentiels de forme tr\`es simple. Par exemple avec un potentiel de la forme:
   
   \be V=\lambda M_P^{4-\alpha}\phi^\alpha,\ee
   si initialement la valeur l'inflaton est tr\`es grande, l'\'equation (\ref{infl2}) implique que $H$ est grand \'egalement, donc le terme de friction $3H\dot{\phi}$ est aussi tr\`es grand, ce qui permet que l'inflaton entre dans un r\'egime de slow-roll et l'inflation peut avoir lieu.  Le nombre des e-foldings est donn\'e par :

\be N={1\over M_P^2}\int_{\phi_f}^{\phi_i} {\phi\over \alpha}d\phi \sim {\phi_i^2\over 2\alpha M_P^2},\ee
donc initialement l'inflaton a une valeur $\phi_i \sim \sqrt{2\alpha}\sqrt{N}M_P\gg M_P$  et
  l'inflation s'arr\^ete lorsque $\epsilon=\alpha^2M_P^2/(2\phi^2)\sim 1$, c'est \`a dire $\phi\sim \alpha M_P$. A la fin de l'inflation le champ $\phi$ commence \`a osciller autour du minimum du potentiel $V(\phi)$ et perd son \'energie en cr\'eant des paires de particules \'el\'ementaires. C'est le m\'ecanisme de r\'echauffement. Une fois que les particules produites se sont thermalis\'ees l'univers peut \^etre d\'ecrit par le mod\`ele cosmologique standard.
  
   Apr\`es  l'inflation la taille de l'univers peut atteindre des valeurs beaucoup plus grandes que la taille de l'univers visible, $10^{28}$ cm. Si l'univers \'etait initialement form\'e de plusieures r\'egions avec des diff\'erentes valeurs pour $\phi$, les r\'egions o\`u la valeur de  l'inflaton n'est pas assez large ne vont pas subir d'inflation et \`a la fin de l'inflation l'univers sera form\'e d'immenses "\^iles" hommog\`enes, de taille beaucoup  plus grande que l'univers observable, qui apparaissent du chaos initial, d'o\`u le nom d'inflation chaotique. 
   
   Il existe \'egalement des mod\`eles d'inflation "hybride"\cite{hybride}, qui utilisent deux champs scalaires, dont un joue le r\^ole de l'inflaton et l'autre permet d'arr\^eter l'inflation.  La motivation de ce mod\`ele est d'\'eviter que l'inflation commence avec une valeur $\phi>M_P$.
   L'exemple le plus simple est le potentiel effectif :
   
   \be V(\sigma,\phi)={1\over 4\lambda}(M^2-\lambda \sigma^2)^2+{m^2\over 2}\phi^2+{g^2\over 2}\phi^2\sigma^2\ee
   
   Le champ $\sigma$ a une masse effective $m_\sigma^2=-M^2+g^2\phi^2$. Pour $\phi>\phi_c=M/g$ le potentiel pour $\sigma$ a un minimum \`a $\sigma=0$. Le champ $\sigma$ est pris au pi\`ege au minimum et le potentiel effectif pour l'inflaton, $\phi$ est:
   
   \be V=V_0+{1\over 2}m^2\phi^2,\ee
   o\`u $V_0={M^2\over 4\lambda}$. Ce potentiel permet que l'inflation a lieu si $m^2\gg M^4/2\lambda$.  Pendant l'inflation le champ $\phi$ d\'ecro\^it lentement jusqu'\`a la valeur $\phi_c$. Le potentiel pour $\sigma$ est alors modifi\'e et deux nouveaux minimas apparaissent en $\sigma=\pm \sqrt{(M^2-g^2\phi^2)/\lambda}$. Ceci modfie le potentiel de l'inflaton et l'inflation prend fin.\\
   
   Les anisotropies du CMB peuvent \^etre une source d'information sur le potentiel de l'inflation. En effet le fond diffus cosmologique n'est pas compl\`etement  isotrope. Les d\'eviations de l'isotropie sont de l'ordre de $10^{-5}$ et repr\'esentent un des outils observationnels les plus pr\'ecis en cosmologie. A l'origine de ces d\'eviations se trouvent les inhomog\'en\'eit\'es dans l'univers au moment de la derni\`ere diffusion. Les variations dans la densit\'e de mati\`ere ont g\'en\'er\'e des potentiels gravitationnels qui ont eu comme cons\'equence l'\'emission des photons avec de l\'eg\`eres variations de longueur d'onde, qui se traduisent dans les anisotropies de temp\'erature du CMB. Les inhomog\'en\'eit\'es initiales se sont accentu\'ees par instabilit\'e gravitationnelle et sont \`a l'origine des grandes structures observ\'ees aujourd'hui, \'etoiles, galaxies et clusters. Dans ce sens les anisotropies du CMB portent l'empreinte des conditions initiales qui ont g\'en\'er\'e la structure de l'univers. 
   
   L'origine des perturbations dans la densit\'e de mati\`ere est un point cl\'e de la cosmologie et l'inflation offre une solution: l'expansion acc\'el\'er\'ee peut convertir des fluctuations quantiques du vide en perturbations cosmologiques classiques. L'inflation dilue la mati\`ere initiale et l'univers se retrouve dans un \'etat du vide. Un \'etat du vide dans un univers en expansion a une temp\'erature non nulle, 
   la temp\'erature de Gibbons-Hawking 
   
   \be T_{\rm GH}={H\over 2\pi}\ee
   et les fluctuations de l'inflaton sont donn\'ees par $\delta \phi_k=T_{\rm GH}$ pour toutes les longueurs d'onde. Ces fluctuations sont reli\'ees \`a celles de la densit\'e par:
   
   \be \delta \rho={dV\over d\phi}\delta\phi.\ee
Le spectre des fluctuations scalaires est donn\'e par:

\be A_S^2(k)\sim {V^3\over M_P^6V'^2}\mid_{k=aH},\ee
o\`u $V^3/V'^2$ est \'evalu\'e au moment o\`u les perturbations sont gel\'ees, c'est \`a dire au moment o\`u la longueur d'onde physique devient \'egale au rayon de l'horizon, $\lambda=a/k=H^{-1}$. D'autre fluctuations possibles sont les fluctuations tensorielles de la m\'etrique:

\be A_T^2(k)\sim{V\over M_P^4}\mid_{k=aH}.\ee

L'amplitude des perturbations du CMB offre des informations sur l'\'echelle d'\'energie de l'inflation.
Si les anisotropies du CMB sont dues en grandes parties aux fluctuations tensorielles on peut d\'eduire rapidement $V_{\rm inflation}=(10^{16}GeV)^4$. Si les anisotropies du CMB sont de nature scalaire on en d\'eduit $V^{1/4}_{\rm inflation}=\epsilon^{1/4}10^{16}GeV$ et si $\epsilon$ n'est pas extr\^emement petit on peut conclure que $V^{1/4}_{\rm inflation}=10^{15}-10^{16}GeV$.

  \chapter{Cosmologie des cordes}
 

  \section{Inflation et alternatives \`a l'inflation en th\'eorie des cordes}
  
  La principale critique qui puisse \^etre formul\'ee \`a l'\'egard des mod\`eles d'inflation est le fait que l'inflaton est ajout\'e "ad hoc" avec l'espoir que ce sc\'enario pourrait trouver une justification dans un cadre plus fondamental. Les modules des cordes qui d\'ecrivent la g\'eom\'etrie des compactifications pourraient \^etre des possibles candidats pour l'inflaton. 
  Le potentiel de l'inflaton doit \^etre assez plat, mais pas compl\`etement. Dans les th\'eories des cordes supersym\'etriques il y a des directions plates, qui pourrait \^etre utiles pour l'inflation, apr\'es brisure de supersym\'etrie.
  
  {\bf Inflation des branes.}
  Il existe  en th\'eorie des cordes des mod\`eles qui essayent de g\'en\'erer l'inflation en utilisant des $D$ -branes. Dans ces mod\`eles la distance entre les $D$ -branes est identifi\'ee \`a l'inflaton. Si la configuration des branes est supersym\'etrique, il n'y a aucune force entre les branes, l'attraction gravitationnelle \'etant compens\'ee par la r\'epulsion  due aux charges de R-R, et l'inflaton n'a pas de potentiel. En revanche, la brisure de la supersym\'etrie peut g\'en\'erer un potentiel attractif de la forme utile pour l'inflation\cite{braneinflation}. Une possibilit\'e est d'utiliser des configurations des $D$ et $\bar{D}$ branes. Dans ce cas, les branes ayant des charges oppos\'ees, la force due aux charges R-R est attractive et s'ajoute \`a la force gravitationnelle. Le potentiel g\'en\'er\'e dans ce cas, apr\`es compactification \`a 4 dimensions,  a la forme d'un potentiel gravitationnel: 
  
  \be V = A-  {B\over z^{d_\perp-2}},\ee
  o\`u  $z$ est la distance entre les branes, associ\'ee \`a l'inflaton, $d_\perp$ est le nombre des dimensions transverses aux branes et $A$ et $B$ sont des coefficients reli\'es \`a la tensions des branes et au volume des  dimensions compactes parall\`eles aux branes.  Les conditions de slow-roll se traduisent par $z\gg r_\perp$, o\`u $r_\perp$ est le rayon des dimensions transverses compactes. Mais les branes ne peuvent pas \^etre s\'epar\'ees par une distance plus grande que la taille des dimensions compactes. Ce probl\`eme peut \^etre contourn\'e  dans certains mod\`eles\cite{KKLMMT}.
  
  Quand la distance entre les branes devient de l'ordre de l'\'echelle de longueur des cordes un mode tachyonique appar\^it dans le spectre des cordes ouvertes \cite{bantib}. Le potentiel du tachyon reproduit les potentiels utilis\'es dans l'inflation hybride. La brisure de sym\'etrie correspond dans ce cas \`a l'annihilation des branes. Le m\'ecanisme doit \^etre control\'e pour que, apr\`es annihilation, il reste des $D$- branes. L'annihilation est encore mal comprise et on ne sait pas comment le r\'echauffement se produit.

  Les cordes permettent aussi la formulation des mod\`eles qui pourraient \^etre des alternatives \`a l'inflation. 
  En th\'eorie des cordes l'action effective prend la forme:
  
  \be S_S={1\over 2\kappa^2_{10}}\int d^{10}x \sqrt{-g}e^{-2\phi}(R+4\partial^\mu\phi\partial_\mu\phi),\ee
  o\`u $\phi$ est le dilaton et $R$ le scalaire de Ricci. Cette action repr\'esente l'interaction graviton-dilaton, qui sont des \'etats de la corde ferm\'ee. Cette contribution vient de la sph\`ere ($\chi=2$), ce qui explique la d\'ependance dans la constante de couplage des cordes. Pour retrouver l'action de Einstein-Hilbert habituelle,
  \be S_E={1\over 16\pi G}\int d^{D}x \sqrt{-g}(R+...),\ee
  il faut effectuer la transformation suivante de la m\'etrique:
  
  \be g_{\mu\nu}^{(S)}=e^{{4\over D-2}\phi}g_{\mu\nu}^{(E)} ,\label{tr} \ee
  en dimension $D$. Les courbures scalaires sont reli\'ees alors par:
  
  \be R_S=e^{-{4\over D-2}\phi}\left(R_E-4{D-1\over D-2}\nabla^2\phi-4{D-1\over D-2}  \partial^\mu\phi\partial_\mu\phi\right).\ee

  L'action de Einstein-Hilbert devient:
  
  \be S_E={1\over 16\pi G}\int d^{D}x \sqrt{-g}\left(R-{4\over D-2}\partial^\mu\phi\partial_\mu\phi\right). \ee
   
   L'action de la relativit\'e g\'en\'erale, $\int \sqrt{-g}R$, est invariante sous la transformation $t\rightarrow -t$, qui se traduit pour le facteur d'\'echelle et la constante de Hubble par:
   
   \be a(t)\rightarrow a(-t), \quad H(t)\rightarrow -H(-t).\ee
   La pr\'esence du dilaton permet d'avoir une nouvelle sym\'etrie:
   
   \be a(t)\rightarrow 1/a(t), \quad \phi(t)\rightarrow \phi(t)-2(D-1)\mbox{ln}a(t).\ee

  La transformation $ a(t)\rightarrow 1/a(t)$ implique $H(t)\rightarrow -H(t)$.  Cette dualit\'e d'\'echelle, combin\'ee avec l'inversion du temps, rend l'espace des solutions plus riche que dans la cosmologie standard. Une solution en cosmologie des cordes contient quatre branches: $a(t), a(-t),a^{-1}(t),a^{-1}(-t)$. Deux branches d\'ecrivent des univers en expansion($H>0$) et les deux autres des univers en contraction($H<0$).  Entre autre une solution, $H(t)$, d\'ecrivant un univers en expansion avec une courbure d\'ecroissante, $\dot{H}(t)<0$, admet une solution duale, $\tilde{H}(-t)$, valable pour $t<0$, qui d\'ecrit un univers en expansion avec courbure croissante, $\dot{\tilde{H}}(-t)>0$(et qui dans le rep\`ere d'Einstein appara\^it comme un univers en contraction).

  Ceci est le point de d\'epart du mod\`ele du {\it pre-Big Bang}\cite{prebigbang}, qui suppose que le Big Bang n'est pas le d\'ebut de l'Univers, mais le passage entre une \'epoque pre-Big Bang et l'univers actuel, post-Big Bang. Dans ce sc\'enario l'univers commence dans un \'etat avec densit\'e d'\'energie et courbure faibles. A cause d'une instabilit\'e l'univers commence \`a se contracter et la courbure cro\^it. Avant d'atteindre la singularit\'e du "Big Crunch", l'univers "s'\'echappe"$\footnote{Le passage entre la contraction et l'expansion reste un ph\'enom\`ene sp\'eculatif, connu dans la litt\'erature sous le nom de "gracefully exit".}$,  de la phase de contraction et entre dans la phase d'expansion, le post-Big Bang.

  Ce sc\'enario n'est pas compl\`etement ind\'ependant de l'inflation, mais permet d'obtenir les avantages de l'inflation avec un \'episode inflationnaire situ\'e dans le pre-Big Bang et non pas apr\`es le Big Bang, l'avantage \'etant que la question de ce qui a pr\'ec\'ed\'e l'inflation dans le mod\`ele standard cosmologique est \'elimin\'ee. Un autre avantage du mod\`ele du pre-Big Bang est le fait que son point de d\'epart est la th\'eorie des cordes, qui pourrait \^etre la th\'eorie fondamentale des interactions, par rapport \`a l'inflation qui n'a pas encore trouv\'e de justification dans une th\'eorie plus fondamentale.   
  
  Le probl\`eme principal du mod\`ele du pre-Big Bang est la phase de transition entre les solutions pre- et post-Big Bang. Ce passage se passe \`a couplage fort et la th\'eorie des cordes n'est pas encore assez d\'evelopp\'ee pour ma\^itriser ce probl\`eme non-perturbatif. 
  
  Un autre sc\'enario qui s'inspire de la th\'eorie des cordes est le mod\`ele de l'univers {\it ekpyrotic}\cite{ekpyrotic}. Son point de d\'epart  est la th\'eorie de Horava-Witten\cite{hw} (M-th\'eorie h\'et\'erotique compactifi\'ee \`a 5 dimensions). L'univers quadri - dimensionnel a comme fronti\`eres deux $D3$ - branes, une avec tension positive et une avec tension n\'egative. Dans ce sc\'enario nous vivons sur la brane avec tension n\'egative, appel\'ee brane "visible" (l'autre brane s'appelle la brane "cach\'ee"). Une troisi\`eme brane se d\'eplace dans le bulk. L'\'etat initial est statique et proche du vide. La dynamique est 
  g\'en\'er\'ee par une faible brisure de la supersym\'etrie qui cr\'ee un potentiel atractif entre la brane visible et la brane dans le bulk. La collision de ces deux branes, appel\'ee "ekpyrosis", correspond au Big Bang. La brane du bulk est absorb\'ee dans la brane visible et une partie de son \'energie cin\'etique g\'en\'ere la mati\`ere sur la brane visible. Au moment de la collision les branes sont presque parall\`eles ce qui assure l'homog\'en\'eit\'e et l'isotropie de l'univers. Les fluctuations quantiques des branes peuvent \^etre la source des anisotropies du CMB. Un observateur sur la brane visible verrait l'\'epoque avant la collision comme un univers en contraction et  l'\'epoque d'apr\`es comme notre univers en expansion. Contrairement au pre-Big Bang ce sc\'enario n'utilise pas l'inflation, mais il reste incomplet.
    
  Un sc\'enario amelior\'e est {\it l'univers cyclique}\cite{cyclic} dans lequel le Big Bang correspond \`a la collision des branes visible et cach\'ee. Les branes sont suppos\'ees pouvoir passer l'une \`a travers l'autre et ensuite revenir pour une nouvelle collision et ainsi de suite. A la fin de chaque cycle, avant la collision, il y a une expansion g\'en\'er\'ee par une constante cosmologique qui prepare les conditions pour un nouveau cycle. D'une certaine mani\`ere ce sc\'enario fait appel lui aussi \`a l'inflation. 
  
  La collision des branes dans les mod\`eles de l'univers ekpyrotic et cyclique a lieu \`a couplage faible, n\'eanmoins le m\'ecanisme n'est pas encore compris et reste sp\'eculatif.
  
 
 \section{Orbifolds Lorentziens}
 
  Les orbifolds d\'ependants du temps permettent de mod\'eliser en th\'eorie des cordes la singularit\'e du Big Bang. Par exemple les mod\`eles ekpyrotic et cyclic utilisent des orbifolds d\'ependants du temps de la th\'eorie M. 
  
  La th\'eorie des cordes est caract\'eris\'ee par deux d\'eveloppements perturbatives:
  
  \begin{description}
\item[ $\bullet$] le d\'eveloppement en puissances du couplage $g_S=e^{<\Phi>}$, comme en th\'eorie des champs
\item[ $\bullet$] le d\'eveloppement en $\alpha'$ qui est typique aux cordes et disppara\^it dans la limite $l_S=(2\pi \alpha')^{-1/2}\rightarrow 0$, o\`u $l_S$ est la longueur des cordes
\end{description}

 Les termes d'ordre sup\'erieur en $\alpha'$ deviennent importants lorsque la courbure de l'espace-temps devient importante, comme dans la r\'egion de la singularit\'e du Big Bang.  Les orbifolds repr\'esentent des solutions exactes en $\alpha'$. Les singularit\'es des orbifolds euclidiens (les points fixes) sont "r\'esolues" en th\'eorie des cordes, dans le sens que la propagation des cordes est bien d\'efini\'e et les cordes voient une g\'eom\'etrie r\'eguli\`ere. On pourrait esp\'erer qu'il soit de m\^eme pour les singularit\'es des orbifolds lorentziens, mais ces mod\`eles sont plus compliqu\'es et encore mal compris.
 
  L'exemple de plus simple d'orbifold lorentzien est le quotient de l'espace de Minkowski $\mathbb{R}^{1,1}: ds^2=-dT^2+dX^2$, par le groupe $\mathbb{Z}$, engendr\'e par le boost: 

\be X^{\pm} \rightarrow e^{\pm 2 \pi \lambda}X^{\pm},\ee
o\`u $X^{\pm}=\frac{1}{\sqrt{2}}(T\pm X)$ sont les coordonn\'ees du c\^one de lumi\`ere. 

$\bullet$ Les r\'egions avec $X^+X^->0$ d\'ecrivent un espace de Milne. Avec le changement des coordonn\'ees 
$X^{\pm}=\frac{\tau}{\sqrt{2}}e^{\pm \lambda  x}$ la m\'etrique se r\'e\'ecrit: $ ds^2=-d\tau^2+\lambda ^2 \tau ^2 dx^2$, avec $x=x+2 \pi  $.  L'espace de Milne pr\'esente une singularit\'e de type espace \`a $\tau=0$ et la m\'etrique d\'epend du temps.

$\bullet$ Les r\'egions $X^+X^-<0$ d\'ecrivent des espaces de Rindler avec la m\'etrique $ ds^2=-\lambda ^2 x^2 d\tau^2+ dx^2$, avec $\tau=\tau+2 \pi $. Ces r\'egions contiennent des courbes ferm\'ees du genre temps. 

 La fonction de partition pour l'orbifold lorentzien est donn\'ee par:
 
 \be \mathcal{T}=\mbox{Tr}\rl\sum_{k=-\infty}^{\infty}g^k\rr \mathcal{T}_{\mbox{untw+tw}},\ee
 o\`u $\sum_{k=-\infty}^{\infty}g^k$ est le projecteur dans les \'etats invariants sous les boosts.
 Le spectre perturbatif de cet orbifold a \'et\'e trouv\'e en \cite{spectre}.
 
  En \cite{hp} un argument de r\'elativit\'e g\'en\'erale a \'et\'e expos\'e pour montrer que les orbifolds d\'ependants du temps sont instables: l'addition d'une particule causerait l'effondrement de tout l'espace-temps dans une singularit\'e. Une particule localis\'ee dans l'orbifold correspond \`a une infinit\'e des particules dans l'espace de Minkowski reli\'ees par des boosts. L'int\'eraction entre les images des particules peut cr\'eer un trou noir si le param\`etre d'impact, $b$, est plus petit que le rayon de Schwarzchild associ\'e \`a l'\'energie du centre de masse, $E$, de la particule et son $n$-i\`eme image :
  
  \be GE>b^{D-3},\ee
  avec $G$ la constante de Newton.
  Pour l'orbifold de l'espace de Minkowski bi-dimensionnel par un boost,  l'\'energie du centre de masse de la particule et son $n$-i\`eme image cro\^it comme $E\sim (\mbox{cosh}(2\pi\lambda n))^{1/2}$, et le param\`etre d'impact est ind\'ependant de $n$. La condition pour la formation d'un trou noir est satisfaite et pour $n$ grand le rayon de Schwarzchild est arbitrairement grande et occupe tout l'espace.
  
  In \cite{kutasov} des amplitudes de diffusion des cordes $2\rightarrow2$ au niveau des arbres ont \'et\'e calcul\'ees dans un espace-temps de Milne et il a \'et\'e montr\'e que ces amplitudes pr\'esentent des divergences associ\'ees \`a l'\'echange du graviton pr\'es de la singularit\'e cosmologique. Ces divergences peuvent \^etre \'evit\'ees par un ajustement fin des conditions initiales. 
  
  Le spectre des cordes dans l'espace-temps de Milne a \'et\'e obtenu en \cite{spectre}.
  
  
  \section{Solutions cosmologiques des cordes nonsupersym\'etriques}
  
\begin{figure}[!h]
\centering
\includegraphics[width=75mm]{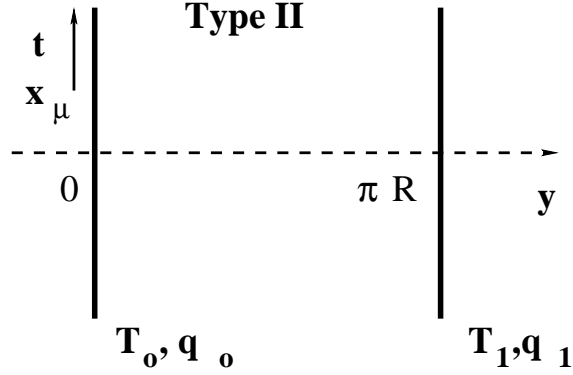}
\caption{Configuration des branes}
\label{model}
\end{figure}

   Les cordes nonsupersym\'etriques pr\'esentent un int\'er\^et particulier pour la cosmologie, car  l'espace de Minkowski n'est plus une solution du vide et on peut obtenir des solutions d\'ependantes du temps, donc, une \'evolution cosmologique. En \cite{ejc} nous avons trouv\'e des solutions d\'ependantes du temps pour un mod\`ele g\'en\'erique d'orientifold de la th\'eorie IIB, qui contient des $D8$- branes et $O8$ - planes.  Les branes sont plac\'ees, comme sur la  figure \ref{model}, \`a l'origine, $y=0$, et en $y=\pi R$, o\`u $y$ est une coordonn\'ee compacte, $y=y+2\pi R$, orthogonale aux branes, et avec la sym\'etrie $y\sim -y$.

    La somme des tensions(charges R-R) des objets localis\'es en $y=0$ est not\'ee $T_0$ ( $q_0$ ) et $T_1$ ( $q_1$ ) pour les objets en $y=\pi R$.  La condition d'annulation des tadpoles impose $q_0+q_1=0$, en revanche la somme des tensions peut \^etre non nulle. Comme la supersym\'etrie est bris\'ee nous incluons une constante cosmologique dans le bulk, $\Lambda_1$. Ce type de configuration peut \^etre obtenue, par exemple, dans le mod\`ele expos\'e \`a la fin de la section \ref{modele}, apr\`es une T-dualit\'e. 
    
      L'action effective est donn\'ee par:
      \ba
S &=& {1 \over 2{\kappa}^2} \int d^{10} x \sqrt {-G} \biggl[ e^{-2 \Phi} ( R+4 ({\partial \Phi})^2)- {1 \over 2 \times 10 !} F_{10}^2 -2 {\kappa}^2 \Lambda_1 
\biggr] \nonumber \\
&-& \int_{y=0} d^9 x (T_0 \ \sqrt{-\gamma} \ e^{-\Phi}
\ + q_0 \ A_9) - \int_{y= \pi R} d^9 x  (T_1 \sqrt{-\gamma} \ e^{-\Phi}
\ +q_1 A_9 ), \nonumber\\
 \label{t1b}
\ea
o\`u $A_9$ est la 9-forme de R-R qui couple aux $D8$ - branes et $O8$ - planes, $\gamma$ la m\'etrique induite et $\kappa^2=1/M_S^8$, avec $M_S$ l'\'echelle des cordes.   Les \'equations classiques n'admettent pas de solution avec la sym\'etrie $SO(9)$\cite{ej}.  Nous avons cherch\'e des solutions avec la plus grande sym\'etrie possible, c'est \`a dire $SO(8)$. Les solutions d\'ependent de la coordonn\'ee compacte $y$ et d'une autre coordonn\'ee qui peut \^etre le temps ou une autre coordonn\'ee d'espace. Nous nous int\'eressons ici \`a une des solutions d\'ependantes du temps qui pr\'esente une relation int\'eressante avec une solution supersym\'etrique. La forme g\'en\'erale d'une solution d\'ependante du temps est :
\ba
ds^2 &=& e^{2 A (t,y)} \left(1+{k x^2 \over
4}\right)^{-2} \delta_{\mu \nu} dx^{\mu} dx^{\nu} + 
e^{2 B (t,y)} (-dt^2 + dy^2) \nonumber \\
F_{10} &=& f(t,y) \ \epsilon_{10} \quad , \quad \Phi = \Phi (t,y) 
\ , \label{t2}  
\ea 
o\`u $\epsilon_{10}$ est la forme de volume en 10 dimensions et $k=-1,0,1$ correspond \`a un 8-hyperboloid, l'espace plat \`a 8 dimensions et la 8-sph\`ere pour la m\'etrique 8-dimensionelle \`a $t$ et $y$ fix\'es.  La 9-forme n'est pas dynamique et peut \^etre absorb\'ee dans la constante cosmologique:

\be
f=-q_0 \ {\kappa}^2 e^{5\Phi/2} \ \epsilon (y) \ , \label{t03}
\ee
avec $\epsilon(y)$ une fonction impaire p\'eriodique de $2\pi R$ et  $\epsilon(y)=1$ pour $y\in[0,\pi R]$.

On d\'efinit dans la suite la constante cosmologique effective:
\be
\Lambda_e \ = \ \Lambda_1+{q_0^2 \over 4} {\kappa}^2 \ .\label{t04}
\ee

  Deux  conditions n\'ecessaires pour avoir des solutions sont:
  
  \be T_0^2\ge q_0^2+4{\Lambda_1\over \kappa^2}, \quad T_1^2\ge q_1^2+ 4{\Lambda_1\over \kappa^2}.  \ee
  Pour $\Lambda_e>0$ les solutions des \'equations d'Einstein et du dilaton sont donn\'ees, dans le rep\`ere d'Einstein, par :
  
  \be
ds^2 =  \biggl[ G_0+ {3 {\kappa} \sqrt{\Lambda_e} \over \lambda} e^{
\lambda t} sh (\lambda |y|+\omega) \biggr]^{1 \over 12} \ \biggl[  
\delta_{\mu \nu} dx^{\mu} dx^{\nu} + e^{2 \lambda t} (-dt^2+dy^2) \biggr]
\ , \label{t5}   
\ee  

\be e^{\Phi} = \biggl[G_0 + {3 {\kappa} \sqrt{\Lambda_e} \over \lambda} e^{
\lambda t } sh (\lambda |y|+\omega) \biggr]^{-{5 \over 6}}\ee
  o\`u les param\`etres $\omega$ et $\lambda$ sont reli\'es aux tensions, \`a la charge et \`a la constante cosmologique:
  
  \be
ch(\omega)= -T_0 {\kappa} /(2\sqrt{\Lambda_e}) \quad , \quad
ch(\pi \lambda R+\omega)= T_1 {\kappa} / (2\sqrt{\Lambda_e}) \ , \label{t7} 
\ee
et $G_0$ est une constante d'int\'egration, choisie positive, pour \'eviter les singularit\'es. 

 Par le changement des coordonn\'ees :
 
 \be
T \ = {1 \over \lambda} \ e^{\lambda t} \ ch (\lambda y+\omega) \quad , \quad 
X \ = {1 \over \lambda} \ e^{\lambda t} \ sh (\lambda y+\omega)  \ , \label{t8} 
\ee
on obtient une m\'etrique ind\'ependante du temps:

\be
ds^2 =  
\biggl[ G_0+ 3{\kappa} \sqrt{\Lambda_e} X   \biggr]^{1 \over 12} \biggl[  
\delta_{\mu \nu} dx^{\mu} dx^{\nu} -dT^2+ dX^2 \biggr]
\ , \label{t9}   
\ee
valable pour $y>0$.  Cette solution a la m\^eme forme que la solution supersym\'etrique de Polchinski-Witten \cite{pw}. Comme on le verra dans la suite la d\'ependance dans le temps de notre solution initiale est report\'ee dans le nouveau rep\`ere sur les fronti\`eres. 

 Dans ce nouveau syst\`eme de coordonn\'ees la sym\'etrie $\mathbb{Z}_2$ devient une parit\'e, $\Pi_X$,  dans la coordonn\'ee $X$, multipli\'ee avec un boost de param\`etre $2\omega$, $K_{2\omega}$. La parit\'e initiale sur la surface d'univers , $\Omega'=\Omega\cdot \Pi_y$, devient $\Omega'' \ = \ \Omega \ \Pi_X \ K_{2 \omega}$. On peut v\'erifier facilement que $\Omega''^2=1$ car $\Pi_X\ K_{\theta}=K_{-\theta}\  \Pi_X$.  Par cons\'equent $\Omega''^2=\ ( \Omega \ \Pi_X \ K_{2 \omega})
\ (\Omega \ \Pi_X \ K_{2 \omega})=\ (\Omega \ \Pi_X \ K_{2 \omega})\ (K_{-2\omega}\  \Pi_X \ \Omega)=1$. 

L'identification sur la cercle $y=y+2\pi R$ se traduit dans les coordonn\'ees $(T,X)$ par un boost, $K_{2\pi \Lambda}$ avec la vitesse $v= th(2 \pi \lambda R)$ determin\'ee par les tensions, la charge et la constante cosmologique: 

\ba 
\left( 
\begin{array}{c} 
T \\ 
X 
\end{array} 
\right)  
& \ \rightarrow \ & 
\left(
\begin{array}{cc} 
ch (2 \pi \lambda R) & sh (2 \pi \lambda R)  \\ 
sh (2 \pi \lambda R)  & ch (2 \pi \lambda R) 
\end{array}
\right)
\left(
\begin{array}{c}
T \\
X
\end{array}
\right) \ , \label{t10}
\ea

  Les points fixes dans l'espace $(T,X)$ sont donn\'es par:
  
  \ba
\Omega'~: \  X &=& th \omega \ T \ ,\nonumber \\
\Omega'~ g : \  X &=& th (\pi \lambda R+ \omega ) \ T \ . \label{t010}
\ea
Les branes et orientifolds situes \`a l'origine se d\'eplacent avec la vitesse 
\be
v_0 = th \ \omega \ , \label{t12}
\ee
dans le fond statique (\ref{t9}), alors que les objets en $y=\pi R$ ont une vitesse
\be
v_1=th \ (\pi\lambda R+\omega) \ . \label{t13}
\ee

Les conditions aux bords 
(\ref{t7}) se r\'e\'ecrivent sous la forme: 
\be
T_0 \sqrt{1-v_0^2} + T_1 \sqrt{1-v_1^2} = 0 \ ,\label{t14} 
\ee
  qui r\'epresente une condition d'annulation des tadpoles de NS "boost\'ee". 
  
\begin{figure}[!h]
\centering
\includegraphics[width=95mm]{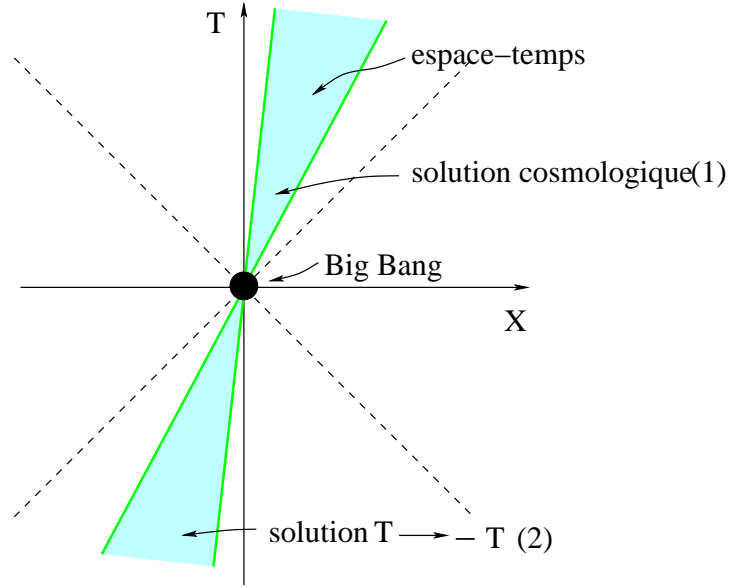}
\caption{Solution cosmologique}
\label{bigbang}
\end{figure}

  L'espace-temps obtenu est pr\'esent\'e dans la figure \ref{bigbang}.
       
  L'espace-temps est contenu entre les deux fronti\`eres en mouvement. A $T=0$ il y a une singularit\'e de type Big Bang et l'approche de th\'eorie effective n'est plus valable.  Nous avons repr\'esent\'e \'egalement la solution $T\rightarrow -T$. En supposant que les corrections des cordes permettraient une transition continue entre les deux solutions on obtient un univers du type pre-Big Bang. 
  
  Il est int\'eressant d'\'etudier la m\'etrique per\c{c}ue par un observateur sur les branes. La m\'etrique induite \`a l'origine est compl\`etement plate et un observateur situ\'e sur les branes en $X=0$ ne verrait pas la singularit\'e \`a $T=0$. Sur l'autre fronti\`ere l'obervateur per\c{c}oit un univers en expansion avec une singularit\'e dans la pass\'e:
  \be
ds^2 =  
\biggl[ G_0+ 3{\kappa} \sqrt{\Lambda_e} \tilde T   
\biggr]^{1 \over 12} \biggl[  
\delta_{\mu \nu} dx^{\mu} dx^{\nu} -d\tilde T^2+ dX^2 \biggr]
\ , \label{t779}   
\ee
 avec  $\tilde T=th(\pi \lambda R)T$. La singularit\'e \`a $\tilde T=-G_0/ 3{\kappa} \sqrt{\Lambda_e}$ est une illusion, car la m\'etrique perd sont validit\'e \`a $\tilde{T}=0$.   \\

  En \cite{ejc2} cette solution a \'et\'e compactifi\'ee \`a 4 dimensions. Nous avons choisi le cas $\omega=0$, c'est \`a dire $\Lambda_1=0$, ce qui est effectivement le cas dans le mod\`ele discut\'e \`a la fin de la section \ref{modele} dans la limite o\`u la distance entre les orientifolds est tr\`es grande. Dans ce cas les points fixes sont donn\'es par:
  
   \be
\Omega'~: \  X = 0\ ,\quad
\Omega'~ g : \  X= th (\pi \lambda R) \ T \ . \label{t010}
\ee
Les branes et orientifolds situ\'es \`a l'origine sont en repos, alors que les objets localis\'es en $y=\pi R$ se d\'eplacent avec la vitesse $v_1=th(\pi \lambda R)$. Les conditions aux bords (\ref{t7})   se r\'e\'ecrivent:
  
  \be T_0=q_0, \quad \ T_1 \sqrt{1-v_1^2}=q_1\ee
   et la condition (\ref{t14}) est alors \'equivalente \`a la condition d'annulation des charges de R-R. 

 Nous consid\`erons que les branes plac\'ees en $y=\pi R$ contiennent des champs de jauges, $F$:
 
 \be S=-\! \int_{y= \pi R} d^9 x \ \sqrt{-\gamma} [\ e^{-\Phi} (T_1+
 {1\over g^2} {\rm tr} F^2) + q_1 A_9 ], \ee
 avec $1/g^2\sim M_S^5$ reli\'e au couplage de Yang-Mills 9d. 
 
  La m\'etrique initiale dans la rep\`ere d'Einstein et le dilaton sont donn\'es par:
  
  \ba
ds_E^2 &=&  \biggl[ G_0+ {3 |q_0| {\kappa}^2 \over 2} |X| \biggr]^{1 \over 12} \ \biggl[  
\delta_{\mu \nu} dx^{\mu} dx^{\nu} + e^{2 \lambda t} (-dt^2+dy^2)
\biggr] \  \nonumber \\
e^{\Phi} &=&  \ \biggl[ G_0+ {3 |q_0| {\kappa}^2 \over 2 } |X| \biggr]^{- {5 \over 6}} \  \label{s03}
\ea
et la m\'etrique se r\'e\'ecrit dans le rep\`ere des cordes:
\be
ds^2 =  \biggl[ G_0+ {3 |q_0| {\kappa}^2 \over 2} |X| \biggr]^{-{1 \over 3}} \ \biggl[  
\delta_{\mu \nu} dx^{\mu} dx^{\nu} + e^{2 \lambda t} (-dt^2+dy^2) \biggr]
\ . \ee

 Nous consid\`erons que cinq dimensions parall\`eles aux branes et orientifolds sont compactifi\'ees sur un tore. On obtient des $D3$ - branes et $O3$ - planes qui se propagent dans un espace-temps cinq-dimensionnel. Les coordonn\'ees sont index\'ees par  
 $M = (\alpha, m)$, o\`u $\alpha = 0 \cdots 4$ sont les coordonn\'ees noncompactes plus la coordonn\'ee $X$ et $m = 5 \cdots 9$ sont les cinq coordonn\'ees parall\`eles aux $D8$ - branes et $O8$ - planes.
 
 La compactification de 10 \`a 5 dimensions implique:

\be
g_{mn}^{(10)} = V_5^{2 \over 5} \delta_{mn} \quad , \quad 
g_{\alpha \beta}^{(10)} =  g_{\alpha \beta}^{(5)} \ , \label{s12}
\ee  
o\`u $V_5 \equiv v_5 \exp (5 \sigma)$ est le volume du tore interne cinq-dimensionnel avec $v_5=r_c^5$ le param\`etre constant de volume et 
$\sigma$ est donn\'ee par:
\be
\langle V_5 \rangle \ = \ e^{5 <\sigma>} \ = \ v_5 \ 
\biggl[ G_0+ {3 |q_0| {\kappa}^2 \over 2 } |X| \biggr]^{-{5 \over 6}}  . \label{s13}
\ee 
La m\'etrique 5d prend la forme suivante, dans le rep\`ere des cordes:
\be
ds_{5}^2 =  \biggl[ G_0+ {3 |q_0| {\kappa}^2 \over 2 } |X|) \biggr]^{-{1 \over 3}} \ \biggl[  
\delta_{\mu \nu} dx^{\mu} dx^{\nu} -dX_0^2 + dX^2
\biggr] \ . \label{s14}  
\ee

 Dans le rep\`ere d'Einstein la m\'etrique et le volume 5d sont donn\'es par:
 
 \be
\langle V_{E,5} \rangle \ = \ v_5 \ 
\biggl[ G_0+ {3 |q_0| {\kappa}^2 \over 2 } |X| \biggr]^{5 \over 24} \ , \label{s16}
\ee 

\be
ds_{E,5}^2 =  \biggl[ G_0+ {3 |q_0| {\kappa}^2 \over 2 } |X| \biggr]^{2 \over 9} \ \biggl[  
\delta_{\mu \nu} dx^{\mu} dx^{\nu} -dX_0^2 + dX^2
\biggr] \ . \label{s117}  
\ee

Cette m\'etrique est une solution classique du lagrangien :

\ba
S_5 &=& {1 \over 2 {\kappa}_5^2} \int d^{5} x \sqrt {-G} \biggl[ 
R^{(5)} - {1 \over 2} ({\partial \Phi})^2 - {40 \over 3} ({\partial
\sigma})^2 - {1 \over 2 \times 5 !} \ e ^{- {5 \Phi \over 2} + {10
\sigma \over 3}} \ F_{5}^2 \biggr] \nonumber \\
&-& \int_{X=0} d^4 x \biggl[ \sqrt{-\gamma} \ 
 T_0 \ e^{ {5 \Phi \over 4} - {5 \sigma \over 3} } \ + q_0 \ A_4 
\biggr]  \nonumber \\
&-& \int_{X=v_1 T} d^4 x \biggl[ \sqrt{-\gamma} \ 
 (T_1 \ e^{ {5 \Phi \over 4} - {5 \sigma \over 3} }+ 
{v_5\over g^2} e^{{\Phi \over 4} + 5 \sigma } {\rm tr} F^2) \ + q_1 \ A_4 
\biggr]
\ , \label{s17}
\ea
o\`u  $\sigma=\tilde{\sigma} -\Phi/4$,  $(1/ {\kappa}_5^2) = v_5 M_s^8$ et $T_{0,1} \ (\ q_{0,1}\ )$ d\'enotent maintenant les tensions et charges des $D3$ - branes.

Si au cours de l'\'evolution temporelle l'espace interne 5d devient plus petit que l'\'echelle des cordes il faut effectuer des T-dualit\'es le long des coordonn\'ees du tore 5d. Apr\`es les T-dualit\'es le volume 5d devient $V_5'=1/(V_5M_S^{10})$ et le couplage des cordes est ${\rm exp}({\Phi'})={\rm exp}(\Phi)/(V_5M_S^5)$. Le lagrangien 5d qui d\'ecrit la solution apr\`es les T-dualit\'es est donn\'e par:

\ba
S_5 &=& {1 \over 2 {\kappa}_5^2} \int d^{5} x \sqrt {-G} \biggl[ 
R^{(5)} - {1 \over 2} ({\partial \Phi'})^2 - {40 \over 3} ({\partial
\sigma'})^2 - {1 \over 2 \times 5 !} \ e ^{ {40
\sigma' \over 3}} \ F_{5}^2 \biggr] \nonumber \\
&-& \int_{X=0} d^4 x \biggl[ \sqrt{-\gamma} \ 
 T_0 \ e^{ - {20 \sigma' \over 3} } \ + q_0 \ A_4 
\biggr]  \nonumber \\
&-& \int_{X=v_1 T} d^4 x \biggl[ \sqrt{-\gamma} \ 
 (T_1 \ e^{ - {20 \sigma' \over 3} }+ 
{v_5\over g^2} e^{-\Phi  } {\rm tr} F^2) \ + q_1 \ A_4 
\biggr],\label{tdual}
\ea

En d\'efinissant le temps propre, dans le rep\`ere des cordes, sur la brane avec vitesse constante $X=v_1 T$ par:

\be
\tau \ = \ {4 \sqrt{1-v_1^2} \over 5 v_1 |q_0| {\kappa}^2} \ ( G_0+ {3 |q_0|
{\kappa}^2 \over 2 } v_1 X_0)^{5 \over 6} \ , \label{h1}
\ee
on obtient un univers en contraction :
\ba
ds_4^2 &=&  ({{5 v_1 |q_0| {\kappa}^2 \over 4 \sqrt{1-v_1^2}} \ \tau})^{-{2
\over 5}} \delta_{\mu \nu} dx^{\mu} dx^{\nu} - d \tau^2 \ , \nonumber \\
R_c &=& ({{5 v_1 |q_0| {\kappa}^2 \over 4 \sqrt{1-v_1^2}} \ \tau})^{-{1 \over
5}} \ r_c \quad , \quad e^{\Phi} \ = \  \ ({{5 v_1 |q_0| {\kappa}^2 \over 4
\sqrt{1-v_1^2}} \ \tau})^{-1} \ . \label{h2}
\ea
avec un couplage de Yang-Mills independant du temps:
\be
{1 \over g_{YM}^2} \ = \ e^{- \Phi } \ (R_c M_s)^5 \
= \ {\rm v}_5  \ , \ \label{h3}
\ee
o\`u ${\rm v_5}$ d\'enote le volume 5d en unit\'es de l'\'echelle des cordes.
 Si $r_c\sim M_S^{-1}$ il faut effectuer des T-dualit\'es le long du tore 5d. La solution T-duale d\'ecrit un univers en expansion. Le couplage des cordes et d'ordre un, ${\rm exp}(\Phi')=1/{\rm v_5}$, et il d\'etermine \'egalement le couplage de Yang-Mills $1/g^2_{YM}\sim {\rm exp }(-\Phi')$.

 Dans le rep\`ere d'Einstein l'univers appara\^it en expansion :
 \be
ds_4^2 =  ({{5 v_1 |q_0| {\kappa}^2 \over 3 \sqrt{1-v_1^2}} \ \tau_E})^{1
\over 5} \delta_{\mu \nu} dx^{\mu} dx^{\nu} - d \tau_E^2\ee

Ceci d\'ecrit un univers FRW avec une \'equation d'\'etat et le param\` etre de Hubble$\footnote{Ce r\'esultat ne contredit pas la cosmologie 4d, car, en fait, il est obtenu \`a partir de la m\'etrique 5d (\ref{s117}) induite sur les branes.}$:
\be
p \ = \ {17 \over 3} \rho \quad , \quad H \ = \ {\dot a \over a} \ = 
\ {1 \over 10 \tau_E} \ . \label{h06}
\ee

 L'\'evolution cosmologique conduit \`a un univers avec trois dimensions spatiales larges, une dimension $X$ de l'ordre du mm et cinq dimensions compactes tr\`es petites et avec un \'evolution cosmologique tr\`es lente. Le couplage des cordes est de l'ordre de $10^{-15}$ et l'\'echelle des cordes vaut $M_S\sim 10^7GeV$. 
 
 En \cite{ec} nous avons \' etudi\'e la stabilisation de la coordonn\'ee noncompacte $X$, appel\'ee $y$ dans la suite, pour le lagrangien (\ref{tdual}). Dans ce but nous avons rajout\'e des potentiels sur les fronti\`eres:

\ba 
S_V= -\int_{y=0}d^4 x \sqrt{-\gamma}\ V_0 (\sigma, \Phi) - 
\int_{y=y_1} d^4 x \sqrt{-\gamma}\ V_1(\sigma, \Phi) \nonumber 
\ea

Ces potentiels peuvent \^etre g\'en\'eres par des effets nonperturbatifs. Par exemple, en prenant en compte l'anomalie de Weyl \`a une boucle, qui intervient lors de la transformation de Weyl qui effectue le passage entre le rep\`ere des cordes et celui d'Einstein, on obtient, apr\`es la condensation des jauginos, des potentiels de la forme:

\be V=\alpha\  e^{20\sigma\over 3}\ , \ {\rm avec} \  \alpha<<1\ee

Nous d\'effinisons les potentiels complets qui incluent ceux d\'ej\`a existants en (\ref{tdual}):
\be V_{i, {\rm tot}} \equiv V_i (\phi_a) \ + \ T_i \ e^{-{20 \sigma
\over 3}}\ee
 et nous cherchons des solution de la forme:
\ba 
ds_5^2 &=& e^{2A(y)} \ \eta_{\mu\nu}dx^{\mu}dx^{\nu} + e^{2B(y)} \
dy^2 \ , \nonumber \\
F_5 &=& \tilde{f}(y) \ \epsilon_{5} \quad , \quad \sigma=\sigma(y) \ , 
\quad \Phi = \Phi_0 = {\rm const.} \nonumber
\ea

 La solution obtenue:

\be 
ds_5^2 \quad = \quad e^{{2\over 3}z|y|} \ \eta_{\mu\nu}dx^\mu
dx^\nu+e^{{20 \over
3}z |y|+{40\over 3} C_\sigma} \ dy^2 \ee
est, en fait, \'equivalente, apr\`es un changement des coordonn\'ees \`a (\ref{s117}). 
On trouve \'egalement les conditions locales suivantes pour les potentiels:
\be
<V_{i, {\rm tot}}> \ = \ <V_{SUSY}>|_{y=y_i} \quad , \quad 
<{\partial V_{i, {\rm tot}} \over \partial \phi_a}> \ = \ <{\partial V_{SUSY}
  \over \partial \phi_a}>|_{y=y_i} \nonumber
\ee
o\`u $\phi_a = \sigma,\Phi$ et $V_{SUSY} = q_i \exp (-20 \sigma / 3)$ est le potentiel supersym\'etrique(BPS: $T_i=q_i$). Ces conditions imposent que les sources, pour les champs du bulk, donn\'ees par les branes non BPS, soient les m\^emes que pour les branes BPS.  

Nous avons \'etudi\'e deux examples pour lesquels 
 $V_0=0\  \ \& \ 
\ V_1 =\alpha_1 \ e^{\beta_1 \sigma} \ + \ 
\alpha_2 \ e^{\beta_2 \sigma} $.

(i) $\alpha_2=0$. Dans ce cas les conditons aux bords impliquent 
$$ \beta_1=-{20\over 3} \quad {\rm et} \quad \alpha_1=|q_0|-T_1 < 0 $$
\be 
V_{1,{\rm tot}} \ = \ T_1 e^{-{20\over 3}\sigma}+\alpha_1 e^{-{20\over
    3}\sigma} \ = \ q_1 \ e^{-{20\over 3}\sigma} \nonumber
\ee
On retrouve donc le potentiel supersym\'etrique et la coordonn\'ee $y$ n'est pas stabilis\'ee.

(ii) $\beta_1={20\over 3}\ , \ \beta_2=0$. Les conditions aux bords d\'eterminent les param\`etres du potentiel:

\be 
\alpha_1 \ <e^{{40\over 3}\sigma}>\ |_{y=y_1}=T_1-|q_0|>0 \quad  {\rm
and} \quad 
 \alpha_2=-2\sqrt{\alpha_1(T_1-|q_0|)}<0\label{alpha1} \nonumber
\ee

En tenant compte du fait que $\alpha_1\ll 1$, la premi\`ere condition implique $<e^{{40\over 3}\sigma}>$, c'est \`a dire que le volume de l'espace compact est tr\`es grand, ce qui permet de r\'ealiser effectivement les hierachies d\'ecrites en \cite{ejc2}. 
On obtient \'egalement une condition pour $y_1$:
\be
e^{{20 z y_1 \over 3}+ {40 C_{\sigma} \over 3}} = {T_1 - q_1 \over
  \alpha_1} >> 1 \nonumber
\ee
 Le potentiel total s'\'ecrit:

\be 
V_{1, {\rm tot}}=\rl \sqrt{T_1-|q_0|}e^{-{10\over
3}\sigma}-\sqrt{\alpha_1}e^{{10\over 3}\sigma}\rr^2+q_1e^{-{20\over
3}\sigma} \nonumber
\ee

 et en int\'egrant sur la coordonn\'ee $y$ on obtient un potentiel quadri-dimensionnel positivement d\'efini, comme dans les th\'eories de supergravit\'e sans \'echelle \cite{noscale}:
 
 \be
V_4 = \bigl(\sqrt{T_1-|q_0|} -\sqrt{\alpha_1} e^{{20\over 3}\sigma}
\bigr)^2  \nonumber 
\ee
 
  Le dilaton peut \^etre stabilis\'e par la suite en ajoutant des flux de R-R et NS-NS\cite{flux1}\cite{flux2}. En \cite{ec} nous avons montr\'e qu'il existe \'egalement des solutions de Sitter dans le voisinage des solutions Minkowski. 
  
   Une fois que le potentiel nonperturbatif appara\^it l'evolution temporelle s'arr\^ete r\'ealisant les hierarchies en \cite{ejc2} et tout en permettant que $y$ soit stabilis\'e \`a une valeur assez petite pour ne pas cr\'eer des probl\`emes ph\'enom\`enologiques comme la d\'eviation de la loi de Newton. Nous avons \'egalement construit en \cite{ec} des  mod\`eles chiraux en 6 et 4 dimensions, en utilisant des orbifolds $\mathbb{Z}_2$ et $\mathbb{Z}_2\times \mathbb{Z}_2$, pour lesquels tout les consid\'erations de cette section s'appliquent.

\section{Conclusion}

Les th\'eories des cordes avec brisure de supersym\'etrie sont un cadre naturel pour l'\'etude de la cosmologie. Nous avons construit des orbifolds non-tachyoniques de la th\'eorie des cordes avec brisure de supersym\'etrie \`a la Scherk-Schwarz en dimension 8, 6 et 4. Les mod\`eles en dimension 6 et 4 contiennent des fermions chiraux. Les th\'eories effectives de ces mod\`eles g\'en\'erent des solutions d\'ependantes du temps qui s'interpr\`etent comme des fronti\`eres en mouvement dans un 
 espace-temps statique soumis \`a une identification par un boost. Ces solutions peuvent \^etre compactifi\'ees \`a 4 dimensions et g\'en\'erent des couplages de jauge ind\'ependants du temps, au niveau des arbres, sur les $D_3$ - branes, qui sont les fronti\`eres. L'anomalie de Weyl \`a une boucle induit une d\'ependance logarithmique dans le temps de ces couplages et  
 apr\`es un temps exponentiellement long le syst\`eme entre dans un r\'egime non-perturbatif
 qui invalide notre solution. 
 Ce temps tr\`es long qui s'\'ecoule avnat le r\'egime non-perturbatif permet une \'evolution cosmologique qui g\'en\'ere des hierarchies entre l'\'echelle des cordes et la masse de Planck. 
 
 Nous avons \'etudi\'e le probl\`eme de stabilisation des modules. Dans ce but 
 nous avons ajout\'e des potentiels sur les branes et nous avons trouv\'e que pour stabiliser les modules internes ces potentiels  doivent reproduire sur les fronti\`eres les m\^emes sources que dans le cas supersym\'etrique. Apr\`es stabilisation la solution classique devient la m\^eme que dans le cas non-supersym\'etrique. Cette proc\'edure ne stabilise pas le dilaton, qui peut \^etre stabilis\'e, par la suite, en ajoutant des flux de R-R et NS-NS. Dans le cas des potentiels non-perturbatifs, induits par condensation de jauginos, on obtient un potentiel d\'efini positif, comme dans les th\'eories de supergravit\'e sans \'echelle. 

Un des r\'esultats principaux du travail effectu\'e pendant cette th\`ese a
\'et\'e de trouver les solutions du vide pour des mod\`eles de cordes avec
brisure de supersym\'etrie et tadpoles de NS, car dans ces cas l'espace-temps
de Minkowski n'est plus une solution. Comme certaines de ces solutions
pr\'esentent une d\'ependance dans le temps, elles s'av\`erent utiles pour
\'etudier des probl\`emes li\'es \`a la cosmologie. N\'eanmoins ces solutions
sont encore loin des mod\`eles cosmologiques r\'ealistes et des questions
comme leur stabilit\'e quantique et la quantification des cordes dans ces
fonds restent \`a \'etudier. D'autre part nous avons montr\'e qu'on peut
construire des mod\`eles de cordes avec des caract\'eristiques r\'ealistes,
brisure de supersym\'etrie, absence des tachyons et pr\'esence des fermions
chiraux en 4d, mais on est encore loin de la r\'ealisation du
Mod\`ele Standard comme limite de basse \'energie de la th\'eorie des
 cordes.

\end{document}